\documentclass[12pt]{article}
\pdfoutput=1

\newlength{\abstractwidth}
\usepackage{epsf}
\usepackage{color}
\usepackage{graphicx}

\usepackage{amssymb}
\usepackage{latexsym}

\renewcommand{\thefootnote}{\fnsymbol{footnote}}
\renewcommand{\thanks}[1]{\footnote{#1}}
\newcommand{\starttext}{
\setcounter{footnote}{0}
\renewcommand{\thefootnote}{\arabic{footnote}}}

\newcommand{\bea}{\begin{eqnarray}}
\newcommand{\eea}{\end{eqnarray}}
\newcommand{\be}{\begin{eqnarray}}
\newcommand{\ee}{\end{eqnarray}}
\newcommand{\<}{\langle}
\renewcommand{\>}{\rangle}

\def\a{\alpha}

\def\g{\gamma}
\def\G{\Gamma}
\def\e{\epsilon}

\def\tet{\vartheta}
\def\ep{\varepsilon}

\def\ev{\mathfrak{e}}
\def\mn{\mathfrak{n}}
\def\mN{\mathfrak{N}}
\def\mS{\mathfrak{S}}

\def\cA{{\cal A}}
\def\cB{{\cal B}}
\def\cC{{\cal C}}
\def\cD{{\cal D}}
\def\cE{{\cal E}}

\def\cL{{\cal L}}
\def\cM{{\cal M}}
\def\cN{{\cal N}}
\def\cO{{\cal O}}
\def\cP{{\cal P}}
\def\cQ{{\cal Q}}
\def\cR{{\cal R}}
\def\cS{{\cal S}}
\def\cT{{\cal T}}

\def\cV{{\cal V}}

\def\cZ{{\cal Z}}
\def\Z{{\cal Z}}
\def\M{{\cal M}}

\def\bC{{\bf C}}

\def\bZ{{\bf Z}}

\def\Re{{\rm Re \,}}
\def\Im{{\rm Im \,}}
\def\tr{{\rm tr}}

\def\det{{\rm det \,}}

\def\half{ {1\over 2}}
\def\p{\partial}

\def\no{\nonumber}
\def\sm{\smallskip}

\begin{document}
\starttext
\baselineskip=16pt
\setcounter{footnote}{0}

\begin{flushright}
2013 July 5
\end{flushright}

\bigskip

\begin{center}
{\Large\bf TWO-LOOP VACUUM ENERGY }

\vskip 0.15in 

{\Large \bf  For Calabi-Yau Orbifold Models}
\footnote{Research supported in part by National Science
Foundation grants PHY-07-57702 and DMS-12-66033.}

\bigskip\bigskip
{\large Eric D'Hoker$^a$ and Duong H. Phong$^b$}

\bigskip
$^a$ \sl Department of Physics and Astronomy \\
\sl University of California, Los Angeles, CA 90095, USA\\
$^b$ \sl Department of Mathematics\\
\sl Columbia University, New York, NY 10027, USA

\end{center}

\bigskip\bigskip

\begin{abstract}

A precise evaluation of the two-loop vacuum energy is provided for certain ${\bf Z}_2\times{\bf Z}_2$ 
Calabi-Yau orbifold models in the Heterotic string. The evaluation is based on the recent general prescription 
for superstring perturbation theory in terms of integration over cycles in supermoduli space, implemented 
at two-loops with the gauge-fixing methods based on the natural projection of supermoduli space onto 
moduli space using the super-period matrix.  It is shown that the contribution from the interior of supermoduli 
space (computed with the procedure that has been used in previous two-loop computations)
vanishes identically for both the $E_8\times E_8$ and $Spin (32)/{\bf Z}_2$ Heterotic strings. 
The contribution from the boundary of supermoduli space is also evaluated, and shown to 
vanish for the $E_8\times E_8$ string but not for the $Spin (32)/{\bf Z}_2$ string, thus breaking 
supersymmetry in this last model. As a byproduct, the vacuum energy in Type II superstrings 
is shown to vanish as well for these orbifolds.

\end{abstract}

\vfill\eject


\newpage

\baselineskip=15pt
\setcounter{equation}{0}
\setcounter{footnote}{0}

\section{Introduction}
\setcounter{equation}{0}

In theories with unbroken supersymmetry the vacuum energy vanishes
since contributions from bosons and fermions cancel one another  identically. When supersymmetry 
is broken, the masses of bosons and fermions are split and a non-zero net vacuum energy 
should be produced. It remains a major challenge to achieve mass splittings large enough 
to fit the Standard Model, accompanied by the production of a vacuum energy small 
enough to fit present cosmological data.

\sm

In superstring theory, mass splittings and vacuum energy should be calculable from first principles. 
Of special interest are  compactifications of Heterotic strings to 4 space-time dimensions in which 
$\cN=1$ supersymmetry is preserved at string tree-level. A large class of such 6-dimensional compactifications
can be constructed using the embedding of the spin connection  into the gauge group 
to guarantee anomaly cancellation \cite{Candelas:1985en}. These compactifications include Calabi-Yau 
manifolds in the large volume limit, and Calabi-Yau orbifolds of flat tori as vacuum solutions to string theory \cite{Dixon:1986jc}.

\sm

Non-renormalization theorems restrict the ways in which space-time supersymmetry can be 
broken by string loop corrections to the appearance of a Fayet-Iliopoulos $D$-term \cite{Dine:1987xk}. 
The Fayet-Iliopoulos (FI) mechanism \cite{Fayet:1974jb} exists only when the unbroken gauge group contains a 
commuting  $U(1)$-factor, in which case the $D$-term is properly gauge invariant. The standard embeddings 
of the  spin connection into the gauge group distinguish the fate of supersymmetry breaking in the two
Heterotic string theories, since we have,
\bea
Spin (32)/ \bZ_2 & \to & SU(3) \times U(1) \times SO(26)
\no \\
E_8 ~ \times ~ E_8 & \to & SU(3) \times ~E_6 ~\times ~ E_8
\eea
With gauge group $E_8 \times E_8$, no commuting $U(1)$ factor remains, 
no $D$-term can be generated, and supersymmetry  remains unbroken by loop corrections.
With gauge group $Spin (32)/ \bZ_2$, a commuting $U(1)$ factor remains, the FI 
mechanism is operative, and supersymmetry will be broken by loop corrections. One-loop corrections
to the $D$-term tadpole and to the masses of scalars (massless at tree-level) were evaluated in \cite{Dine:1987gj,Atick:1987gy} 
and found to be non-zero.  One-loop corrections to the vacuum energy, however, vanish for either case,
as contributions arise solely from the tree-level spectra of the theories, which are supersymmetric in both cases.
Two loops is the lowest order for which the vacuum energy is sensitive to perturbative supersymmetry breaking
in theories with tree-level supersymmetry \cite{Atick:1987qy}. 

\sm

Two loops is also the lowest order of superstring perturbation theory for which odd moduli start 
playing a non-trivial role \cite{D'Hoker:1988ta}, and where the global structure of supermoduli space 
must be taken into account when pairing left and right chiral amplitudes \cite{Witten:2012bh}. 
Witten's prescription for effecting this pairing in the 
case of Heterotic strings on  a genus $h\geq 2$ worldsheet may be briefly summarized as follows. 
Left chiral amplitudes depend on a supermoduli space $\mathfrak{M}_h$ of dimensions $(3-3|2h-2)$,
while right chiral amplitudes depend on a bosonic moduli space $\cM_{hR}$ of dimension $(3h-3|0)$. 
The pairing of left with right chiral amplitudes is realized by integrating their product over a cycle 
$\Gamma  \subset \mathfrak{M}_h \times \cM_{hR}$ of dimension $(3h-3|2h-2)$, subject to certain 
conditions at the Deligne-Mumford compactification divisor.\footnote{In terms of sets of
local coordinates $(m, \bar m, \zeta)$ for $\mathfrak{M}_h$,  and $(m_R, \bar m_R)$
for $\cM_{hR}$ respectively, with $m, \bar m, m_R, \bar m_R$ even and $\zeta $ odd moduli, 
a choice of $\Gamma$ may be realized locally by a relation $\bar m_R = m +$ terms even and nilpotent in $\zeta $.}
For genus $h \geq 5$, no holomorphic projection $\mathfrak{M}_h \to \cM_h$ exists \cite{Donagi:2013dua}, 
and no natural choice of $\Gamma$ is available, but a superspace generalization of 
Stokes's theorem \cite{Witten:2012ga,Witten:2012bh},  guarantees that the 
full amplitude is independent  of the choice of $\Gamma$. As explained in \cite{Witten:2012bh}, 
supersymmetric Ward identities have to be realized by integrals of closed forms over supermoduli space.

\sm

At two loops, however, there does exist a natural holomorphic projection of the interior of supermoduli 
space $\mathfrak{M}_2$ onto the interior of moduli space $\cM_2$ (more precisely onto spin moduli 
space $\cM _{2 ,{\rm spin}}$ for even spin structures). This projection may be realized concretely 
in terms of the genus 2 super-period matrix $\hat \Omega$, whose components may be used as locally 
supersymmetric even moduli. Odd moduli may then be naturally integrated out while keeping $\hat \Omega$ fixed.
The super-period matrix prescription was used for Type II and Heterotic theories in flat Minkowski space-time, 
where space-time supersymmetry is unbroken,
to compute the superstring measure for even spin structures \cite{D'Hoker:2001zp} (see \cite{D'Hoker:2002gw} 
for a survey). In turn, the measure  was used to evaluate scattering amplitudes of up to four 
massless NS bosons, and to prove various non-renormalization theorems \cite{D'Hoker:2005jc}. 

\sm

However, subtleties arise, even at genus 2, as one considers extending the natural projection via the 
super-period matrix, and the pairing of left and right chiralities, to the Deligne-Mumford 
compactification divisor of supermoduli space $\mathfrak{M}_2$ (often referred to 
as the boundary of supermoduli space). In particular, even if a natural holomorphic projection exists, 
it does not necessarily lead to a natural cycle $\Gamma$ that behaves as one would like at infinity \cite{Witten:2013cia}.
These subtleties appear to be inconsequential in  background space-times with unbroken supersymmetry, 
when the string spectrum is sufficiently simple (see section 3.2.5 of \cite{Witten:2013cia}).
But they do have physical implications, for example, when supersymmetry is broken. 
In particular, it was conjectured in \cite{Witten:2013cia} that the two-loop contribution to the vacuum energy 
from the {\sl interior (or bulk) of supermoduli space} vanishes for both Heterotic $Spin (32)/\bZ_2$ and 
$E_8 \times E_8$ theories for any compactification that preserves space-time supersymmetry at tree level. 
The totality of the two-loop vacuum energy then arises from the boundary of supermoduli space which was 
shown in \cite{Witten:2013cia} to be given by $\cV_G = 2 \pi g_s^2 \< V_D\>^2$,
where $\< V_D\>$ is the $D$-term tadpole vacuum expectation value, and $g_s$ the string coupling.

\sm

In the present paper, we shall compute these two-loop contributions from first principles, 
both from the interior, as well as from the boundary of supermoduli space, for the special case 
of $G=\bZ_2 \times \bZ_2$ Calabi-Yau orbifolds. These orbifolds are constructed so that the 
holonomy group $G$ of the spin connection is a subgroup of $SU(3)$,  and space-time 
supersymmetry is preserved at tree level.

\sm

We discuss now the contents of the present paper in somewhat greater detail. The models under consideration 
are orbifolds of $3$-dimensional complex tori, parametrized by complex coordinates $z^\gamma$ with $\g =1,2,3$.
Each non-trivial element of the orbifold group $G$ acts by a $\bZ_2$ twist of two of the coordinates $z^\gamma$,
leaving the third coordinate untwisted, and with corresponding twists also on the RNS worldsheet fermions. 

\sm

We consider genus 2 worldsheets $\Sigma$ with a fixed  homology basis, $A_I,B_I$, satisfying canonical intersection pairing 
$\#(A_I\cap A_J)=\#(B_I\cap B_J)=0$, $\#(A_I\cap B_J)=\delta_{IJ}$ for $I,J=1,2$.
Each $\bZ_2$ twist of a field $z^\g$ gives rise to $2^4$ sectors, collectively indexed by a half-characteristic $e^\g$
for $\g =1,2,3$.  Thus the twist sectors of the $G$ orbifold theory with all three fields $z^\g$ may be indexed by
vectors $\ev=(e^1,e^2,e^3)$ of three half-characteristics, satisfying the condition,
\bea
e^1+e^2+e^3=0 \quad (\rm mod\ 1)
\eea
required by the group relations in $G$. The sectors arrange into 6 irreducible orbits under the 
action of the modular group $Sp(4,\bZ)$ on the twists. 
The key novel orbit, which distinguishes the $\bZ_2\times \bZ_2$ model from $\bZ_2$ models as well as the two-loop 
order from the one-loop order, is the orbit $\cO_+$ to be defined in (\ref{1e1a}).

\sm

Our starting point is the gauge-fixed measure on supermoduli space obtained for general superstring 
compactifications in \cite{D'Hoker:2001nj}, and worked out explicitly for $\bZ_2$ orbifold compactifications in \cite{ADP}. 
The left chiral measure arises as a superfield in the odd moduli $\zeta ^1, \zeta ^2$, 
\bea
\label{2a3}
d\cA _L [\delta ; \ev] (p_{L \ev} ; \hat \Omega ,  \zeta) 
= \bigg ( d\mu^{(0)}_L  [\delta; \ev](p_{L \ev} ; \hat\Omega  ) 
+ \zeta ^1 \zeta ^2  d \mu _L [\delta ; \ev] (p_{L \ev}; \hat\Omega ) \bigg )  d\zeta ^1 d \zeta ^2
\eea
where $\delta$ denotes an even spin structure on the worldsheet, and $p_{L\ev}$ refers to the internal loop momenta.
For the Heterotic strings, the right chiral measure depends on internal loop momenta $p_{R\ev}$, 
on even moduli $\Omega_R$, but there are no right odd moduli.

\sm

As explained in  \cite{Witten:2013cia}, the bulk contribution to the vacuum energy arises from the top component 
$d \mu _L [\delta ; \ev] (p_{L \ev}; \hat\Omega )$ of the superfield, while the boundary contribution arises from 
a regularized limit  to the boundary of supermoduli space of the bottom component
$d\mu^{(0)}_L  [\delta; \ev](p_{L \ev}, \hat\Omega  )$, paired suitably with the contributions of the right sector. 
In both cases, the main difficulty will be to determine the contributions from twists $\ev$ in the above-mentioned orbit $\cO_+$.

\sm

A first main result of this paper is the vanishing, pointwise on supermoduli space, of the bulk contribution from each 
twist in the orbit $\cO_+$, upon summing over the spin structures $\delta$, as required for the GSO projection. 
An essential ingredient in this vanishing is the subtle correction term $\Gamma[\delta;\ep]$ uncovered in \cite{ADP}, 
which distinguishes between the Prym period matrix of the super Riemann surface, and the supersymmetric covariance 
matrix arising from the chiral splitting of the super matter fields. Taking this correction term into account, the bulk 
contribution for sectors in $\cO_+$ can then be shown to vanish by a seemingly new identity (\ref{F1}) 
between $\tet$-constants. Mathematically, this identity is more subtle than other more familiar identities, 
since it is only covariant with respect to a coset subgroup $Sp(4,\bZ)/\bZ_4$ rather than the full modular 
group $Sp(4,\bZ)$. Thus the identity cannot be established just from standard structure theorems for the ring of 
genus 2 modular forms and examining its behavior in the degeneration limit. 
It should be an interesting mathematical problem to  develop structure theorems 
for genus 2  modular forms with respect to natural modular subgroups such as $Sp(4,\bZ)/\bZ_4$,
and use such theorems to understand identities of the type proven in (\ref{F1}).

\sm

The second main result of the paper is the evaluation of the boundary contributions to the vacuum energy. 
Following the prescription of \cite{Witten:2013cia}, the left and right sectors are paired together not by setting 
$\hat \Omega=\Omega_R$ as would be done for the bulk contribution, but rather by a nilpotent regularization 
of the conditionally convergent integrals that arise  at the separating degenerating node. 
With this prescription, the contributions of twists in each orbit can be calculated using degeneration formulas for 
$\tet$-functions and some simplifications derived from earlier work in \cite{D'Hoker:2001qp}. 
In the approach of \cite{D'Hoker:2001nj}, 
the bottom component of the chiral measure has an intermediate dependence on the choice of slice in the gauge-fixing 
procedure. With the help of the regularization of \cite{Witten:2013cia} near the separating node, proper
gauge slice independence is restored for the full boundary contribution.

\sm

With this preparation, we find then that the boundary contributions to the vacuum energy vanish in all models for all 
twists not belonging to $\cO_+$. But for twists in $\cO_+$,  the structure of the GSO summation over right spin 
structures $\delta _R$ of the  internal fermion partition function factors $\overline{\tet[\delta_R]}^{4\mn}$ in the 
right sector differs for the two Heterotic theories. Since for the $E_8\times E_8$ string we have $\mn=1$, while for the 
$Spin(32)/\bZ_2$ string we have $\mn=3$,  we will find that the vacuum energy contribution vanishes for the 
$E_8\times E_8$ string, while it is strictly positive for the  $Spin(32)/\bZ_2$ string. Its value in the latter case 
will be evaluated explicitly.

\sm

As a byproduct, we confirm that contributions from the interior and from the boundary of supermoduli
space vanish for Type II strings compactified on the same $\bZ_2 \times \bZ_2$ Calabi-Yau orbifolds.

\subsection{Organization}

The paper is organized as follows. 
In Section 2, we describe the $\bZ_2 \times \bZ_2$ Calabi-Yau orbifold model,  the indexing of the orbifold 
sectors by vectors $\ev$ of twists, and the orbit structure of the vectors $\ev$ under the modular group. 
In Section 3, we calculate the partition function for each fixed spin structure. This begins with a description of the 
results of \cite{ADP} for $d\cA _L [\delta ; \ev]$, together with a description of the contributions of the matter fields, 
whether they are compactified, twisted, or a combination of both. The matter fields contributions are then worked 
out orbit by orbit.  The section concludes with the explicit evaluation of the contributions of the right sector for 
both $E_8\times E_8$ and $Spin(32)/\bZ_2$ Heterotic theories, which is straightforward. 
Sections 4 and 5 are devoted to the identification of the GSO phases and the summation over spin structures, 
for sectors in the orbits distinct from $\cO_+$ and sectors in $\cO_+$ respectively. For sectors in all orbits distinct 
from $\cO_+$, the contributions are found to vanish by the $\tet$-function identities already established in 
\cite{D'Hoker:2001nj} and \cite{ADP}. The contribution of sectors from the orbit $\cO_+$ is found to vanish by the 
new identity (\ref{F1}), the proof of which is the subject of Section~6. 
Sections 7 and 8 are devoted to the evaluation of the boundary contributions, again for sectors in the 
orbits distinct from $\cO_+$ and sectors in $\cO_+$ respectively. The former all vanish by genus $1$ Riemann 
identities, while the latter exhibits the different behavior explained above for the both $E_8\times E_8$ and 
$Spin(32)/\bZ_2$ Heterotic strings.

\sm

Some useful items have been gathered in the appendices for the convenience of the reader. 
Basic formulas for genus one $\tet$-functions are listed in Appendix A. 
Appendix B contains similar formulas for genus two $\tet$-functions, together with a detailed account of 
modular transformations acting on characteristics. In Appendix C, a new and simplified evaluation of the 
sign factor for the term $\Gamma[\delta;\ep]$ is provided in detail. This factor was not given correctly in \cite{ADP}; 
it ended up being immaterial there, but will play a crucial role in the present paper.  

\vskip 0.3in

\noindent
{\bf \large Acknowledgments} 

\medskip

We are very grateful to Edward Witten for suggesting the orbifold problem to us, for sharing his paper \cite{Witten:2013cia}
with us prior to publication, and for detailed and very helpful comments 
all along the project as well as on a draft of this paper. One of us (E.D.) 
is happy to acknowledge the generous hospitality of the Columbia University Department of Mathematics,
where part of this work was completed.

\newpage

\section{$\bZ_2 \times \bZ_2$ Calabi-Yau Orbifold Compactifications}
\setcounter{equation}{0}
\label{sec1}

We shall consider Heterotic and Type II superstring theories compactified on 
6-dimensional ${\bf Z}_2\times {\bf Z}_2$ Calabi-Yau orbifolds, 
as described for example in \cite{Antoniadis:1987,Donagi:2004ht}. Ten-dimensional space-time is 
of the form $M_4\times Y$ where $M_4$ is 4-dimensional Minkowski space-time
and the internal space $Y$ is an orbifold of a 6-dimensional torus, 
\bea
\label{1a1}
Y=(T_1\times T_2\times T_3)/G.
\eea
The orbifold group $G$ is isomorphic to $\bZ_2 \times \bZ_2$.
The complex tori $T_\g$ are given by $T_\g={\bf C}/\Lambda_\g$ for $\g =1,2,3$,
where the lattices  $\Lambda_\g$ are defined by,
\bea
\label{1a2}
\Lambda_\g=\{m_\g +n_\g  t_\g ~ \hbox{with} ~  m_\g ,n_\g \in {\bf Z}\}
\eea
for some fixed moduli $t_\g \in{\bf C}$ with $\Im t_\g >0$. Setting 
$\Lambda=\Lambda_1\times\Lambda_2\times\Lambda_3$, we may also view the torus as
$T_1 \times T_2 \times T_3 = \bC^3 /\Lambda$, and use local complex coordinates 
$(z^1, z^2, z^3)$.  

\sm

For the Heterotic strings with either gauge group $Spin(32)/Z_2$ or $E_8\times E_8$, 
the orbifold group  $G = \bZ_2 \times \bZ_2$ is chosen to be a subgroup of 
the $SU(3) \subset SO(6)$ acting on $T_1 \times T_2 \times T_3$. Thus, $Y$ is
a Calabi-Yau orbifold and $\cN=1$ supersymmetry is  preserved in the effective 
4-dimensional theory. The holonomy group of the spin connection is embedded into the gauge group
to assure proper anomaly cancellation. For Type II superstrings, compactification on the orbifold $Y$
will preserve $\cN=2$ supersymmetry in the effective 4-dimensional theory.

\subsection{Fields}

The worldsheet fields in the RNS formulation will be denoted as follows. Bosonic fields for Minkowski $M_4$ 
are real and denoted by $x^\mu$ with $\mu =0,1,2,3$, while those for the internal space $Y$ are complex fields 
$z^\gamma, z ^{\bar \gamma}$ with $\g, \bar \g =1,2,3$. The left chirality fermionic fields  are
similarly split into $\psi _+^\mu$ and $\psi ^\gamma, \psi  ^{\bar \gamma}$. For the Type II strings, 
the right chirality fermions are split into $\psi  _- ^\mu$ and $\tilde \psi ^\g, \tilde \psi ^{\bar \g}$.
For Heterotic strings, we shall use the fermionic representation of the internal degrees of freedom
in terms of 32 right chirality fermions $\psi _- ^A$ with $A=1,\cdots, 32$. Upon embedding the 
spin connection with holonomy group $G \subset SU(3)$ into the gauge group, we split also these
fermions in a manner natural to this $SU(3)$ action, into $\psi _- ^\alpha$ with $\alpha =1, \cdots , 26$
and $\xi ^\gamma, \xi ^{\bar \gamma}$ with $\g, \bar \g =1,2,3$. 
In summary, the fields $z^\g, \psi ^\g, \xi ^\g$ transform under 
a ${\bf 3}$ of $SU(3)$ while $z^{\bar \g}, \psi ^{\bar \g}, \xi ^{\bar \g}$ transform under the $\bar {\bf 3}$.

\sm

For gauge group $Spin(32)/\bZ_2$, an additional commuting $U(1)$ factor arises in the embedding 
$SU(3) \times U(1) \subset SO(6) \subset SO(32)$ under which the fields $\psi ^\g$ and $\xi ^\g$ have 
charge 1, while $\psi ^{\bar \g}$ and $\xi ^{\bar \g}$ have charge $-1$ for all $\g, \bar \g =1,2,3$.
The associated conserved $U(1)$ current is given by (repeated indices are summed), 
$J_z = \delta _{\g \bar \g } \, \psi ^\g \psi ^{\bar \g}$ and $J_{\bar z} = \delta _{\g \bar \g} \, \xi ^\g \xi ^{\bar \g}$.

\sm

The spin structure assignments in this orbifold model are as follows. The left chirality fermions 
$\psi ^\mu _+$ and $\psi ^\g, \psi ^{\bar \g}$ couple to the worldsheet gravitino field, so they all 
must have the  same spin structure, which we denote by  $\delta$. 
For the $Spin(32)/\bZ_2$ string, all 32 internal fermions $\psi _-^\alpha, \xi ^\g, \xi ^{\bar \g}$ 
have common spin structure $\delta_R$, which is summed over to carry out the GSO projection 
and guarantee  modular invariance. 
For the $E_8\times E_8$ string, the 32 fermions are grouped into two sets of 16 fermions each.
Within each set, the 16 fermions are assigned the same spin structure, $\delta_R^1$ for the first set, 
and $\delta_R^2$ for the second set. The summation over spin structures is performed over all
$\delta_R^1$ and $\delta_R^2$, independently of each other.

\sm

As was explained in the Introduction, the orbifold compactification breaks the gauge symmetries
of the Heterotic strings in different fashions. For the $Spin(32)/\bZ_2$ string, the gauge symmetry is broken to,
\bea
Spin (32)/\bZ_2 \to SU(3) \times U(1) \times SO(26) \to U(1) \times SO(26)
\eea
For the $E_8\times E_8$ string, the embedding of $G\subset SU(3)$ will be restricted to only one of the
$E_8$ factors, so that the  twisted fermions $\xi^\gamma,\xi^{\bar\gamma}$ with $\gamma,\bar\gamma=1,2,3$ 
belong to the first group of 16 right fermions. With this assignment, the gauge symmetry is broken to,
\bea
E_8\times E_8\to SU(3)\times E_6\times E_8\to E_6\times E_8
\eea
In both theories, the $SU(3) $ itself is broken by the spin connection when the orbifold group is $G = \bZ_2 \times \bZ_2$.
(When the orbifold group is the center $G=\bZ_3$ of $SU(3)$, however, an unbroken $SU(3)$ will remain as well;
this was the situation analyzed in \cite{Dine:1987gj,Atick:1987gy}.)

\subsection{Action of the orbifold group on the fields}

The action of the group $G= \{ 1, \lambda _1, \lambda _2, \lambda _3 \}$ 
on $z^\g$,  $\psi ^\g$, and $\xi^\g$ for $\g=1,2,3$ is given by,
\bea
\label{1c3}
\lambda _\beta \, z^{\gamma}  & = & (2 \delta _{\beta,\gamma} -1) z^\g
\no \\
\lambda _\beta \, \psi ^\gamma & = & (2 \delta _{\beta,\gamma} -1) \psi ^\g
\no \\
\lambda _\beta \, \xi ^\gamma & = & (2 \delta _{\beta,\gamma} -1) \xi^\g
\eea
and similarly for the fields $z^{\bar \g}, \psi ^{\bar \g}$ and $\xi^{\bar \g}$. The remaining fields,
$x^\mu$, $\psi ^\mu_+$  for $\mu =0,1,2,3$ and $\psi _- ^\alpha$ for 
$\alpha = 1,\cdots, 26$ are invariant under $G$.

\subsection{Twisted sectors in terms of characteristics}
\label{sec13}

In the previous section, we have described the action of the orbifold group $G$ on the fields. 
We now identify all the twisted sectors which arise in the orbifold theory, and we express each 
twist sector in terms of a vector $\ev$ of three characteristics $\ev=(e^1,e^2,e^3)$.
\footnote{The tori $T_\g $ will generically be inequivalent, as their moduli $t_\g$ 
will be different from one another. This justifies the notation of the triplet of twists as
a vector $\ev = (e^1, e^2, e^3)$ instead of as a set. 
We note, however, that permutations of the twists $e^\g$ in $\ev = (e^1, e^2, e^3)$ have a 
geometrical significance. They  form a discrete subgroup $\mS_3$ of the $SU(3)$ acting on the 
torus $T_1 \times T_2 \times T_3$. Any $SU(3)$-singlet, such as the $U(1)$-current 
for the Heterotic $Spin (32)/\bZ_2$ theory, will be invariant under $\mS_3$, and depend on $\ev$ only as  a set.
It is precisely such singlets that we shall be interested in when studying the vacuum energy in these theories.}


\begin{figure}[htb]
\begin{center}
\includegraphics[width=4in]{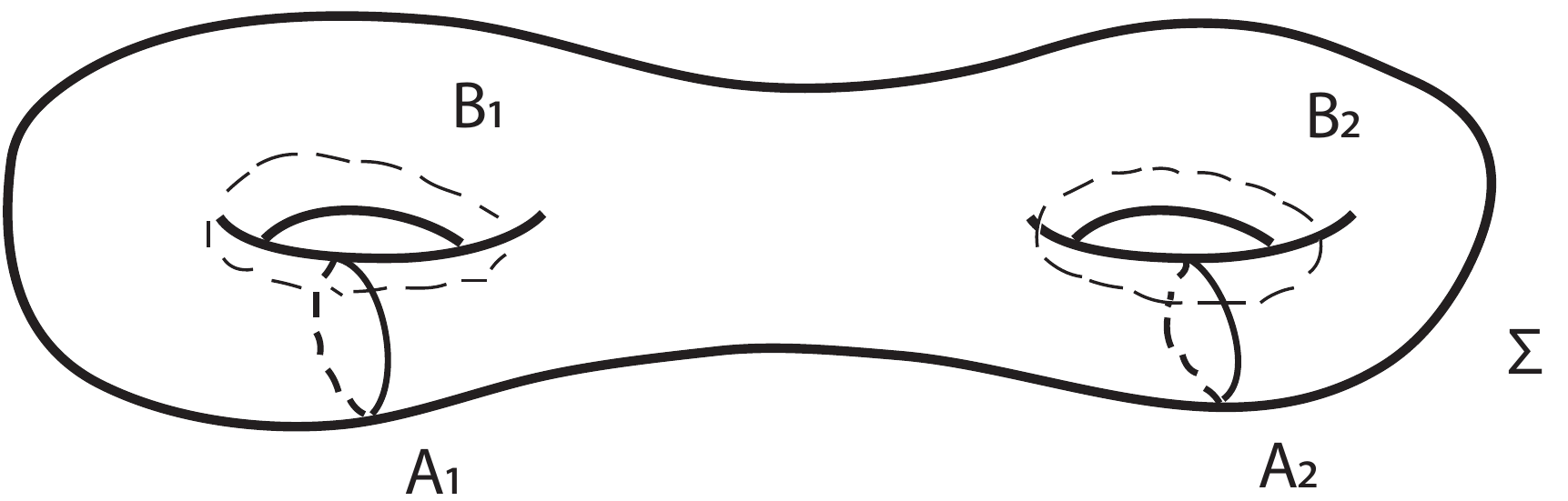}
\caption{Canonical homology basis for genus 2.}
\end{center}
\label{fig:1}
\end{figure}


We choose a basis of homology cycles $A_I, B_I$ in $H^1(\Sigma, \bZ)$ with canonical normalization 
$\# (A_I, A_J)= \# (B_I, B_J) =0$ and $\# (A_I, B_J ) = \delta _{IJ}$, as shown in Figure 1. 
For each value of $\g=\bar \g=1,2,3$, the fields $z^\g, z^{\bar \g}, \psi ^\g, \psi ^{\bar \g}$ and 
$\xi^\g, \xi ^{\bar \g}$  are twisted in the same manner by a single $\bZ_2$ twist. We shall label this twist by 
a genus 2 half-characteristic $e^\g$ using the standard notation with $(e^\g)'_I, (e^\g)''_I \in \{ 0, \half \}$ 
for $I=1,2$, 
\bea
\label{1d1}
e ^\g = \left (\matrix{(e^\g )'_1  \cr (e^\g )'_2 \cr} \bigg | \matrix{(e^\g )''_1 \cr (e^\g )'' _2 \cr} \right ) 
\eea
Similarly, we label the spin structures $\delta $ and  $\delta_R$ by half-characteristics $\delta =(\delta_I' | \delta_I'')$ and 
$\delta_R=((\delta_R)_I' | (\delta_R)_I'')$ respectively. Taking into account the spin structure 
assignments of the fermion fields,  the monodromy relations are as follows.  Around  $A_I$-cycles we have,
\bea
\label{1d3}
z^\g (w+ A_I) & = & (-)^{2 (e^\g )'_I } z^\g (w)
\no \\
\psi^\g (w+ A_I) & = & (-)^{2 (e^\g)'_I + 2 \delta _I'} \, \psi ^\g (w) 
\no \\
\xi ^\g (w+ A_I) & = & (-)^{2 (e^\g )'_I + 2 (\delta _R)_I'} \, \xi ^\g(w)
\eea
and for $B_I$-cycles,
\bea
\label{1d4}
z^\g (w+ B_I) & = & (-)^{2 (e^\g)''_I } z^\g (w)
\no \\
\psi ^\g (w+ B_I) & = & (-)^{2 (e^\g)''_I + 2 \delta _I''} \, \psi ^\g (w)
\no \\
\xi ^\g (w+ B_I) & = & (-)^{2 (e^\g)''_I + 2 (\delta _R)_I''} \, \xi ^\g (w)
\eea
The combined twist $\ev= (e^1, e^2, e^3)$ of all compactified fields, 
$z^\g, z^{\bar \g}, \psi ^\g, \psi ^{\bar \g}, \xi^\g, \xi ^{\bar \g}$
for $\g=1,2,3 $, represents a group element of $G=\bZ_2 \times \bZ_2$ 
provided we have the relation,
\bea
\label{1d2}
e^1 + e^2 + e^3 \equiv 0 \qquad (\hbox{mod} ~ 1)
\eea
and all twists by $G$ may be implemented  uniquely on $\Sigma$ in this manner.

\sm

A different, but equivalent, way of implementing the representation of $G$ on the fields is by working with 
the two group factors of $G=\bZ_2 \times \bZ_2$ separately. Let  $\lambda _1$ and $\lambda _2$ 
generate the first and second $\bZ_2$ factors respectively. All possible twist sectors may be 
labelled by the cycles $D_\eta$ and $D_\ep$ in $H^1 (\Sigma, Z_2)$ around which the twisting by 
$\lambda _1 $ and $\lambda_2$ is carried out. Each cycle $D_\ep$ (resp. $D_\eta$) may be represented  in terms
of a half-integer characteristic $\ep$ (resp. $\eta$),  
\bea
\label{1d5}
D_\ep = \sum _I \left ( 2 \ep ' _I A_I + 2 \ep ''_I B_I \right )
\eea
The action of the element $\lambda_3$ is now fixed, since
it is already realized as the product  $\lambda _3 = \lambda_1 \lambda _2$.
This representation coincides with the one given in (\ref{1d3})
and (\ref{1d4}), provided we identify, 
\bea
\ev = (e^1, e^2, e^3)  = ( \ep , \eta, \ep + \eta) \qquad (\hbox{mod} ~ 1)
\eea
a realization which automatically satisfies the condition (\ref{1d2}).

\subsection{Modular orbits of the twists}
\label{sec:MO}

The modular transformation properties of a single twist $\ep$ are listed in Appendix \ref{secB}.
To summarize:  of the 16 twists, one is invariant under modular 
transformations (corresponding to no twisting), while the remaining 15 transform in a single 
irreducible modular orbit, which we denote by $\cE$. The subgroup $H_\ep \subset Sp(4,\bZ)$ 
which leaves a non-zero twist $\ep$ invariant may be determined for any convenient twist, 
for example $\ep = \ep_2$. By inspection of table \ref{table:5}), we see that the group $H_{\ep_2}$ 
is generated by the elements $M_1, M_2, M_3, T_2 = \Sigma T \Sigma$, and $S_2 = S M_1 S M_1$, 
where $M_1, M_2, M_3, S, T, \Sigma$ are defined in (\ref{B12}). The groups $H_\ep$ for general $\ep$
may be obtained from $H_{\ep_2}$ by conjugation $H_\ep = M H_{\ep_2} M^{-1}$ by any 
modular transformation $M$ which maps the reference twist $\ep_2$ to the general twist $\ep = M \ep_2$.

\sm

We denote by $\cO_{\rm tot}$ the set of all possible twisted sectors of our full orbifold theory. 
Following the description of the previous section,  $\cO_{\rm tot}$ may be identified with the set of all 
triplets of half characteristics $\ev=(e^1,e^2,e^3)$ with $e^1+e^2+e^3\equiv 0$ (mod 1). 
The orbifold quantum field theory requires a summation over 
all sectors, and thus over all triplets of twists $\ev$. It will be convenient to organize this summation 
in terms of orbits which are irreducible under the action of the modular group $Sp(4,\bZ)$. 
To do so, it will be convenient to parametrize 
$\ev=(\ep,\eta,\ep+\eta)$  by a pair $(\ep,\eta)$ of independent twists, as was explained in the preceding section.
The set of all  pairs of twists $(\ep, \eta)$ may be distinguished by their transformation properties, 
and arranged into the following cases.
\begin{enumerate}
\itemsep -0.04in
\setcounter{enumi}{-1}
\item $\ep = \eta =0$ gives the untwisted sector;
\item  $\ep =0$ and $\eta \not=0$ is the sector twisted by $\lambda_1$;
\item  $\eta =0$ and $\ep \not=0$ is the sector twisted by $\lambda_2$;
\item  $\eta =\ep$ and $\ep \not=0$ is the sector twisted by $\lambda_3$;
\item  $0 \not= \ep \not = \eta \not= 0$.
\end{enumerate}
We  readily deduce the decomposition of $\cO_{\rm tot}$ into irreducible modular orbits.

\sm

$\bullet $ Case 0. above corresponds to the orbit $\cO_0$ with a single point, the zero twist,
\bea
\cO_0 = \{ (\ep, \eta), ~ \ep = \eta =0 \}
\eea
This is the untwisted sector, and it is invariant under the full modular group $Sp(4,\bZ)$. 

\sm

$\bullet$ Cases 1. 2. and 3. correspond to the irreducible orbits of a single $\bZ_2$ subgroup of $G$.
Each case is isomorphic to the non-zero irreducible orbit $\cE$ of a single twist, and we have,  
\bea
\cO_1 & = & \{ \ev = (0 , \ep, \ep), ~ \ep  \in \cE \} 
\no \\
\cO_2 & = & \{ \ev = (\ep, 0,  \ep), ~ \ep \in \cE \} 
\no \\
\cO_3 & = & \{ \ev = (\ep, \ep, 0), ~ \ep \in \cE \} 
\eea

\sm

$\bullet $ Case 4. above comprises two irreducible modular orbits $\cO_\pm$,
distinguished as follows,
\bea
\label{1e1a}
\cO _\pm = \{ \ev = (\ep , \eta, \ep+\eta), ~ \ep, \eta \in \cE, ~ \ep \not=\eta, ~   \< \ep | \eta \> = \pm 1 \}
\eea
where $\< \ep | \eta\>$ is the standard mod 2 symplectic invariant, which takes the form,
\bea
\label{1e1}
\< \ep | \eta \> = \exp \{ 4 \pi i ( \ep ' \cdot \eta '' - \eta ' \cdot \ep '' ) \}
\eea
The fact that $\cO_+$ and $\cO_-$ each forms a single irreducible modular orbit may be proven 
by explicit construction of the orbits. To do so, we again use the fact that for any twist $\ep$ there 
exists a modular transformation $M$ such that $\ep = M\ep_2$. It is now straightforward to  
distinguish the pairs $(\ep, \eta)$ that belong to $\cO_\pm$, and we find, 
\bea
\cO_+ & = & \left \{ \ev = M (\ep_2, \eta, \ep_2+\eta),  ~ \eta \in \{ \ep_3, \, \ep_4, \, \ep_7, \, \ep_8, \, \ep_{12}, \, \ep_{14} \},
~ M \in Sp(4, \bZ)  \right \}
\\
\cO_- & = & \left \{ \ev = M(\ep_2, \eta, \ep_2+\eta), ~ \eta \in \{ \ep_5, \, \ep_6, \, \ep_9, \, \ep_{10}, \, \ep_{11}, \, 
\ep_{13}, \, \ep_{15}, \, \ep_{16} \}, ~ M \in Sp(4, \bZ)  \right \}
\no \eea
By applying the subgroup $H_{\ep_2}$ respectively to the pairs, $(\ep _2, \ep _3) \in \cO_+ $, and 
$(\ep _2, \ep _5) \in \cO_- $, one verifies that each orbit $\cO_\pm$ transforms irreducibly under $Sp(4, \bZ)$. 
Note that these two cases have clear geometrical interpretations.
Given that $\ep =\ep_2$ corresponds to a twist around the cycle $B_2$, the case
$\eta \in \cO_+$ corresponds to a twist around the cycle $B_1$ for $\eta = \ep_3$,
while the case $\eta \in \cO_-$ corresponds to a twist around the cycle $A_2$ for $\eta = \ep_5$. 

\sm

$\bullet$
The union of all orbits equals $\cO_{{\rm tot}}$.  The cardinalities indeed check, 
using the fact that the cardinalities of $\cO_0, \cO_i, \cO_+, \cO_-$ are respectively
given by 1, 15 (with multiplicity 3), 90, and 120, adding up to $256=16^2$, as expected.

\newpage

\section{The Two-Loop Vacuum Energy}
\setcounter{equation}{0}
\label{sec2}

Following \cite{D'Hoker:2001nj,ADP}, the vacuum energy of a superstring compactification 
is built from the holomorphic blocks of the ghost and super ghost system as in flat space-time, 
and from the holomorphic blocks of the matter fields of the compactification. For orbifold 
models, the contributions from the matter fields  from all twisted sectors must be included. 
As shown in Section \ref{sec1}, the sectors are labeled the twists $\ev = (e^1, e^2, e^3)$, so that the sum
over all sectors can be viewed as the sum over the set $\cO_{\rm tot}$ of all 256 possible twists $\ev$.

\subsection{General structure of the two-loop vacuum energy}

Specializing the expressions of \cite{ADP} (which were written in all generality,
so as to include both symmetric and asymmetric orbifold constructions) to the case of the 
symmetric $G=\bZ_2 \times \bZ_2$ orbifolds discussed in Section \ref{sec1}, the vacuum energy 
$\cV_G$ takes the form,\footnote{Throughout, we shall choose units in which $\alpha '=2$,
unless otherwise stated.} 
\bea
\label{2a1}
\cV_G = g_s^2 \, \mN \, \int _{\Gamma}  \sum_{\ev \, \in \cO_{{\rm tot}} } \sum _{p_{L \ev}, p_{R \ev} } 
C_\delta [\ev] \, d\cA _L[\delta;\ev ](p_{L \ev} ; \hat\Omega,  \zeta ) \wedge 
\overline{d\mu_R [\ev ](p_{R \ev}; \Omega _R)}  
\eea
In particular, the pairing factor $K(\ev, p_{L \ev}, p_{R \ev}) $, which was needed for the general
formulation of \cite{ADP}, may be set to equal an overall normalization factor $\mN$, 
multiplied by the coupling $g_s^2$ for two loops, as was done  in (\ref{2a1}).
The notation for the remaining ingredients is as follows:

\sm

$\bullet$ The first sum, running over all twists $\ev\in\cO_{\rm tot}$, represents  the sum over all sectors of the orbifold theory. 
In view of the decomposition of $\cO_{\rm tot}$ into modular orbits  $\cO_\alpha$ in the previous section, the sum 
over $\cO_{{\rm tot}}$ may be recast in terms of a sum over  irreducible orbits, 
\bea
\sum_{\ev \, \in \cO_{\rm tot}}
\ =\
\sum_\alpha\sum_{\ev \, \in \cO_\alpha}
\eea
with the label $\alpha$ taking values in $\{0, 1,2,3,\pm\}$.

\smallskip

$\bullet$ The second sum runs over all  left and right momenta $(p_{L\ev},p_{R\ev})$ for fixed twist $\ev$.
Note that left and right-momenta are in general different since we compactify on a 6-dimensional torus. 
We shall describe their range  at the end of Section \ref{sec16}.

\smallskip

$\bullet$
Following \cite{Witten:2012bh} for the Heterotic string,  the integration is 
over an arbitrary  cycle $\Gamma \subset \mathfrak{M}_2 \times \cM_{2R}$  (subject to certain 
asymptotic and reality conditions). Here $\cM_{2R}$ is the moduli space of all Riemann surfaces of 
genus $2$ used for right chiral amplitudes, and $\mathfrak{M}_2$ is the supermoduli space of 
all super Riemann surfaces of genus $2$, used for left chiral amplitudes. 
The independence of the integral on the choice of cycle $\Gamma$ is guaranteed by a super Stokes theorem. 
The sum over spin structures is an integral part of the integration over $\mathfrak{M}_2$ and thus $\Gamma$. 
For general genus, the summation over spin structures cannot be separated in this process
from the integration over odd moduli. 

\smallskip

$\bullet$ 
In genus 2, we have a natural projection from $\mathfrak{M}_2$ onto 
$\cM_2$ provided by the super-period matrix \cite{D'Hoker:2002gw,D'Hoker:2001nj}. 
Thus, we may  parametrize $\mathfrak{M}_2$ by 
$(\hat\Omega,\zeta;\delta)$  where $\hat\Omega$ is the super-period matrix of the underlying super Riemann surface,
$\zeta$ the two odd moduli of genus 2, and $\delta$ the spin structure. 
The GSO phases $C_\delta [\ev]$ are determined  so as to guarantee modular invariance of the integrand. 
After integration over the odd moduli, the spin structures $\delta$ are summed according to the GSO projection.
Parametrizing  $\cM_{2R}$ 
by a period matrix $\Omega_R$, the choice of the cycle $\Gamma$ corresponds to the choice of a 
relation between $\hat\Omega$ and  $\Omega_R$. The general  form of such relations is dictated by complex 
conjugation, up to the addition of nilpotent terms bilinear in the odd moduli $\zeta$ \cite{Witten:2013cia},
\bea
\label{LR}
\hat\Omega=\Omega_R+ \cO (\zeta^1\zeta^2). 
\eea

\smallskip
$\bullet$ 
The bulk contribution of supermoduli space is obtained from the top component of $d\cA_L[\delta;\ev]$ 
in an expansion in the odd moduli $\zeta^1, \zeta^2$. For this contribution, the term $\cO(\zeta^1\zeta^2)$ 
in (\ref{LR}) is immaterial, and the natural choice is to set $\hat\Omega=\Omega_R$.
But, as was shown in \cite{Witten:2013cia}, for the boundary contribution of supermoduli space 
a regularization of conditionally convergent integrals may produce a non-vanishing contribution from the bottom 
component of $d\cA_L[\delta;\ev]$, and the term $\cO(\zeta^1\zeta^2)$ does matter. 
More specifically, if $\hat\Omega$ is viewed as the super-period matrix of a supergeometry $(g_{mn},\chi_{\bar z}{}^+)$, 
and $\Omega$ is the period matrix of the metric $g_{mn}$, then the correct relation (\ref{LR}) for the boundary contributions 
amounts essentially to a regularized version of setting $\Omega=\Omega_R$ near the boundary of supermoduli space. 

\smallskip

$\bullet$ The left block $d\cA _L[\delta;\ev](p_{L \ev} ; \hat \Omega, \zeta) $ depends on the left spin 
structure $\delta$, the  super-period matrix $\hat \Omega$ and the odd moduli $\zeta $. The right block 
$d \mu _R [\ev] (p_{R \ev}; \Omega _R)$ depends on moduli  $\Omega _R$.

\sm

Concretely, we begin by  making explicit the dependence on odd moduli $\zeta$,
\bea
\label{2a3a}
d\cA _L [\delta ; \ev] (p_{L \ev} ; \hat \Omega ,  \zeta) 
= \bigg ( d\mu^{(0)}_L  [\delta; \ev](p_{L \ev}, \hat\Omega  ) 
+ \zeta ^1 \zeta ^2  d \mu _L [\delta ; \ev] (p_{L \ev}; \hat\Omega ) \bigg )  d\zeta ^1 d \zeta ^2
\eea
Carrying out the integration over $\zeta$ will then produce the following contributions,
\bea
\cV_G = \cV_G ^{\rm bdy} + \cV_G ^{\rm bulk}
\eea
where the term $\cV^{\rm bulk}_G$ refers to the contribution from the bulk of supermoduli space,
while $\cV_G^{\rm bdy}$ refers to the contributions from conditionally convergent integrals
arising from the boundary of supermoduli space. The bulk term is the contribution of the top component 
$d\mu_L[\delta;\ev]$ in which we may set $\hat \Omega=\Omega_R\equiv\Omega$, as was explained earlier.
Thus the bulk term is given by,\footnote{The holomorphic volume form 
$d^3 \Omega = d\Omega _{11} \, d\Omega _{12} \, d \Omega _{22}$  on $\cM_2$ is included in both measures 
$d \mu_L$ and $d \mu _R$. } 
\bea
\cV_G^{\rm bulk} = g_s ^2 \mN \, \int _{\cM_2}  \sum_{\ev \, \in \cO_{{\rm tot}} } 
\sum _{p_{L \ev},p_{R \ev}} 
\sum _{\delta } C_\delta [\ev] \, d\mu _L[\delta  ; \ev ](p_{L \ev} ; \Omega  ) \wedge  
\overline{d\mu_R [\ev ](p_{R \ev}; \Omega )}.  
\eea
The boundary contribution is from the term $d\mu^{(0)}_L[\delta;\ev]$ and will be schematically denoted  by,
\bea
\label{2t1}
\cV_G ^{\rm bdy} = 
g_s ^2 \mN \, \int _{ \p \Gamma}  \sum_{\ev \, \in \cO_{{\rm tot}} } \, 
\sum _{p_{L \ev},p_{R \ev}}  \, 
C_\delta [\ev] \, d\mu^{(0)} _L[\delta  ; \ev](p_{L \ev} ; \hat\Omega ) d \zeta ^1 d \zeta ^2 \wedge 
\overline{d\mu_R [\ev ](p_{R \ev}; \Omega _R)}  
\eea
with the understanding that the regularization procedure of \cite{Witten:2012bh} must be used
to parametrize and relate $\hat\Omega, \zeta $, and $\Omega _R$ at the boundary $\p \Gamma$ 
of the cycle $\Gamma$.  It will be seen that, 
with the proper choice of cycle $\Gamma$ and after the regularization procedure, 
the term $\cV_G^{\rm bdy}$ reduces to an integral over the separating node divisor
part of the boundary of moduli space.

\subsection{Internal loop momenta}
\label{sec16}

The range of the internal loop momenta $(p_{L\ev}, p_{R\ev})$ depends on whether the 
corresponding fields are uncompactified, compactified but untwisted, or compactified with a non-zero twist.

\medskip

For each value of $\g=1,2,3$, the fields $z^\g, z^{\bar \g}, \psi ^\g, \psi ^{\bar \g}, \xi^\g, \xi ^{\bar \g}$
are untwisted when $e^\g=0$, and twisted by a common $\bZ_2$ when $e^\g \not= 0$.
A field subject to a $\bZ_2$ twist $\ep = e^\g$ may be viewed as defined on the surface $\Sigma$ 
with a quadratic branch cut along a cycle $C_\ep$, as represented in Figure 2. 
The $\bZ_2$-twisted field is then double-valued around the conjugate cycle $D_\ep$,
defined earlier in (\ref{1d5}). The remaining two cycles $A_\ep, B_\ep$ are defined so that 
$\# (A_\ep , B_\ep ) = \# (C_\ep, D_\ep )=1$  with all other intersection numbers vanishing.


\begin{figure}[tbph]
\begin{center}
\includegraphics[width=4in]{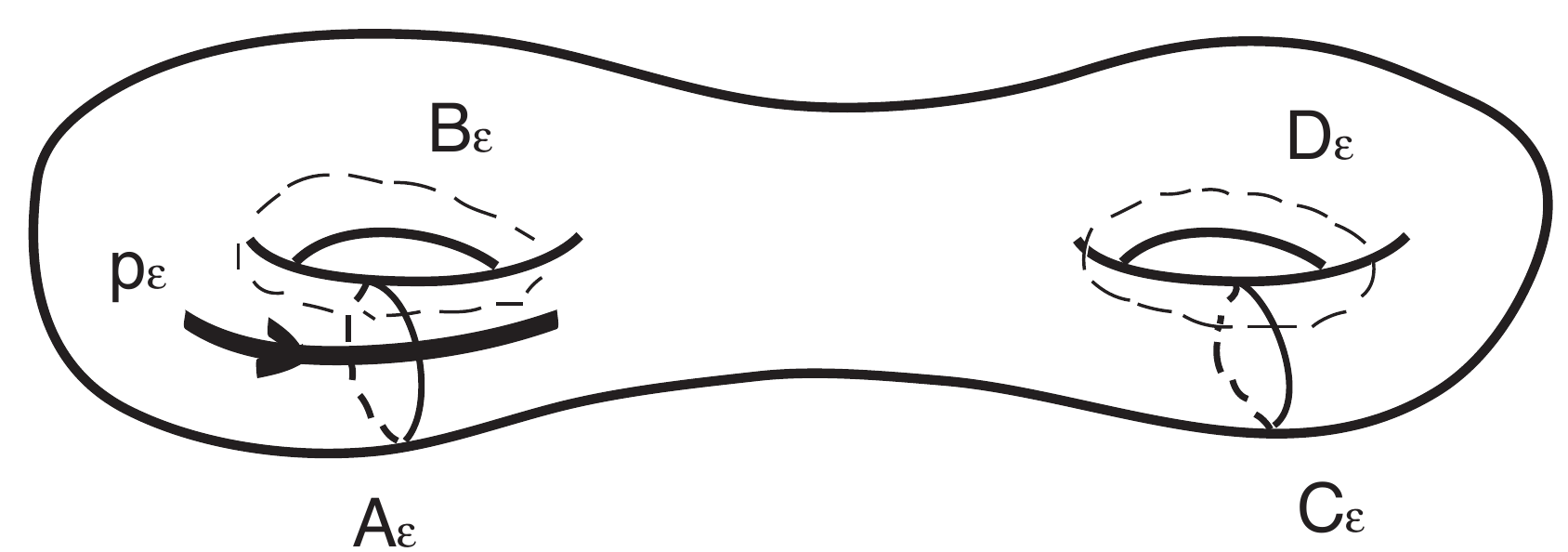}
\caption{Canonical cycles and internal loop momenta $p_\ep=(p_\ep^\g, p_\ep ^{\bar \g})$ for twist 
$\ep = e^\g \not= 0$.}
\end{center}
\label{fig:2}
\end{figure}


$\bullet$ 
In the uncompactified dimensions, $\mu =0,1,2,3$, left and right internal loop momenta are equal,
and denoted by $p^\mu_I$. They may be viewed as traversing cycles $A_I$ in Figure 1. 

$\bullet$
In the compactified dimensions, we distinguish twisted from untwisted directions
(this distinction will depend on the modular orbit of the twist).
For the untwisted fields $z^\g, z ^{\bar \g}$, we have internal loop momenta 
$(p^\g _{LI}, p^{\bar \g} _{LI})$ and $(p^\g _{RI}, p^{\bar \g} _{RI})$ with $I=1,2$, which may be 
viewed as traversing cycles $A_I$ in Figure 1. They correspond to the torus compactification,
and may be parametrized by the lattice $\Lambda$ of (\ref{1a2}), and its dual $\Lambda ^*$.
The result is standard \cite{Narain:1986am},
\bea
\label{1g1}
p_{LI}  = k_I - \half \ell_I
\hskip 1in 
p_{RI}  = k_I + \half \ell_I
\eea
with the vectors $k_I \in \Lambda^*$ and $\ell_I \in \Lambda$ for $I=1,2$, in units where $\alpha '=2$.

\sm

$\bullet$
For the twisted  fields $z^\g, z ^{\bar \g}$, the $\bZ_2$ twist $\ep =e^\g \not= 0$ would reverse the sign 
of internal loop momenta crossing the cycle $C_\ep$, so that such loop momenta must vanish. 
The remaining loop momenta  $(p_{L \ep}^\g, p_{L \ep} ^{\bar \g}$) and 
$(p_{R \ep}^\g, p_{R \ep} ^{\bar \g}$)  are across the cycle $A_\ep$. The range of the momenta
is again dictated by torus compactification, and we have, 
\bea
\label{1g2}
p_{L \ep}  = k_\ep - \half \ell_\ep
\hskip 1in 
p_{R \ep}  = k_\ep + \half \ell_\ep
\eea
with the vectors $k_\ep \in \Lambda^*$ and $\ell_\ep \in \Lambda$.

\subsection{The top component $d\mu_L[\delta;\ev]$ in the left chiral measure}
\label{sec24}

The left chiral factor $d\cA_L[\delta;\ev]$ can be derived  from the earlier results of \cite{D'Hoker:2001nj,ADP}. 
In \cite{D'Hoker:2001nj} a general prescription is given for adapting the form of the 
measure for Minkowski space derived there to a general compactification. In \cite{ADP}, the chiral blocks 
for $\bZ_2$ twisted fields were derived. In the present case, the only additional modification is that the 
$\bZ_2$ twisting can be applied as well to fields valued in a torus. This results only in restricting the 
corresponding left and right momenta to the dual torus, which we already described in detail in the 
previous section. Putting together these various ingredients, we obtain the formula below. 
(As explained in the previous section, in the top component we may set $\hat \Omega = \Omega_R=\Omega$).

\medskip

For given even spin structure $\delta$ and twist $\ev$, 
the measure $d \mu_L[\delta;\ev]$ is given by,
\bea
\label{2c2}
d\mu _L [\delta ;\ev  ](p_{L\ev}; \Omega)
&=&
d^3\Omega \, \cQ[\ev]( p_{L \ev})  \, 
{Z_C[\delta ; \ev ](\Omega) \over Z_M[\delta] (\Omega) } \, 
\left \{ 
{\Xi_6[\delta](\Omega)  \tet [\delta](0,\Omega)  ^4
\over 16\pi^6 \Psi_{10} (\Omega) } \right .
\no \\ &&  
+ \sum _{\g=1}^3 (1 - \delta _{e^\g}) \Big ( i\pi p^\g_{L \ev}  p^{\bar \g}_{L \ev}
- 2 \p_{\tau_\g }\ln \tet _i (0, \tau_\g ) \Big ) \Gamma [\delta ;e^\g] \Bigg  \}
\eea
where the various terms in the formula are as follows.

\medskip
(a) The factor $\cQ[\ev] ( p_{L\ev})$ represents the internal loop momenta, and is defined by,
\bea
\label{2c1}
\cQ [\ev] ( p_{L\ev}) =\exp \bigg \{ \pi i  \Omega_{IJ} p_I^\mu p_J ^\mu + \pi i  
\sum _{\g=1}^3 \left ( \delta _{e^\g,0} \Omega _{IJ} p^i _{LI} p^{\bar \g} _{LJ} +
(1-\delta _{e^\g,0}) \tau_\g  p^\g_L  p^{\bar \g } _L  \right ) \bigg \}
\eea
The index $\mu=0,1,2,3$ is summed over the uncompactified directions.
The Kronecker symbol $\delta _{e^\g}$ used in (\ref{2c1}), is defined by, 
\bea
\delta _{e^\g,0} = \left \{ \matrix{ 1 & \hbox{if} &  e^\g=0 \cr 0 & \hbox{if} & e^\g \not = 0 \cr} \right .
\eea 
The (super) Prym period $ \tau_{e^\g}$ associated with twist $e^\g$ is the genus 1
modulus of the Prym variety. It will be abbreviated by $\tau _\g = \tau _{e^\g}$.
Its relation to the genus 2 period matrix $\Omega$
and the twist $e^\g$ will be provided in (\ref{2d5}) below; it was introduced in \cite{ADP}. 
Thus the expression $\cQ[\ev] (p_{L\ev} )$ reflects the fact that, for untwisted diections, whether 
compactified or not, the covariance matrix in the internal loop momenta is the super-period 
matrix $\Omega_{IJ}$, while for twisted directions, it is the Prym period matrix $\tau_\g$.
This dependence on $\Omega$ and $\tau_\g$ of $\cQ [\ev] $ will always be understood.

\medskip
(b) The factor $Z_C/Z_M$ is the ratio of the contributions of the matter fields of the given 
compactification to those of Minkowski space. It can be constructed as follows:

\begin{itemize}
\itemsep 0in
\item A pair of untwisted chiral bosons contributes a factor $1/Z(\Omega)^2 $, with $Z$ given by \cite{Verlinde:1986kw}, 
\bea
\label{4a4}
Z^3 & = & {\tet (z_1+z_2- w_0- \Delta) E(z_1, z_2) \sigma (z_1) \sigma z_2) 
\over \sigma (w_0) E(z_1, w_0) E(z_2, w_0) \det \omega _I (z_J)}
\no \\
{ \sigma (z) \over \sigma (w)} & = & {\tet (z-w_1-w_2+\Delta) E(w,w_1) E(w,w_2) 
\over 
\tet (w-w_1-w_2+\Delta) E(z,w_1) E(z,w_2) }
\eea
where $z_1, z_2, w_0, w_1, w_2$ are arbitrary points on $\Sigma$. A brief summary  of genus 2 
characteristics, Jacobi $\tet$-function, and related functions and forms is provided in Appendix B.  
In particular, the form $\sigma (z)$ and the prime form $E(z,w)$ were introduced in \cite{Fay}.
\item A pair of untwisted  fermions with spin structure $\delta$ contributes a factor
$\tet [\delta ](0,\Omega)/Z(\Omega)$. 
\item A pair of twisted fermions with spin structure $\delta$ and twist $ e^\g$ contributes  a factor 
$\tet [\delta + e^\g ] (0, \Omega) / Z(\Omega) $ (see for example \cite{DVV}, \cite{Bernard:1987cc}).
This factor is non-vanishing when the spin structure $\delta + e^\g$ is even, but vanishes when $\delta + e^\g$ is  odd.
\item A pair of bosons with twist $e^\g$ contributes  a factor,
\bea
\label{2c3}
 { \tet [\delta ^{\g+} _i ](0,\Omega)   \tet [\delta ^{\g-} _i ](0,\Omega)
\over Z (\Omega)^2 \tet _i (0, \tau _\g)^2} 
\eea
\end{itemize}
The notation is as follows. For any  twist $e^\g \not= 0$, the set of the 10 even spin 
structures splits into two sets, depending on whether $\delta+e^\g$ is an even 
or an odd spin structure. The set 
\bea
\label{2d20}
\cD[e^\g] = \{ \delta ~ \hbox{even, such that} ~ \delta + e^\g ~ \hbox{ is also even} \}
\eea
consists of 6 elements. The spin structures in $\cD[e^\g]$ may be grouped in three pairs, 
$( \delta ^{\g +}_i, \delta ^{\g -} _i  )$ labelled by  $i  \in \{ 2, 3,4 \} $, where 
each pair sums to the original twist, $\delta ^{\g +}_i +  \delta ^{\g -} _i = e^\g$. 
Each pair $( \delta ^{\g +}_i, \delta ^{\g -} _i  )$ is in one-to-one correspondence with 
an even spin structures $\mu_i$ of the Prym variety, and corresponding genus one
$\tet$-function $\tet _i$,  following the conventions of (\ref{A3}).

\sm

The partition function of (\ref{2c3}) is independent (possibly up to a sign) 
of the label $i$ in view of the Schottky relations, which hold for any pair  $i, j \in \{ 2, 3,4 \}$,
\bea
\label{2d5}
{\tet_j (0, \tau_\g)^4 \over \tet_i (0, \tau_\g)^4}
= 
{\tet [\delta_j ^{\g +} ]^2(0, \Omega) \, \tet [\delta_j ^{\g -}]^2(0, \Omega)^2
\over 
\tet [\delta_i ^{\g +} ]^2(0, \Omega) \, \tet [\delta_i ^{\g -} ]^2(0, \Omega)^2}
\eea
This formula gives $\tau_\g$ in terms of $\Omega$ and $e^\g$, a relation that 
we shall abbreviate as $\tau_\g = R_{e^\g} (\Omega)$. Geometrically, $\tau_\g$ is the 
genus one Prym period associated with the genus two surface $\Sigma$ with period 
matrix $\Omega$, endowed with a  quadratic branch cut across cycle $C_{e^\g}$
(see Figure 2).

\medskip
(c) The factor 
\bea
{\Xi_6[\delta](\Omega) \, \tet[\delta](0,\Omega)^4\over 16\pi^6 \, \Psi_{10}(\Omega)}
\eea
 is the chiral superstring measure for Minkowski space-time. To define $\Xi_6[\delta] (\Omega)$, 
 we make use of some properties special to genus 2. Specifically, 
 there are 6 odd spin structures $\nu_j$, $j =1,\cdots, 6$. Each even spin structure can be 
 identified with a partition of the 6 odd spin structures into two sets of 3 in 
 each. The sum of the odd spin structures in each set adds to the given even spin 
 structure. If $\delta=\nu_1+\nu_2+\nu_3$, then $\Xi_6[\delta]$ is defined by
\bea
\label{Xi6}
\Xi_6[\delta](\Omega)=\sum_{1\leq i<j\leq 3}\<\nu_i|\nu_j\>
\prod_{k=4,5,6}\tet[\nu_i+\nu_j+\nu_k](\Omega)^4.
\eea
Alternative forms of $\Xi_6[\delta]$ have been given in \cite{D'Hoker:2004}.
The expression $\Psi_{10}(\Omega)$ is the weight 10 modular form  in genus two \cite{Igusa},
\bea
\Psi_{10}(\Omega)=\prod_{\kappa\ {\rm even}}\tet[\delta](0,\Omega)^2
\eea

\medskip
(d) The term $\Gamma[\delta;\ev]$ is due to twisting, and provides a key correction 
to the super Prym matrix $\tau_\ep$ arising from chiral splitting \cite{ADP}, eq. (5.44). 
It may be defined as follows.
Since $\delta \in \cD [e^\g]$, we may identify it uniquely with one of its 
elements $\delta = \delta ^{\g \alpha } _i$, and we then have, 
\bea
\label{2d6}
\Gamma [\delta^{\g \a} _i ; e^\g] 
= - i  {\< \nu_0 | \mu_i \> \,   \tet [\delta _i ^{\g \a} ](0, \Omega) ^4 \over (2 \pi )^7 \eta (\tau_\g )^{12}  }\times 
{ \tet _j (0, \tau_\g) ^8 \over  \tet [\delta _j ^{\g +} ](0, \Omega)^4 \tet [\delta _j ^{\g -} ](0, \Omega)^4} 
\eea
Here, $\mu_i$ is the genus 1 spin structure associated with $\delta ^{\g \a} _i$,
the symplectic pairing mod 2 is denoted $\< \nu_0 | \mu_i \>$, $\G$ is 
independent of the choice of $j$; and we abbreviate 
$\tet [\delta _j ^{\g \pm} ] \equiv \tet [\delta _j ^{\g \pm} ](0,\Omega)$.
The overall sign was computed in \cite{ADP}, but the final
expression given there is not correct. A~simplified and corrected 
calculation is presented here in  Appendix \ref{secQ}, resulting in (\ref{2d6}).

\subsection{The bottom component $d\mu_L^{(0)}[\delta;\ev]$ in the left chiral measure}

The bottom component $d\mu^{(0)}_L [ \delta ; \ev]$ was evaluated in \cite{ADP}, and  is given by
\bea
\label{4a2}
d \mu ^{(0)} _L [\delta ; \ev] (p_\ev; \hat \Omega ) = 
\cZ [\delta ] (\hat \Omega) {Z_C [\delta  ; \ev ] (\hat \Omega) \over Z_M [\delta ] (\hat \Omega) } 
\cQ [\ev] (p_{L\ev}) d^3 \hat \Omega
\eea
Here $Z_C [\delta  ; \ev ]/  Z_M [\delta ] $ is the contribution of the compactified fields relative to 
the contribution of the uncompactified fields in Minkowski space-time. 
The factor $\cZ[\delta ] $ is given by,
\bea
\label{4a3}
\cZ [\delta ]  =  { \tet [\delta ]^5 \tet (p_1+p_2+p_3-3 \Delta) \prod _{a<b} E(p_a, p_b) \prod _a \sigma (p_a)^2
\over Z^{15} \tet [\delta ] (q_1+q_2-2 \Delta) E(q_1, q_2) \sigma (q_1)^2 \sigma (q_2)^2 
\det (\omega_I \omega_J (p_a))}
\eea
and $\cZ[\delta]$ is common to all sectors and orbits.
It is independent of the choice of points $p_1,p_2,p_3, z_1, z_2, w_0, w_1, w_2$, but 
does depend on $q_1, q_2$. The dependence on $q_1,q_2$ can be traced back to the choice 
of gravitino gauge slice in the space of all two-dimensional supergeometries,
\bea
\chi_{\bar z}^+
=
\zeta^1\delta(z,q_1)+\zeta^2\delta(z,q_2)
\eea
so that the factor $\cZ[\delta]$ has a dependence on the choice of gauge slice.

\subsection{Calculation of $Z_C[\delta;\ev]/Z_M[\delta]$ orbit by orbit}

Both components $d\mu_L^{(0)}[\delta;\ev]$ and $d \mu_L[\delta;\ev]$ of the left chiral measure 
depend on the ratio $Z_C[\delta;\ev]/Z_M[\delta]$ of the matter fields of the compactification to the matter fields 
of Minkowski space. These ratios
depend quantitatively and qualitatively on the orbit to which $\ev$ belongs. 
We proceed to their calculation, orbit by orbit. 

\subsubsection{The orbit $\cO_0$}

The orbit $\cO_0$ corresponds to the untwisted sector. The effects of some of the fields being 
compactified on tori, resulting in a discretization of the corresponding internal loop momenta, 
have already been incorporated in the factor $\cQ[\ev](p_L)$. Thus we have for $\cO_0$
\bea
Z_C[\delta;\ev]/Z_M[\delta]=1
\eea

\subsubsection{The orbits $\cO_1$, $\cO_2$, and $\cO_3$}

The orbits  $\cO_1, \cO_2$ and $\cO_3$ are merely permutations of one another.
Twisting in one of the orbits $\cO_\g$ for $\g=1,2,3$ is effectively by a $\bZ_2$ subgroup 
generated by $\lambda _\g$. The twist vector $\ev$ in orbit $\cO_\g$ is given by $e^\g=0$ and $e^{\g'}=\ep$ 
for $\g' \not = \g$, and we will express the left chiral amplitudes in terms of this twist $\ep$. 
Again, ignoring the effects of compactification on tori already incorporated in the factor $\cQ[\ev]( p_\ev)$, 
the blocks for the orbits $\cO_1, \cO_2, \cO_3$ are exactly the same 
blocks as in the supersymmetric $\bZ_2$ orbifolds studied in \cite{ADP}.
In the case of $\cO_1,\cO_2,\cO_3$, we have two pairs of twisted fields and hence,
\bea
\label{2d3}
{Z_C[\delta;\ev ]
\over
Z_M[\delta]  }
=
{\tet [\delta_i^+]^2 \, \tet [\delta_i^-]^2 \, \tet [\delta+\ep ] ^2
\over
\tet_i  (0, \tau_\ep)^4 \, \tet [\delta]^2}
\eea
where $\delta ^+ _i + \delta ^- _i = \ep$.  We note that the incorporation of the factor
$\Gamma [\delta, \ep]$ given by (\ref{2d6}) with $Z_C[\delta;\ev]/Z_M[\delta]$ produces,
\bea
{Z_C [\delta ; \ev] \over Z_M [\delta] } \, \Gamma [\delta , \ep]
=  - i  \< \nu_0 | \mu_i \> { \tet _i (0, \tau_\ep) ^4 \over \eta (\tau_\ep)^{12}}
\eea
The identification of  $\delta$ with one of the 6 elements $\delta_i^\pm$  in $\cD [\ep]$ determines the index $i$.

\subsubsection{The orbit $\cO_-$}
\label{sec:AB}

For any twist $\ev \in \cO_-$, all compactified fields $z^\g, \psi ^\g, \xi^\g$  are twisted,
leaving 4 untwisted bosons $x^\mu$, four untwisted left fermions $\psi _+^\mu$ and 26 
untwisted Heterotic fermions $\psi_- ^\alpha$.  We shall now show that all contributions
from twists in orbit $\cO_-$ vanish identically.

\sm

The ratio $Z_C[\delta;\ev]/Z_M[\delta]$ will enter both the left and the right blocks with spin structures 
$\delta$ and $\delta _R$ respectively.  It may be calculated using the ingredients of Section \ref{sec24}, 
and we find, 
\bea
{Z_C[ \delta; \ev]  \over Z_M [\delta] }
= \prod _{\gamma=1}^3 
\left ( { \tet [ \delta ^{\g +} _i ]  \, \tet [ \delta ^{\g -} _i ] \, 
\tet [\delta + e^\g] \over  \tet _i (0,\tau_\g)^2 \, \tet [\delta ] }
\right )
\eea
We focus on the factors involving the twisted fermions with spin structure $\delta$, 
\bea
\label{2d10}
\prod _{\g =1}^3 { \tet [\delta + e^\g]  \over \tet [\delta ] }
\eea
The factor (\ref{2d10}) will vanish unless  the spin structures $\delta + e^\g$ are even for all $\g=1,2,3$.
Actually,  for $\ev \in \cO_-$ and $\delta$ an even spin structure,
at least one of the spin structures $\delta + e^\g$ must be odd. 
A basis-independent argument may be given as follows. As usual, 
we define,
\bea
\sigma (\kappa) = e^{4 \pi \kappa ' \cdot \kappa ''}
\eea
The product (\ref{2d10}) will vanish unless $\sigma (\delta+e^\g) = 1$ for each $\g=1,2,3$. 
A necessary condition is that the product of all three $\sigma (\delta + e^\g)$ must be 1. 
Using the fact that $\delta$ is even, and that $e^1+e^2+e^3=0$, it is readily shown that,
\bea
\prod _{\g=1}^3 \sigma (\delta + e^\g) = \prod _{\g=1} ^3 \sigma (e^\g) = \< e^1 | e^2 \>
\eea
By the definition of the orbit $\cO_-$ we have  $\< e^1 | e^2 \>=-1$, and hence 
there can be no even spin structures $\delta$ such that all $\delta + e^\g$ are even.

\sm

The above general argument may be checked by using the explicit representation
of twists given in Appendix \ref{secB}. Choosing the element 
$(\ep , \eta) = (\ep _2, \ep_5) \in \cO_-$, we have $(e^1, e^2, e^3) = (\ep_2 , \ep _5, \ep_{11})$,
in the notation of (\ref{1f1}).  Requiring $\delta $ even and $\delta + e^1=\delta +\ep_2$ even as well
restricts to $\delta \in \{ \delta _1, \delta _2, \delta _3, \delta _4, \delta _7, \delta _8\}$. 
Requiring in addition $\delta + e^2= \delta + \ep_5$  even restricts further to 
$\delta \in \{ \delta _1, \delta _3,  \delta _7\}$. But for each of these, $\delta + e^3 = \delta + \ep_{11}$ is odd.
Thus, the factor of (\ref{2d10}) is identically zero for all $\delta$ as long as $\ev$ belongs to $\cO_-$.
The contribution from the right blocks vanishes  for $\ev \in \cO_-$ as well, and thus so do  the contributions 
to the vacuum energy from the entire orbit $\cO_-$.

\subsubsection{The orbit $\cO_+$}

Finally, we present here a preliminary analysis of the contributions from twists $\ev $ in the orbit $\cO_+$,
with a full analysis deferred to Section \ref{sec:R}. Following the rules of Section \ref{sec24}, the ratio $Z_C/Z_M$ of 
partition functions for the left chiral amplitudes  is given by,
\bea
\label{5a1}
{Z_C[ \delta; \ev ]  \over Z_M [\delta]  }
= Z_B [\ev] \prod _{\g=1}^3 
{ \tet [\delta + e^\g]  \over   \tet [\delta ] }
\eea
where the contribution $Z_B [\ev] $ from the twisted bosons is given by, 
\bea
\label{5a3}
Z_B [\ev ] = \prod _{\g=1}^3 { \tet [\delta ^{\g +} _i ] \tet [\delta ^{\g -} _i] \over \tet _i (0, \tau _\g)^2}
\eea
Inspecting the product of factors involving the twisted fermions as in (\ref{2d10}), we see
that $Z_C/Z_M$ will vanish for $\ev \in \cO_+$ unless the spin structures $\delta$ satisfies,  
\bea
\label{5a2}
\delta \in \cD [\ev] = \cD [e^1] \cap \cD [e^2] \cap \cD [e^3] 
\eea
where $\cD [e^\g]$ was defined in (\ref{2d20}) as the set of even spin structures $\delta$
such that $\delta + e^\g$ is even.

\medskip

We observe that there will be no contributions from $\delta + e^\g$ odd. To establish this is slightly subtle.
Even though $Z_C/Z_M$ vanishes when one of the $\delta + e^\g$ is odd, there might still arise  a 
contribution  provided the Dirac zero modes corresponding to odd $\delta +e^\g$
are absorbed by the two-point function of the supercurrent correlator. For genus 2,  the supercurrent 
can absorb at most two chiral zero modes. Thus, we conclude that if $\delta + e^\g$ is odd for two
different $\g \in \{ 1,2,3 \}$, then there are not enough field insertions to soak up all the zero modes,
and the contribution must vanish. It is easy to check that the orbit $\cO_+$ never allows for spin 
structures $\delta$ such that only a single $\delta + e^\g$ is odd. As a result, all contributions from 
$\delta \not \in \cD[\ev]$ cancel identically, as expected.

\subsection{The right chiral measure $d\mu_R[\ev]$}

Right chiral amplitudes are constructed in a similar fashion. 
We have 6 twisted pairs of fields $(z^\g, z^{\bar \g})$ and $\xi^\g, \xi ^{\bar \g}$ with $\g, \bar \g = 1,2,3$, 
four untwisted bosons $x^\mu$ with $\mu =0,1,2,3$,  and 26 untwisted fermions $\psi _- ^\alpha$ 
with $\alpha =1, \cdots, 26$. The spin structure assignment for the GSO projection in the 
right chiral amplitudes distinguishes the $Spin (32)/Z_2$ from the $E_8 \times E_8$ Heterotic strings, and we have,
\begin{itemize}
\item 
For the $Spin (32)/Z_2$  string,  all 32 components of $\xi$ and $\psi _-$ have the same 
even spin  structure $\delta_R$. The GSO projection requires that $\delta_R$ be summed
over all even spin structures with a suitable phase factor $C_{\delta_R}^{(1)}[\ev]$. The final result is,
\bea
\label{5d1}
d \mu _R [\ev ] = d^3 \Omega \, \cQ [\ev] (p_{R\ev}) \, { Z_B [\ev] \over \Psi_{10} } \sum _{\delta _R} C_{\delta_R} ^{(1)}[\ev]
\tet [\delta_R ] ^{16}  \prod _{\g=1}^3 
{ \tet [\delta_R + e^\g]  \over  \tet [\delta_R ] }
\eea
where we have suppressed the dependence on $\Omega$, and denoted $\tet [\delta] (0, \Omega)$ 
simply by $\tet [\delta ]$.
The presence of the product factor involving $\tet [\delta_R + e^\g]$ makes contributions for 
$\delta_R$ vanish unless $\delta_R \in \cD [\ev ]$. The GSO phases $C_{\delta_R}^{(1)}[\ev]$ 
will be determined later by modular invariance and will turn out to be sign factors.
\item 
For the $E_8 \times E_8$  string, the 32 right fermions are grouped into two sets of
16 fermions. Within each set, all 16 fermions are endowed with the same spin structure, 
$\delta _R^1$ and $\delta _R^2$ respectively,  and summed independently over 
$\delta _R^1$ and $\delta _R^2$ with modular covariant GSO phase factors. 
The twisting is performed in the first set of 16, corresponding to the 
embedding $SU(3) \times E_6 \times E_8 \subset E_8 \times E_8$. 
The GSO phases for the spin structure summation over $\delta _R^2$ must then
all be equal, and may be set to 1. The final result is,
\bea
\label{5d2}
d \mu _R [\ev ] = d^3 \Omega \, \cQ [\ev] (p_{R\ev})   Z_B [\ev]  {\Psi _4 \over \Psi_{10}} 
 \sum _{\delta _R^1} C_{\delta_R^1}^{(2)} [\ev] \, 
\tet [\delta_R^1 ] ^{8} \, \prod _{\g=1}^3  { \tet [\delta_R^1 + e^\g]  \over  \tet [\delta_R^1 ] } 
\eea
where the sum over $\delta _R^2$ has produced the genus 2 modular form $\Psi _4$ defined by
\bea
\label{5j1}
\Psi _4 (\Omega) =  \sum _{\delta ^2_R}  \tet [\delta_R^2 ] (0,\Omega) ^{8} 
\eea
The GSO projection signs $C_{\delta_R^1}^{(2)}[\ev]$  will be determined later by modular invariance. 
They will also turn out to be all sign factors.
\end{itemize}

\newpage

\section{Interior of Supermoduli Space, Part I}
\setcounter{equation}{0}
\label{sec:R}

In this section and the next, we shall establish the vanishing of the contribution to the two-loop
vacuum energy arising from the interior of supermoduli space (computed with the procedure 
that has been used in previous two-loop superstring calculations \cite{D'Hoker:2002gw,D'Hoker:2001nj}) 
for both Heterotic and Type II superstrings. Specifically, we shall show that the GSO summation
over spin structures of the left chiral measure, integrated over odd moduli, vanishes point-wise in the 
interior of moduli space for every twist $\ev \in \cO_{\rm tot}$.

\sm

In this section, we briefly review the analogous cancellation of the vacuum energy at genus 1.
We then go on to show that a first part of  contributions to the vacuum energy vanish,
specifically those arising from the orbits $\cO_\g$ with $\g =0,1,2,3$, and well as $\cO_-$. 
The calculation of the part arising from the orbit $\cO_+$ will be deferred to Section \ref{sec:S}.

\subsection{Vanishing contribution from 1-loop}

We verify that the vacuum energy for the $\bZ_2 \times \bZ_2$ orbifold models 
under consideration vanishes pointwise on moduli at genus $1$. 
Contributions from the untwisted sector vanish by the  Riemann identity.
There are three distinct non-zero twists  $\ep_2 =  [ 0  |  \half ]$,
$ \ep_3 =  [ \half  |  0  ] $, and $\ep_4 =  [ \half |  \half  ] $.
The contributions arising from the orbits $\cO_1, \cO_2, \cO_3$
are all equivalent to one another. The contribution of the sum over 
(even) spin structures $\mu$ to the partition function in an orbit with twist $\ep$
is given by, 
\bea
\sum _\mu \< \nu_0 | \mu \> \tet [\mu ] (0, \tau)^2 \tet [\mu + \ep ](0,\tau )^2 
\eea
where $\nu_0$ is the  odd spin structure.
For the example $\ep = \ep_2$, only the spin structures 
$\mu = \mu_3$ and $\mu = \mu_4 $ contribute; the factor $\< \nu_0 | \mu \>$
is opposite for those, so that the sum vanishes. 

\sm

To study the orbits $\cO_\pm$, we proceed as follows.
For any pair of distinct twists, we have $\< \ep _\g| \ep _{\g'} \>=-1$, so that
at genus 1 the orbit $\cO_+$ is empty. The orbit $\cO_-$ contains all
three pairs $(\ep _\g, \ep_{\g'})$ for $\g' \not=\g$. But there are no even spin structures
$\delta$ such that $\delta + \ep_\g$ is even for all $\g$. Hence, the contribution 
from orbit $\cO_-$ vanishes at genus one.

\subsection{Two-loop GSO projected left chiral amplitudes}

At two-loop order, we exploit the existence of a holomorphic projection from $\mathfrak{M}_2$ to 
$\cM_2$ in order to parametrize the points in $\mathfrak{M}_2$ by coordinates $(\hat\Omega, \zeta^1, \zeta ^2, \delta)$. 
An integral over $\mathfrak{M}_2$ reduces to  integrals over $\cM_2$ and $\zeta^\alpha$, 
and a sum over the spin structures $\delta$.  This last sum is the GSO projection. 
In this section, and the next, we shall show that for any vector $\ev$ of twists, the GSO summation over
spin structures of the left chiral amplitudes  vanishes point-wise in the interior of supermoduli space,
namely,
\bea
\label{bulkvanishing}
\sum_\delta C_\delta [\ev] \, d\mu_L[\delta;\ev](p_{L\ev};\hat\Omega)=0.
\eea
Since the integrand of the contribution to the vacuum energy from the interior  of supernoduli space is 
proportional to $d\mu_L$, it will vanishes pointwise on $\cM_2$.
Therefore,  the entire bulk contribution to the vacuum energy will vanish for the Heterotic and 
Type II superstrings.

\smallskip

To show that (\ref{bulkvanishing}) holds, we shall proceed orbit by orbit 
$\cO_\alpha$, with $\alpha=0,1,2,3,\pm$. Since the calculation for the orbit $\cO_+$
is significantly more involved than for the other orbits, we shall postpone to the next section the 
discussion for the orbit $\cO_+$. 

\subsection{Vanishing contribution from orbit $\cO_0$}

For the orbit $\cO_0$, there is no twisting, so the contribution to the vacuum energy from this 
orbit vanishes since the vacuum energy for Minkowski space vanishes, as established 
in \cite{D'Hoker:2001nj}. Specifically, we have $Z_C[\delta;\ev]/Z_M[\delta]=1$,
and  all GSO phases are equal  \cite{D'Hoker:2001nj}. The contribution to the vacuum energy 
from the orbit $\cO_0$  is proportional to the following left chiral factor, 
\bea
\sum_\delta \Xi_6[\delta]\tet[\delta]^4=0
\eea
which vanishes point-wise on moduli space \cite{D'Hoker:2001nj}.

\subsection{Vanishing contribution from orbits $\cO_1,\cO_2,\cO_3$}

For $\ev$ in orbits $\cO_1$, $\cO_2$, and $\cO_3$, two pairs of fields are twisted by a single twist, 
which we denote by $\ep$.  Thus the vanishing of the vacuum energy 
reduces in this case to the vanishing of the $\bZ_2$ orbifold models established in \cite{ADP}. 
For a given twist $\ep$,  the GSO phases have to be equal to insure modular covariance
\cite{ADP}. In view of (\ref{2c2}) and (\ref{2d3}), we have then, 
\bea
\sum_\delta d\mu_L[\delta;\ev](p_{L\ev};\hat\Omega)
=
d^3\hat\Omega \cQ [\ev] ( p_{L\ev})
{\tet[\delta_i^+]^2\tet[\delta_i^-]^2
\over
16\pi^6\Psi_{10}\tet_i(0,\tau_\ep)^4}
\sum_\delta \Xi_6[\delta]\tet[\delta]^2\tet[\delta+\ep]^2
\eea
The relevant identity  was established in \cite{ADP} using the Fay tri-secant formula \cite{Fay}, 
\bea
\sum_\delta\Xi_6[\delta]\tet[\delta]^2\tet[\delta+\ep]^2=0
\eea
It guarantees the vanishing of contributions from orbits $\cO_1, \cO_2, \cO_3$.

\subsection{Vanishing of the bulk energy from $\cO_-$}

We have established in Section \ref{sec:AB} that $Z_C[\delta; \ev]/Z_M[\delta]=0$ for any $\ev$ in the orbit $\cO_-$. 
As a result,  the orbit $\cO_-$ does not contribute to the vacuum energy.

\newpage

\section{Interior  of Supermoduli Space, Part II}
\setcounter{equation}{0}
\label{sec:S}

In this section, we shall show  that the contributions to the vacuum energy  from the interior of 
supermoduli space (computed with the procedure of \cite{D'Hoker:2002gw,D'Hoker:2001nj}) 
of the sectors with twists $\ev$ in the orbit  $\cO_+$ also vanish.  We do so by showing that the 
GSO sum of the top component of the left chiral measure vanishes point-wise on moduli space. 
This part in the study of bulk contributions is the most delicate one, as the orbit $\cO_+$ has a 
counterpart neither in the $\bZ_2$ orbifold  theories studied in \cite{ADP}, nor at genus one. 

\subsection{Isolating the $\delta$-dependence in the left chiral measure}

To organize the proof of the vanishing of the GSO sum over spin structures of the top component 
of the  left chiral measure, we begin by isolating the part of the measure with spin structure $\delta$
dependence from the part that is independent of $\delta$. The starting point will be the expression 
(\ref{2c2}) for the top component of the left chiral measure, restricted to twists $\ev \in \cO_+$. 
The ratio $Z_C/Z_M$ is then provided by (\ref{5a1}) in terms of a $\delta$-dependent product over 
$\tet$-constants, and a $\delta$-independent factor $Z_B [\ev]$, given in (\ref{5a3}). Furthermore,
for any $\ev \in \cO_+$, we have $\delta _{e^\g}=0$ in (\ref{2c2}) and (\ref{2c1}). Using these facts, 
the left chiral measure may  be expressed in the following factorized way,
\bea
\label{5k1}
d\mu _L [\delta ;\ev  ](p_{L\ev}; \Omega)
=
{d^3\Omega \, \cQ[\ev] ( p_{L \ev})  \over 16 \pi^6} Z_B[\ev ](\Omega) 
\Bigg \{ 
{\Xi_6[\delta](\Omega) \over  \Psi_{10} (\Omega) }  \prod _{\kappa \in \cD [\ev]} \! \! \tet [\kappa ](0,\Omega)  
+ \cB [\delta; \ev] (p_{L\ev}; \Omega) \Bigg \} 
\eea
where the combination $\cB$ is given by,
\bea
\label{5k2}
\cB [\delta; \ev] (p_{L\ev}; \Omega)= 
16 \pi^6   \sum _{\g=1}^3  \Big ( i\pi p^\g_{L \ev} \, p^{\bar \g}_{L \ev}
- 2 \p_{\tau_\g }\ln \tet _i (0, \tau_\g ) \Big ) \Gamma [\delta ;e^\g] 
\prod _{\lambda =2,3,4 } {\tet [\delta + e^\lambda ](0, \Omega) \over \tet [\delta ] (0,\Omega)}
\eea
We recall that the index $i$ on the genus one $\tet$-function in (\ref{5k2}) is determined by identifying
the spin structure $\delta$ with one of the six spin structures $\delta _i ^{\g \alpha}$ in the set $ \cD [e^\gamma]$.
All $\delta$-dependence in (\ref{5k1}) has been confined to the terms within the braces.

\subsection{Determining the GSO phases for the left chiral measure}

The GSO projection for both the Heterotic and Type II superstrings requires summation over all spin structures 
$\delta$, for each fixed twist $\ev \in \cO_+$, of the left chiral amplitude multiplied by GSO phase factors $C_\delta [\ev]$.
Thus, the sum to be performed, for fixed $\ev \in \cO_+$, is as follows,  
\bea
\label{5k3}
 \sum _{\delta} C_\delta [\ev] \, d \mu_L [\delta; \ev] (p_\ev, \Omega)
\eea
We shall now determine the phase factors $C_\delta [\ev]$ by modular invariance. It will turn out that their
values are restricted to $\pm 1$ only.

\sm

Given a twist $\ev \in \cO_+$ the only spin structures that produce a non-zero contribution are in the set 
$\cD [\ev]$  introduced in (\ref{5a2}). Here, we shall need the structure of the set $\cD[\ev]$ more explicitly.
In Table 1 below, we have listed the set of all vectors $\ev$ in $\cO_+$, together with the corresponding 
sets $\cD[\ev]$. Direct inspection shows that $\cD[ \ev]$ contains four distinct elements for each $\ev\in\cO_+$.
For example, for the twist $\ev = (\ep_2, \ep_3, \ep_4) \in \cO_+$,  the set $\cD[\ev]$
is given by $\cD [\ev]= \{ \delta _1, \delta _2, \delta _3, \delta _4 \}$, following the notations for
twists and spin structures of Appendix \ref{secB}.

\sm

To determine the GSO phases $C_\delta[\ev]$, we concentrate on the term involving $\Xi_6 [ \delta]$
in (\ref{5k1}).  (The term proportional to $\Gamma [\delta; \ev]$ in (\ref{5k1}) will be assigned the same GSO phases.)
By inspection of  (\ref{5a1}),  we see  that the left chiral measure contains the product of  $\tet [\kappa ]$ 
over all $\kappa \in \cD[\ev]$. 
This product is determined entirely by the twist $\ev$, and is independent of the specific
spin structures $\delta \in \cD [\ev]$. As a result, the remaining spin structure sum for the 
part of the left chiral measure involving $\Xi_6 [\delta]$ reduces to,
\bea
\label{5e2}
\sum _{\delta \in \cD[\ev]} C_\delta [\ev] \, \Xi_6[\delta] (\Omega)
\eea
The modular transformations properties of $\Xi_6 [\delta ] (\Omega)$ coincide with those of $\tet [\delta ] (0, \Omega)^{12}$,
which may be obtained from  (\ref{B10}) and (\ref{B11}) of Appendix B. As a result, 
modular covariance may be realized in terms of the following GSO phase factor assignment, 
\bea
\label{5e5}
C_\delta [\ev] = C_{ \delta_* } [\ev] \, \<  \delta_* | \delta \>
\eea
where $\delta _*$ is any reference spin spin structure in $\cD [\ev]$, and 
$\< \delta_* | \delta \>$ denotes the symplectic invariant mod 2, defined in (\ref{1e1}). In Table \ref{table:1}, the signs 
$C_\delta [\ev]$ have been listed in the same order as the spin structures in $\cD[\ev]$, 
and we have chosen $\delta_*$ to be the first spin structure listed in $\cD[\ev]$.
It is easily seen by inspection that the assignment rule (\ref{5e5}) holds.

\subsection{Determining the GSO phases for the right chiral measure}

In section \ref{sec6}, we shall also need the GSO phase assignments for the right chiral
measure, since they occur in the GSO sums for the right chiral measure in  (\ref{5d1}) and (\ref{5d2}), 
\bea
\label{5e3}
\sum _{\delta _R} C_{\delta_R} ^{(1)}[\ev] \tet [\delta_R ] ^{12}
\hskip 1in 
\sum _{\delta _R^1} C_{\delta_R^1}^{(2)} [\ev]
\tet [\delta_R^1 ] ^{4} 
\eea
The modular transformation  signs of the quantities $\Xi_6 [\delta]$, $\tet [\delta ]^4$ and $\tet [\delta ]^{12}$ are 
all equal to one another, so that we may set,
\bea
\label{5e4}
C_\delta ^{(1)} [\ev] = C_\delta ^{(2)} [\ev] = C_\delta [\ev]
\eea
up to an overall sign factor, not determined by modular invariance alone.
Given the equality of the GSO signs for left and right measures, we may set $C_{ \delta _*}[\ev]=1$
without loss of generality.

\begin{table}[htb]
\begin{center}
\begin{tabular}{|c|c||c|c|c|c|}   \hline
$\ev \in \cO_+$ & $\cO_+^{ e / o} $ & $\cD[\ev]$ & $C_\delta [\ev]$ & $M$ & $\lambda [ \delta_* ; \ev]$
\\ \hline \hline
$(\ep_2, \ep_3, \ep_4)$ & $e$ & $(\delta _1, \delta_2, \delta _3, \delta _4)$ & $ (+, +, +, +)$ 
& $I$ & $+$ 
\\ \hline
$(\ep_2, \ep_7, \ep_8)$ & $e$ & $(\delta _1, \delta_2, \delta _7, \delta _8)$ & $ (+, +, +, +)$ 
& $M_1SM_1S$ & $-$ 
\\ \hline
$(\ep_3, \ep_5, \ep_6)$ & $e$ & $(\delta _1, \delta_3, \delta _5, \delta _6)$ & $ (+, +, +, +)$ 
& $M_2SM_2S$ & $-$ 
\\ \hline
$(\ep_5, \ep_7, \ep_9)$ & $e$ & $(\delta _1, \delta_5, \delta _7, \delta _9)$ & $ (+, +, +, +)$ 
& $S$ & $+$
\\ \hline 
$(\ep_2, \ep_{12}, \ep_{14})$ & $e$ & $(\delta _3, \delta_4, \delta _7, \delta _8)$ & $ (+, +, -, -)$ 
& $SM_1S$ & $+$
\\ \hline
$(\ep_3, \ep_{11}, \ep_{13})$ & $e$ & $(\delta _2, \delta_4, \delta _5, \delta _6)$ & $ (+, +, -, -)$ 
& $SM_2S $ & $+$
\\ \hline
$(\ep_5, \ep_{12}, \ep_{16})$ & $e$ & $(\delta _3, \delta_6, \delta _7, \delta _9)$ & $ (+, +, -, -)$ 
& $M_1S$ & $-$
\\ \hline
$(\ep_7, \ep_{11}, \ep_{15})$ & $e$ & $(\delta _2, \delta_5, \delta _8, \delta _9)$ & $ (+, -, +, -)$ 
& $M_2S$ & $-$
\\ \hline
$(\ep_{10}, \ep_{11}, \ep_{12})$ & $e$ & $(\delta _4, \delta_6, \delta _8, \delta _9)$ & $ (+, -, -, +)$ 
& $M_2M_1S$ & $+$
\\ \hline
$(\ep_4, \ep_9, \ep_{10})$ & $o$ & $(\delta _1, \delta_4, \delta _9, \delta _0)$ & $ (+, +, +, +)$ 
& $TM_1SM_1S$ & $-$
\\ \hline
$(\ep_6, \ep_8, \ep_{10})$ & $o$ & $(\delta _1, \delta_6, \delta _8, \delta _0)$ & $ (+, +, +, +)$ 
& $M_3S$ & $-$
\\ \hline
$(\ep_4, \ep_{15}, \ep_{16})$ & $o$ & $(\delta _2, \delta_3, \delta _9, \delta _0)$ & $ (+, +, -, -)$ 
& $SM_3M_2M_1S$ & $+$
\\ \hline
$(\ep_6, \ep_{14}, \ep_{15})$ & $o$ & $(\delta _3, \delta_5, \delta _8, \delta _0)$ & $ (+, +, -, -)$ 
& $M_3M_1S$ & $+$
\\ \hline
$(\ep_8, \ep_{13}, \ep_{16})$ & $o$ & $(\delta _2, \delta_6, \delta _7, \delta _0)$ & $ (+, -, +, -)$ 
& $M_3M_2S$ & $+$
\\ \hline
$(\ep_9, \ep_{13}, \ep_{14})$ & $o$ & $(\delta _4, \delta_5, \delta _7, \delta _0)$ & $ (+, -, -, +)$ 
& $M_3M_2M_1S$ & $-$
\\ \hline 
\end{tabular}
\end{center}
\caption{Listed are: twists $\ev $ in the orbit $ \cO_+$;  the set $\cD [\ev]$; 
the GSO phases $C_\delta [\ev]$; the signs $\lambda$ defined in (\ref{F1}); and the modular transformation 
$M$ such that $\ev=M(\ep_2, \ep_3, \ep_4)$. For use in Section \ref{sec6}, we distinguish
the  orbits $\cO_+^{e / o}$ under the $SL(2,\bZ) \times SL(2,\bZ)$ which leaves the separating 
node invariant. Finally,  $\delta_*$ is chosen to be the first entry in the array of $\cD[\ev]$.}
\label{table:1}
\end{table}

\subsection{Vanishing contribution from the $p_{L\ev}$ term}

The first step in the evaluation of the contributions from twists $\ev\in\cO_+$ 
is to show that the GSO spin structure sum over $\delta$ of the contribution proportional to
$ i \pi p_{L\ev}^\gamma \, p_{L\ev}^\gamma$ in (\ref{5k2}) vanishes identically.
For this, we  isolate all $\delta$-dependence in its coefficient by casting the product
over $\lambda $ in (\ref{5k2}) in terms of a product of $\tet [\kappa] $ over all $\kappa \in \cD [\ev]$,
divided by $\tet [\delta ]^4 $. After some minor simplifications, we obtain with the help of (\ref{2d6}), 
\bea
\label{5m1}
16 \pi^6 \Gamma [\delta^{\g \pm} _i ; e^\g] 
\prod _{\lambda =2,3,4 } {\tet [\delta + e^\lambda ](0, \Omega) \over \tet [\delta ] (0,\Omega)}
= -  i {  \< \nu_0 | \mu_i \>  \over 8 \pi \, \eta (\tau_\g)^{12} }
{ \tet _j (0, \tau _\g)^8 \over  \tet [\delta ^{\g + } _j ]^4 \, \tet [\delta ^{\g - } _j ]^4}
\prod _{\kappa \in \cD [\ev]} \tet [\kappa] 
\eea
Recall that $\delta $ is to be identified with one the  six even spin structures  $\delta ^{\g \pm} _i \in \cD [e^\g]$, 
which are such that $\delta ^{\g +} _i  + e^\g = \delta ^{\g -} _i$ with $i=2,3,4$. 

\sm

Formula (\ref{5m1})  is independent of $j$ by the Schottky relations of (\ref{2d5}). As was pointed out earlier, the 
product over $\kappa \in \cD[\ev]$  in (\ref{5m1}) only depends on the twist $\ev$, and not on $\delta$. Thus, 
the only dependence on $\delta $ on the rhs of (\ref{5m1}) is through the  symplectic pairing $\< \nu _0 | \mu_i \>$.

\sm

By modular covariance, it suffices to evaluate the sum over $\delta$ for any fixed $\ev\in\cO_+$. 
We choose $\ev_0=(\ep_2,\ep_3,\ep_4)$. By inspection of the Table, we see that $C_\delta[\ev_0]=1$  
for all $\delta\in \cD[\ev_0 ]$. Next, we work out an explicit parametrization of the spin structures in $\cD [\ev_0]$,
in the conventions of Appendix B, and we find, 
\bea
\label{5m2}
\delta ^{1+} _i = \left [ \matrix{  \mu _i  \cr \mu_3 \cr } \right ]
\hskip 1in 
\delta ^{1-} _i = \left [ \matrix{  \mu _i  \cr \mu_4 \cr } \right ]
\eea
where $i=3,4$, yielding a total of 4 spin structures. Finally, we recast the summation over spin structures
of (\ref{5m1}) with the help of this explicit parametrization of $\delta$, and we find, 
\bea
\label{5m3}
\sum _\delta 16 \pi^6 \Gamma [\delta ; e^\g] \!
\prod _{\lambda =2,3,4 } \! {\tet [\delta + e^\lambda ] \over \tet [\delta ]}
=    { - i \over 8 \pi \, \eta (\tau_\g)^{12} }
{ \tet _j (0, \tau _\g)^8 \over  \tet [\delta ^{\g + } _j ]^4 \, \tet [\delta ^{\g - } _j ]^4}
\prod _{\kappa \in \cD [\ev]} \tet [\kappa]  \sum _{ i=3,4} \sum _{\alpha =\pm}  \< \nu_0 | \mu_i \> 
\eea
The sum over $i$ vanishes since $\< \nu _0 | \mu_4 \> = - \< \nu _0 | \mu_3 \> $. This concludes our
proof of the vanishing of the GSO sum of the part in $\cB$ of  (\ref{5k1}) involving $p_{L \ev}$ for 
the special choice of twist $\ev_0$. Modular covariance then automatically guarantees the vanishing of these
terms for all $\ev \in \cO_+$.

\subsection{The contribution in $\p_\tau \ln \tet _i$}

Given the vanishing of the contribution of the terms in $p_{L \ev}$, established in the preceding
subsection, the GSO sum of $\cB$ in (\ref{5k2}) simplifies to the following expression, 
\bea
\label{5n1}
\sum _\delta C_\delta [\ev] \, \cB [\delta; \ev] (p_{L\ev}; \Omega)= 
16 \pi^6 \sum _\delta C_\delta [\ev] \,   \sum _{\g=1}^3 V_i ^\g \Gamma [\delta ;e^\g] 
\prod _{\lambda =2,3,4 } {\tet [\delta + e^\lambda ](0, \Omega) \over \tet [\delta ] (0,\Omega)}
\eea
where we have defined, 
\bea
\label{5n2}
V_i^\g = -  \p_{\tau_\g } \ln { \tet _i (0, \tau_\g )^2 \over \eta (\tau_\g)^2 } 
\eea
and where the indices $\g$ and $i$ are again determined by $\delta = \delta _i ^{\g \pm}$.
We have used the cancellation already shown for the $p_{L \ev}$ contribution to freely insert 
the $\eta$-function term in (\ref{5n2}).
We  evaluate the derivative terms using (\ref{A13}),  and we find in the notations of Appendix A, 
\bea
\label{5n3}
V_2^\g & = & - {i \pi \over 6} \left ( \tet _3(0, \tau_\g)^4 + \tet _4(0, \tau_\g)^4 \right )
\no \\
V_3^\g & = & - {i \pi \over 6} \left ( \tet _2(0, \tau_\g)^4 - \tet _4(0, \tau_\g)^4 \right )
\no \\
V_4^\g & = & + {i \pi \over 6} \left ( \tet _2(0, \tau_\g)^4 + \tet _3(0, \tau_\g)^4 \right )
\eea
Next, we parametrize  the summation over $\delta$ by setting $\delta = \delta ^{\g \a} _i$, 
and rearranging the summation so as to expose the sum over $\g$,
\bea
\label{5n4}
\sum _\delta \cB [\delta; \ev]( p_{L \ev}; \Omega) = \sum _{\g=1}^3 \cB^\g  [\ev] (\Omega)
\eea
where the reduced amplitude $\cB^\g$ for fixed $\g$ is given by, 
\bea
\label{5n5}
\cB ^\g [\ev] (\Omega) = \sum _{\a = \pm} \sum _{i=2,3,4} 
16 \pi ^6 \, C_\delta [\ev] \,  V_i ^\g  \Gamma [\delta^{\g \a} _i  ;e^\g] \prod _{\lambda =2,3,4 } 
{\tet [\delta + e^\lambda ](0, \Omega) \over \tet [\delta ] (0,\Omega)}
\eea
We shall now calculate the contribution $\cB ^\g$ for each value of $\g=1,2,3$, and fixed $\ev \in \cO_+$.

\sm

We begin by evaluating $\cB^\g [\ev] $ for the special choice of twist $\ev_0=(\ep_2, \ep_3, \ep_4)$. 
The GSO phases $C_\delta [\ev_0]$ for all $\delta \in \cD[\ev_0]$ are equal for this twist, and may be set to~1.
Below, we present the even spin structures $\delta ^{\g \alpha} _i$ in the basis provided by the
sets $\cD [e^\g]$ as a function of  $\g \in \{ 1,2,3 \} $, $\alpha \in \{ \pm \} $ and $i \in \{ 2,3,4 \}$, 
\bea
\label{5n6}
\delta ^{1+} _i = \left [ \matrix{  \mu _i  \cr \mu_3 \cr } \right ]
& \qquad
\delta ^{2+} _i = \left [ \matrix{  \mu_3 \cr \mu _i  \cr } \right ]
& \qquad \quad
\delta ^{3+} _2 = \left [ \matrix{  \mu_2  \cr \mu_2 \cr } \right ]
\quad 
\delta ^{3+} _3 = \left [ \matrix{  \mu_3  \cr \mu_3 \cr } \right ]
\quad 
\delta ^{3+} _4 = \left [ \matrix{  \mu_3  \cr \mu_4 \cr } \right ]
\no \\ && \no \\
\delta ^{1-} _i = \left [ \matrix{  \mu _i  \cr \mu_4 \cr } \right ]
& \qquad
\delta ^{2-} _i = \left [ \matrix{  \mu_4 \cr \mu _i  \cr } \right ]
& \qquad \quad
\delta ^{3-} _2 = \left [ \matrix{  \nu_0  \cr \nu_0 \cr } \right ]
\quad 
\delta ^{3-} _3 = \left [ \matrix{  \mu_4  \cr  \mu_4  \cr } \right ]
\quad
\delta ^{3-} _4 = \left [ \matrix{  \mu_4  \cr  \mu_3  \cr } \right ]
\qquad
\eea
Recall that, by construction, we have $\delta ^{\g + } _i + \delta ^{\g -} _i = e^\g$ for all $\g$ and all $i$.
Also, the spin structure $\delta _2 ^{\g \pm} +e^ {\g '} $ is  odd when $\g'  \not=\g$, and will not enter 
into the products of $\tet$-constants in (\ref{5n5}). As a result, for each $\g$, the spin structures $\delta ^{\g \alpha}_i$
contribute only when $i=3,4$. Putting all together,  the contribution $\cB ^ \g [\ev_0]$ is given by, 
\bea
\label{5n7}
\cB^\g [\ev_0 ] = - { i \over 8 \pi \eta (\tau_\g)^{12}} \, { \tet _j (0,\tau_\g)^8 \over \tet [\delta _j ^{\g +}]^4  
\tet [\delta _j ^{\g -}]^4} \prod _{\kappa \in \cD [\ev_0]} \tet [\kappa ] (0, \Omega) 
\sum _\alpha \sum _{i=3,4}\< \nu _0 | \mu _i \> V_i^\g
\eea
The sum over $\alpha$ gives a factor of 2, while the sum over $i$ may be evaluated using (\ref{5n3}),
\bea
\label{5n8}
\sum _{i=3,4}  \< \nu _0 | \mu _i \> V_i ^\g  = V_3^\g - V_4^\g = - { i \pi \over 2} \tet _2 (0, \tau_\g)^4
\eea
As a result, we find, 
\bea
\label{5n9}
\cB^\g [\ev_0 ] = - { \tet _2 (0, \tau_\g)^4 \over 8  \eta (\tau_g)^{12}} \, { \tet _j (0,\tau_\g)^8 \over \tet [\delta _j ^{\g +}]^4  
\tet [\delta _j ^{\g -}]^4} \prod _{\kappa \in \cD [\ev_0]} \tet [\kappa ] (0, \Omega) 
\eea
Using the Schottky relation (\ref{2d5}) we carry out the following rearrangement,
\bea
\label{5n10}
{ \tet _j (0,\tau_\g)^8 \over \tet [\delta _j ^{\g +}]^4 \,  \tet [\delta _j ^{\g -}]^4}
=
{ \tet _3 (0,\tau_\g)^4 \,  \tet _4 (0,\tau_\g)^4 \over \tet [\delta _3 ^{\g +}]^2 \,  \tet [\delta _3 ^{\g -}]^2 \, 
 \tet [\delta _4 ^{\g +}]^2 \,  \tet [\delta _4 ^{\g -}]^2}
\eea
Using the  formula $\tet _2 \tet _3 \tet _4 = 2 \eta ^3$ of (\ref{A12a}), we see that the product of 
genus one $\tet$-functions in the numerator of (\ref{5n9}) cancels against a factor of $ 16 \eta ^{12}$ 
in the denominator, leaving the following simplified result,
\bea
\label{5n11}
\cB^\g [\ev_0 ] = - 2 { \prod _{\kappa \in \cD [\ev_0]}
 \tet [\kappa ] (0, \Omega)  \over  \tet [\delta _3 ^{\g +}]^2  
 \tet [\delta _3 ^{\g -}]^2 \tet [\delta _4 ^{\g +}]^2  \tet [\delta _4 ^{\g -}]^2} 
 = - 2 \prod _{\kappa \in \cD[\ev_0]}  \tet [\kappa] (0, \Omega )^{-1} 
\eea
Noticing that the right hand side of formula (\ref{5n11})  is independent of 
the index $\g$, we readily obtain the final form for the GSO sum of $\cB [\ev_0]$, 
\bea
\label{5n12}
\sum _\delta \cB [\delta; \ev_0]( p_{L \ev}, \Omega) =  - 6 \prod _{\kappa \in \cD[\ev_0]}  \tet [\kappa] (0, \Omega )^{-1} 
\eea
The corresponding result for arbitrary twist $\ev \in \cO_+$ may be obtained from (\ref{5n12}) 
by modular transformation.

\subsection{Vanishing contribution from the bulk for orbit $\cO_+$}

Collecting all the GSO projected contributions to the top component of the left chiral measure 
$d\mu_L$ for the choice of twist $\ev_0 = (\ep_2, \ep_3, \ep_4) \in \cO_+$ gives the following expression, 
\bea
\label{3g1}
\sum _\delta   d\mu_L[\delta;\ev ] (p_{L\ev}; \Omega) 
=
{d^3\Omega \, \cQ [\ev] ( p_{L \ev}) \, Z_B [\ev ]  \over 16\pi^6 \prod _{\delta \in \cD[\ev ]} \tet [\delta ] }
\left (  \sum _{\delta \in \cD[\ev ]} { \Xi_6[\delta]  \over \Psi _{10} }
\prod _{\kappa \in \cD [\ev ]} \tet [\kappa ]^2 -6 \right )
\eea
for $\ev= \ev_0$. 
In the subsequent section, we shall prove a modular identity (\ref{F1}) for all $\ev \in \cO_+$ from which it follows that
the factor in parentheses on the right hand side vanishes pointwise on moduli space when $\ev = \ev _0$.
At the same time the identity (\ref{F1}) will also automatically provide the proper modular generalization 
of (\ref{3g1}) to all $\ev \in \cO_+$.

\newpage

\section{Modular Factorization Identity}
\setcounter{equation}{0}
\label{secF}

In this section, we shall prove the fundamental modular identity responsible for the vanishing
of the contribution to the vacuum energy arising from the orbit $\cO_+$ and from the interior
of supermoduli space. We begin by stating the identity, and then prove its validity and 
key properties in a series of steps.

\sm

For any triplet of twists $\ev   \in \cO_+$, we have the following  factorization identity, 
\bea
\label{F1}
\Bigg ( \sum _{\delta \in \cD [\ev]} \<  \delta_a | \delta \>  \, \Xi _6 [\delta ] (\Omega) \Bigg ) 
\prod _{\delta \in \cD [\ev]} \tet [\delta ](0,\Omega) ^2
=  6 \lambda [\delta_a, \ev ] \Psi _{10}(\Omega)
\eea
where $\delta _a$ is any spin structure in $\cD[\ev]$, and $\lambda [ \delta_a, \ev]$ 
takes values $\pm 1$. We shall prove formula (\ref{F1}), and some of its properties using 
the following steps, 

\begin{enumerate}
\itemsep = -0.05in
\item The identity (\ref{F1})  is covariant under any change of choice of $\delta _a \in \cD[\ev]$; 
\item The identity transforms as a modular form of weight 6 under the  subgroup $Sp(4, \bZ)/\bZ_4$ 
of the full modular group $Sp(4, \bZ)$;
\item Using the hyper-elliptic representation, we prove that the square of  (\ref{F1}) holds;
\item Using degeneration limits we determine  the sign $\lambda$, and 
verify its modular covariance. 

\end{enumerate}

\subsection{Covariance under change of reference spin structure $\delta_a$}

Any triplet of even spin structures $\{ \delta_a, \delta_b, \delta_c \}  \subset \cD[\ev]$ obeys the relation,  
\bea
\label{F1a}
\< \delta _a | \delta _b \> \< \delta _b | \delta _c \> \< \delta _c | \delta _a \> =+1
\eea
or, in the terminology of \cite{Igusa}, is {\sl syzygous}. The relation (\ref{F1a}) is trivially satisfied 
if two spin structures coincide. When the three spin structures are mutually distinct,  we use the fact that 
they  are related by the non-zero twists $e^1, e^2 \in \ev$, so that 
$\delta _b = \delta _a + e^1$ and $\delta _c = \delta _a + e^2$. Simplifying the product 
of (\ref{F1a}), we find, $\< \delta _a | \delta _b \> \< \delta _b | \delta _c \> \< \delta _c | \delta _a \>
= \< e^1 | e^2 \>$. For twists in $\cO_+$, this evaluates to $+1$, as announced.

\sm

Next, consider the sum of $\Xi_6$ in (\ref{F1}), and substitute $\< \delta _a | \delta _b \> 
\< \delta _b | \delta\>$ for $\< \delta _a | \delta\>$,
using the result of (\ref{F1a}). It is then manifest that we have,
\bea
\sum _{\delta \in \cD [\ev]} \<  \delta_a | \delta \>  \, \Xi _6 [\delta ] (\Omega)
= 
\< \delta _a | \delta _b \> \sum _{\delta \in \cD [\ev]} \<  \delta_b | \delta \>  \, \Xi _6 [\delta ] (\Omega)
\eea
As a result, the transformation law for $\lambda$ must be  given by,
\bea
\lambda [\delta _a ; \ev] = \< \delta _a | \delta _b \> \lambda [\delta _b ; \ev] 
\eea
for any reference spin structures $\delta _a, \delta_b \in \cD [\ev]$.

\subsection{Covariance under the modular subgroup $Sp(4,\bZ)/\bZ_4$}

The factorization identity (\ref{F1}) will be shown to hold for any $\ev \in \cO_+$ but the representative $\ev$ 
in $\cO_+$ transforms non-trivially under the  modular group $Sp(4,\bZ)$. Thus, (\ref{F1}) is 
not covariant under the full $Sp(4,\bZ)$, but only under the subgroup that leaves the triplet of
twists $\ev$ invariant. Actually, the identity (\ref{F1}) is invariant under  permutations of the 
3 components $e^\gamma$ of the vector $\ev$. Thus, (\ref{F1}) will be invariant under the 
subgroup of $Sp(4,\bZ)$ that leaves $\ev$ invariant as a set. 

\sm

To determine this group explicitly may be done for a convenient triplet of twists, 
which we take to be $\ev = (\ep_2, \ep_3, \ep_4)$.
By inspection of the action of the modular generators in Appendix \ref{secB}, 
we see that the generators $M_1, M_2, M_3, \Sigma, T$ leave the set $\{ \ep_2, \ep_3, \ep_4 \}$ 
invariant, while $S$ does not. The subgroup generated by
$S$ is given by $\bZ_4 = \{ I, S, -I, -S\}$. Since $S$ coincides with the symplectic matrix, this subgroup  is normal,
and $Sp(4,\bZ)/\bZ_4$ is itself a group. The identity (\ref{F1}) is covariant under this group.

\subsection{The hyper-elliptic representation of $\delta$ and $\tet [\delta]^4$}

We shall prove that the square of (\ref{F1}) holds, using the hyper-elliptic representation for genus 2.
The correspondence between $\tet$-constants and the hyper-elliptic parametrization is
provided by the Thomae formulas. We denote the branch points by $p_a$
for $a=1, \cdots, 6$, and the corresponding  odd spin structures by $\nu_a$.
For genus 2, there is a one-to-one map between branch points and spin structures, given by
the Abel map, $\nu_a = p_a - \Delta$, where $\Delta $ is the Riemann vector (see Appendix B).
Each even spin structure $\delta $ corresponds to a partition of the set of branch points 
into two sets of 3 branch points. Equivalently $\delta$ may be written as the sum of
the corresponding three odd spin structures in two different ways,
\bea
\label{F2}
\delta = \nu_{a_1} + \nu _{a_2} + \nu _{a_3} = \nu_{b_1} + \nu _{b_2} + \nu _{b_3}
\to ( p_{a_1} , p_{a_2}, p_{a_3}) \cup ( p_{b_1} , p_{b_2}, p_{b_3})
\eea
with $(a_1, a_2, a_3, b_1, b_2, b_3)$ a permutation of $(1,2,3,4,5,6)$. There are 
10 different such partitions, namely the total number of even spin structures.
The Thomae formulas give \cite{Fay},
\bea
\label{F3}
\tet [\delta ] ^8 = c^2 \prod _{i < j} (p_{a_i} - p_{a_j})^2 (p_{b_i} - p_{b_j})^2
\eea
where $i,j \in \{1,2,3\}$, and $c$ is a $\delta$-independent function of moduli (the expression 
of which will not be needed here). 

\sm

While  (\ref{F1}) involves factors of $\tet [\delta]^2$, the square of (\ref{F1}) involves only 
the 4-th powers $\tet [\delta]^4$ (recall that $\Xi_6[\delta]$ involves only 4-th powers as well).
Thus we need the hyperelliptic representation of $\tet [\delta]^4$, including its precise overall signs.
To determine the signs, we us the following explicit parametrization of  the square roots of 
the Thomae relations (\ref{F3}), 
\bea
\label{F4}
\tet [\delta _0]^4 & = & c\, u_0 (p_1-p_3)(p_3-p_5)(p_5-p_1) \cdot (p_2-p_4)(p_4-p_6)(p_6-p_2)
\no \\
\tet [\delta _1]^4 & = & c\, u_1 (p_1-p_4)(p_4-p_6)(p_6-p_1) \cdot (p_2-p_3)(p_3-p_5)(p_5-p_2)
\no \\
\tet [\delta _2]^4 & = & c\, u_2 (p_1-p_2)(p_2-p_6)(p_6-p_1) \cdot (p_3-p_4)(p_4-p_5)(p_5-p_3)
\no \\
\tet [\delta _3]^4 & = & c\, u_3 (p_1-p_2)(p_2-p_5)(p_5-p_1) \cdot (p_3-p_4)(p_4-p_6)(p_6-p_3)
\no \\
\tet [\delta _4]^4 & = & c\, u_4 (p_1-p_4)(p_4-p_5)(p_5-p_1) \cdot (p_2-p_3)(p_3-p_6)(p_6-p_2)
\no \\
\tet [\delta _5]^4 & = & c\, u_5 (p_1-p_2)(p_2-p_4)(p_4-p_1) \cdot (p_3-p_5)(p_5-p_6)(p_6-p_3)
\no \\
\tet [\delta _6]^4 & = & c\, u_6 (p_1-p_5)(p_5-p_6)(p_6-p_1) \cdot (p_2-p_3)(p_3-p_4)(p_4-p_2)
\no \\
\tet [\delta _7]^4 & = & c\, u_7 (p_1-p_2)(p_2-p_3)(p_3-p_1) \cdot (p_4-p_5)(p_5-p_6)(p_6-p_4)
\no \\
\tet [\delta _8]^4 & = & c\, u_8 (p_1-p_3)(p_3-p_4)(p_4-p_1) \cdot (p_2-p_5)(p_5-p_6)(p_6-p_2)
\no \\
\tet [\delta _9]^4 & = & c\, u_9 (p_1-p_3)(p_3-p_6)(p_6-p_1) \cdot (p_2-p_4)(p_4-p_5)(p_5-p_2)
\eea
Here the coefficients $u_\a$ can take the values $\pm 1$. We may set $u_0=+1$
without loss of generality; the remaining signs are then uniquely determined by the Riemann
identities, and are found to be, (computed here using MAPLE),
\bea
\label{F5}
u _\a = \left \{ \matrix{ +1 & \hbox{when} & \a= 0,1,4,6,8,9 \cr -1 & \hbox{when} & \a=2,3,5,7 \cr} \right .
\eea
Finally, the form $\Psi _{10}$ is the discriminant, and corresponds to, 
\bea
\label{F6}
(\Psi _{10}) ^2 = \prod _{k=0}^9 \tet [\delta_k ]^4 = c^{10} \prod _{a<b} (p_a-p_b)^4
\eea

\subsection{The hyperelliptic representation of $ \ev \in \cO_+$}

Next, we parametrize triplets of twists $\ev= ( e^1, e^2, e^3) \in \cO_+$  in the hyperelliptic representation.
There is a one-to-one correspondence between the 15 non-zero single twists and sums of two
odd spin structures (mod 1), as simple counting confirms. Thus, every single twist $e^\g$
with $\g =1,2,3$ in $\ev$ may be uniquely expressed  as follows, 
\bea
\label{F7}
e^\g=\nu_{c^\g}+ \nu_{d^\g}
\eea
To guarantee that $\ev \in \cO_+$, we calculate the symplectic invariant between any two twists, 
\bea
\label{F8}
\< e^1 | e^2 \> = 
\< \nu_{c^1} | \nu _{c^2} \> \< \nu_{c^1} | \nu _{d^2} \> 
\< \nu_{d^1} | \nu _{c^2} \> \< \nu_{d^1} | \nu _{d^2} \> 
\eea
Since $e^1 \not= e^2$ for $\ev \in \cO_+$, the sets $\{ \nu _{c^\g}, \nu_{d^\g} \}$ of spin structures 
must be distinct for $\g=1$ and $\g=2$, so that their intersection may either contain one element, 
or be empty. 

\sm

Supposing first that one element is common, say $\nu_{d^1}=\nu_{d^2}$, equation (\ref{F8}) will simplify to
$\< e^1 | e^2 \> =  \< \nu_{c^1} | \nu _{c^2} \> \< \nu_{c^1} | \nu _{d^1} \>  \< \nu_{d^1} | \nu _{c^2} \>  $.
Since $\nu_{c^1} , \nu _{c^2} , \nu_{d^1}$ are all distinct, this product equals $-1$, 
so that the corresponding $\ev$ must belong to $\cO_-$. In the contrary case, we have $\< e^1 | e^2 \>=1$,
and $\ev \in \cO_+$. 
Thus, in the hyperelliptic representation, a twist $\ev \in \cO_+$ uniquely corresponds to a partition of the set of six
branch points into three sets of two branch points. The number of such partitions is $6!/(2!)^3=90$, which 
agrees with the result $\# \cO_+=90$ recorded at the end of Section \ref{sec:MO}.
Note that formula (\ref{F1}) is invariant under permutations of the entries $e^\g$ in $\ev$;
the number of such symmetric partitions  is $6! / (2!)^3/3! = 15$.

\sm

The set $\cD [\ev]$ of even spin structures associated with $\ev \in \cO_+$  
is characterized as follows,
\bea
\label{F9}
\cD [\ev] = \left \{ \delta = \nu_{a^1}+\nu_{a^2} + \nu_{a^3} ~ \hbox{with} ~ 
\# \Big ( \{ \nu_{a^1}, \nu_{a^2}, \nu_{a^3} \}
\cap \{ \nu _{c^\g} , \nu _{d^\g} \} \Big ) =1 ~ \hbox{for} ~ \g=1,2,3 \right \}
\no
\eea
One verifies that $\# \cD [\ev]=4$.

\subsection{Proving the square of equation {\rm (\ref{F1})}}

We shall prove the square of equation (\ref{F1}) first for the twist $\ev_0 = (\ep_2, \ep_3, \ep_4) \in \cO_+$, 
in the conventions of (\ref{1f1}), and then use the modular covariance of the identity to deduce 
the result for arbitrary $\ev \in \cO_+$. The hyperelliptic representation for the twist $\ev_0 $
in terms of a partition of the branch points, is given as follows,
\bea
\label{F10}
\ep_2 = \nu_2+\nu_4 \hskip 1in \ep_3 = \nu_1 + \nu _3 \hskip 1in \ep_4= \nu_5+\nu_6
\eea
The associated set of spin structures is $\cD [\ev_0] = \{ \delta _1, \delta _2, \delta _3, \delta _4 \}$,
in the conventions of (\ref{B3}). Using the Thomae formulas of (\ref{F4}), we then have,
\bea
\label{F11}
\prod _{\delta \in \cD[\ev_0 ] } \tet [\delta ]^4 
& = &
c^4 (p_1-p_2)^2   (p_1-p_4)^2 (p_1-p_5)^2 (p_1-p_6)^2
(p_2-p_3)^2 (p_2-p_5)^2 
\no \\ && \times 
(p_2-p_6)^2 (p_3-p_4)^2
(p_3-p_5)^2 (p_3-p_6)^2 (p_4-p_5)^2 (p_4-p_6)^2
\eea
Note that this result coincides with the product over all pairs, excluding those corresponding to the twist $\ev_0$ which,
according to (\ref{F10}), is given by  the factor $(p_1-p_3)^2 (p_2-p_4)^2(p_5-p_6)^2$. 
The  contributions $\Xi_6[\delta]$ to the identity  take the following  form in terms of $\tet$-constants,
\bea
\label{F12}
\Xi_6 [\delta _1] & = & 
- \tet [\delta _0]^4 \tet [\delta _3]^4 \tet [\delta _7]^4
- \tet [\delta _2]^4 \tet [\delta _6]^4 \tet [\delta _9]^4
- \tet [\delta _4]^4 \tet [\delta _5]^4 \tet [\delta _8]^4
\no \\
\Xi_6 [\delta _2] & = & 
+ \tet [\delta _0]^4 \tet [\delta _4]^4 \tet [\delta _8]^4
- \tet [\delta _1]^4 \tet [\delta _6]^4 \tet [\delta _9]^4
+ \tet [\delta _3]^4 \tet [\delta _5]^4 \tet [\delta _7]^4
\no \\
\Xi_6 [\delta _3] & = & 
+ \tet [\delta _0]^4 \tet [\delta _4]^4 \tet [\delta _6]^4
- \tet [\delta _1]^4 \tet [\delta _8]^4 \tet [\delta _9]^4
+ \tet [\delta _2]^4 \tet [\delta _5]^4 \tet [\delta _7]^4
\no \\
\Xi_6 [\delta _4] & = & 
+ \tet [\delta _0]^4 \tet [\delta _3]^4 \tet [\delta _6]^4
- \tet [\delta _1]^4 \tet [\delta _5]^4 \tet [\delta _8]^4
+ \tet [\delta _2]^4 \tet [\delta _7]^4 \tet [\delta _9]^4
\eea
By inspection of Table \ref{table:1}, the GSO signs for the twist $\ev_0 = (\ep_2, \ep_3, \ep_4) $ 
are all equal to 1, for any choice of reference spin structure $\delta _a \in \cD[\ev_0]$. 
 The sum over these four $\Xi_6 [\delta]$ may now be carried out and, 
using  (\ref{F4}), converted to the hyperelliptic representation. 
Each term $\Xi_6[\delta]$ in the sum over $\delta \in \cD [\ev_0]$ is a polynomial of total degree 18 in $p$,
divisible by  $\prod _{a<b} (p_a-p_b) $. One finds (using MAPLE),
\bea
\label{F13}
\sum _{\delta \in \cD [\ev_0 ] } \Xi_6 [\delta] = 6 c^3 (p_1-p_3)(p_2-p_4)(p_5-p_6) \prod _{a<b} (p_a-p_b)
\eea
Combining  (\ref{F11}) and (\ref{F13}), we prove the square of  (\ref{F1}) for the twist $\ev_0  $, 
\bea
\label{F14}
\bigg ( \sum _{\delta \in \cD [\ev_0 ]}  \Xi _6 [\delta ] \bigg )^2 \prod _{\delta \in \cD [\ev_0]} \tet [\delta ]^4
= 36 c^{10} \prod _{a<b} (p_a-p_b)^4 = 36 (\Psi _{10})^2
\eea
Its validity for arbitrary $\ev \in \cO_+$ follows from modular covariance. Having established the validity of (\ref{F14})
confirms that the factor $\lambda [\delta_a ; \ev]$ in identity (\ref{F1}) can take values $\pm 1$ only.

\subsection{Sign $\lambda$ from separating degeneration}

The sign factor $\lambda [\delta_a ; \ev]$ in the identity (\ref{F1}) may be determined 
from the asymptotic behavior in the separating degeneration limit. This limit is achieved by
letting the off-diagonal entry $\tau$  of the genus 2 period matrix, 
\bea
\label{5c1}
\Omega = \left ( \matrix{ \tau_1 & \tau \cr \tau & \tau_2 \cr} \right )
\eea
tend to 0, while keeping $\tau_1$ and $\tau_2$ fixed. 
In this limit, both sides of the relation (\ref{F1}) tend to 0. In particular, we have \cite{D'Hoker:2001qp},
\bea
\label{psi10}
\Psi _{10} (\Omega) =  - 2^{14} \pi^2 \tau^2 \eta (\tau_1)^{24} \eta (\tau_2)^{24} + \cO(\tau^4)
\eea
Thus, we shall need to retain all terms in the expansion, up to order $\cO(\tau^2)$ included.
In particular, we shall need the asymptotics to this order of the modular objects $\Xi_6 [\delta]$
in the separating degeneration limit. They are given as follows \cite{D'Hoker:2001qp},
\bea
\label{5c2}
\Xi _6 \left [\matrix{\mu_i \cr \mu_j} \right ] (\Omega)
& = &
2^8 \<\mu_i|\nu_0\> \<\mu_j|\nu_0\>
\eta (\tau_1)^{12} \eta (\tau_2)^{12}
\biggl \{ -1
+ {\tau ^2 \over 2} \biggl [
3\p  \ln \tet _i^4 (0,\tau_1) \p \ln \tet _j^4 (0,\tau_2)
\nonumber \\ && 
- \p \ln \eta (\tau _1)^{12} \p \ln \tet _j^4 (0,\tau_2)
-  \p \ln \tet _i ^4 (0,\tau_1) \p \ln \eta (\tau _2)^{12}
\biggr ] \biggr \} + \cO(\tau^4) \qquad
\no \\
\Xi _6 [\delta _0]  (\Omega)
& = & - 3 \times 2^8 \eta (\tau_1)^{12} \eta (\tau_2)^{12} + \cO(\tau^2)
\eea
Here $\mu_i$ refers to the three even spins structures on each genus 1 component of the separating
degeneration, while $\delta_0$ refers to the unique odd -- odd decomposition. The derivatives with 
respect to moduli $\tau_1$ and $\tau_2$ in (\ref{5c2}) may be evaluated using  formula (\ref{A13}).
Note that the expansion of $\Xi _6 [\delta _0] (\Omega)$ has been retained
only to lowest order, as its coefficient in the sum over all $\Xi_6$ will already be of order $\cO (\tau^2)$.

\sm

Concentrating again on the twist $\ev_0 = (\ep_2, \ep_3, \ep_4)$, the separating degeneration 
asymptotics of the sum of the $\Xi_6 [\delta ]$ terms over the four spin structures $\delta \in \cD [\ev_0]$ 
may be simplified and yields the following formula, 
\bea
\label{5c4}
\sum _{\delta \in \cD[\ev _0 ]} \Xi_6 [\delta]
= - 24 \, (4 \pi \tau)^2 \, \eta (\tau_1)^{12} \eta (\tau_2)^{12} \tet_2(0,\tau_1)^4 \tet _2 (0,\tau_2)^4
+ \cO(\tau^4)
\eea
Our final ingredient is the separating degeneration asymptotics of the product of $\tet [\delta]$, 
\bea
\label{3g2}
\prod _{\delta \in \cD [\ev_0 ]} \tet [\delta ]^2 
=
\tet _3 (0, \tau_1)^4 \tet _4 (0, \tau_1)^4 \tet _3 (0, \tau_2)^4 \tet _4 (0, \tau_2)^4 + \cO(\tau^2)
\eea
Combining all these, we find, 
\bea
\label{3g3}
\sum _{\delta \in \cD[\ev_0]}  \Xi_6 [\delta]  \prod _{\delta \in \cD [\ev_0]} \tet [\delta]^2  = 6 \Psi _{10}  
\eea
Hence we have 
\bea
\lambda [\delta _1, \ev_0]=1
\eea
The values of $\lambda [\delta _a, \ev]$ for general $\ev \in \cO_+$ may be readily deduced by 
modular transformation, and are listed in Table \ref{table:1}. We have also double-checked the 
values of  $\lambda$ for general $\ev$ by explicit  calculation in the separating degeneration limit.

\newpage

\section{Boundary of Supermoduli Space, Part I}
\setcounter{equation}{0}
\label{sec4}

In this section and the next, we shall derive the contributions to the vacuum energy 
from the boundary of supermoduli space.  The procedure for their determination 
has been laid out in \cite{Witten:2013cia}. The boundary contributions are due to the regularization of the pairing, 
near the boundary of supermoduli space, of the bottom component $d\mu_L^{(0)}$ in the left chiral 
measure with the right chiral measure along a suitable integration cycle $\Gamma$. As was explained in \cite{Witten:2013cia}, 
(see also section 5 of \cite{Witten:2013tpa}), only separating degenerations contribute. Henceforth, 
we shall restrict attention to this case. 

\sm

In this section, we begin by developing detailed formulas for the separating degeneration asymptotics
of the bottom component $d\mu_L^{(0)}$ in the left chiral measure. We give a preliminary account of the 
regularization procedure
of \cite{Witten:2013cia}, and apply it to show that contributions from orbits $\cO_i$ for $i=0,1,2,3, -$ 
to the vacuum energy from the boundary of supermoduli space vanish for both Heterotic strings.
To do so, we make use of the $Sp(2,\bZ)_1 \times Sp(2,\bZ)_2$ modular subgroup  which leaves the separating
node invariant. In the next section, a more detailed account of the regularization procedure
of \cite{Witten:2013cia} will be given, and the contributions from the remaining orbit $\cO_+$ will be evaluated.

\subsection{Preliminaries}

For convenience, we reproduce here the formula for the bottom component 
of the measure $d\mu_L^{(0)}$ which was written down already in (\ref{4a2}),\footnote{
We stress that the moduli argument of $d\mu_L^{(0)}$ is the super-period matrix $\hat \Omega$.
Contrarily to its top counterpart $d \mu _L$, the bottom component $d\mu_L^{(0)}$ does not allow for 
its argument $\hat \Omega$ to be freely altered to $\Omega$.}
\bea
\label{4z1}
d\mu_L^{(0)}[\delta;\ev](p_{L\ev}; \hat \Omega)
=
\cZ[\delta] (\hat \Omega) {Z_C[\delta;\ev] (\hat \Omega) \over Z_M[\delta] (\hat \Omega) } 
\, \cQ [\ev] (p_{L\ev}) d^3 \hat \Omega
\eea
The factor $\cZ[\delta]$ is given by (\ref{4a3}), and is independent of the twist $\ev$. 
It will be helpful to separate its spin structure dependent part as follows,
\bea
\label{4g6}
{\cal Z}[\delta] (\hat \Omega) \equiv  {\cZ_0 (\hat \Omega)  \, 
\tet[\delta](0, \hat \Omega)^5 \over \tet[\delta](q_1+q_2-2\Delta, \hat \Omega)}
\eea
The spin structure independent factor $\cZ_0$ may be considerably simplified 
using the hyperelliptic representation, with a judicious
choices of the  points $p_1,p_2,p_3$ and $z_1, z_2, w_0, w_1, w_2$ in (\ref{4a3}) and (\ref{4a4}).
The result was derived in \cite{D'Hoker:2001qp}, formula (3.15), and we have,  
\bea
\label{4g7a}
\cZ_0={\cC_1 E(p_1,p_2)^4 \sigma(p_1)^2\sigma(p_2)^2
\over
(\cM_{\nu_1\nu_2})^2 E(q_1,q_2) \sigma(q_1)^2\sigma(q_2)^2}.
\eea
Here, the points  $p_a$ with $a=1,2$  are arbitrary distinct branch points corresponding to the odd spin 
structures $\nu_a$ by the relation $\nu_a = p_a - \Delta$. The combination $\cM_{\nu_1 \nu_2}$ is 
given by \cite{D'Hoker:2001qp},
\bea
\label{4d3}
\cM_{\nu_1 \nu_2} = 
\p_1 \tet [\nu_1 ] (0, \hat \Omega) \p_2 \tet [\nu_2 ] (0, \hat \Omega) 
- \p_1 \tet [\nu_2 ] (0, \hat \Omega) \p_2 \tet [\nu_1 ] (0, \hat \Omega)
\eea
while the prefactor $\cC_1$ is given by,
\bea
\label{4d4}
\cC_1 = \exp \left \{ 2 \pi i \Big ( \nu_1 ' \hat \Omega \nu_1'  
+ \nu_2 ' \Omega \nu_2 ' - 4  \nu_1' \hat \Omega \nu_2' \Big )  \right \}
\eea
The combination $\cZ_0$ is independent of the choice of the distinct 
odd spin structures $\nu_1, \nu_2$.

\subsection{General formulas for the separating degeneration limit}
\label{sec72}

In this section, we shall review some basic formulas for holomorphic Abelian differentials, the period matrix,
$\tet$-functions,  the Riemann vector, the $\tet$-divisor, and the prime form, in the separating 
degeneration limit.\footnote{To lighten notations,
the hats on the super-period matrix and its components will not be exhibited of Sections \ref{sec72}, 
\ref{sec73}, and \ref{sec74}, but will be restored in the final formulas in Section \ref{sec74}.}
The general reference for this material is \cite{Fay}, with additional information
for Green functions in \cite{went}.

\subsubsection{Holomorphic Abelian differentials}

We consider a genus 2 surface $\Sigma$ and a choice of canonical homology basis $A_I, B_I$ with $I=1,2$.
The separating degeneration is taken along a trivial homology cycle which separates $\Sigma$ into two 
genus 1 components, which we denote $\Sigma_I$ with respective homology bases given by the cycles 
$A_I, B_I$. Following \cite{Fay}, we parametrize the degeneration with a complex parameter $t$,  
and degeneration  points $s_I$ on the surfaces surface $\Sigma _I$. The  holomorphic Abelian differentials
with canonical normalization behave as follows,
\bea
\label{4b1}
\omega_1  & = &  \varpi _1(z) +{t \over 4}  \varpi _1 (s_1) \varpi ^{(1)} _{s_1} (z)  \hskip 0.36in  z \in \Sigma _1 
\no \\ & = & 
{t \over 4} \varpi _1 (p_1) \varpi ^{(2)} _{s_2} (z) \hskip 1in  z \in \Sigma _2
\no \\ && \no \\
\omega_2 & = &  {t \over 4} \varpi _2 (s_2) \varpi ^{(1)} _{s_1} (z) \hskip 1in  z \in \Sigma _1
\no \\ & = & 
\varpi _2(z) +{t \over 4} \varpi _2 (s_2) \varpi ^{(2)} _{s_2} (z)  \hskip 0.36in  z \in \Sigma _2 
\eea
up to order $t^2$.
Here $\varpi _I(z)$ are the holomorphic Abelian differentials on component $I=1,2$, and 
$\varpi _s ^{(I)} (z)$ is the second kind Abelian differential on component $I$ with double
pole at $s$ with unit residue. The normalizations are as follows,
\bea
\label{4b2}
\oint _{A_I} \varpi _J = \delta _{IJ} & \hskip 1in & \oint _{B_I} \varpi _J = \delta _{IJ} \tau_I
\no \\
\oint _{A_I} \varpi _s^{(J)} = 0 ~ & \hskip 1in & \oint _{B_I} \varpi _s^{(J)}  = 2 \pi i \delta _{IJ} \varpi _I(s)
\eea

\subsubsection{The period matrix}

As a result, the period matrix  admits the following expansion,
\bea
\label{4b3}
\Omega = \left (
\matrix{
\tau_1 + 
 \frac{\pi i}{2} \, t\,  \varpi_1^2(s_1) &\qquad  \frac{\pi i}{2} \, t \, \varpi_1(s_1)\, \varpi_2(s_2) \cr
& \cr
 \frac{\pi i}{2} \, t \, \varpi_1(s_1)\, \varpi_2(s_2) & \qquad \ \tau_2 + 
 \frac{\pi i}{2} \, t\, \varpi_2^2(s_2)\cr } \right ) + \cO(t^2)
\label{4b4}
\eea
The parameters  $\tau_I$  on the diagonal are the moduli of each genus 1 components $\Sigma _I$.
The off-diagonal entry is proportional to the degeneration parameter $t$ via the relation, 
\bea
\Omega_{12} = \tau =  \frac{\pi i}{2} \, t \, \varpi_1(s_1)\, \varpi_2(s_2) + \cO (t^2)
\eea
Choosing each $\Sigma _I$ to be flat allows us to set $\varpi_I(y_I) = dy_I$, so that $\tau = i \pi t/2$.

\subsubsection{$\tet$-functions}

Genus two  half-integer characteristics are denoted as follows  (see also Appendix B),
\bea
\kappa =  \left [ \matrix{ \kappa _1  \cr \kappa _2  \cr} \right ]
\hskip 1in 
\kappa _I = \left [ \kappa _I ' ~ \kappa _I '' \right ]
\eea
where $\kappa _I ', \kappa _I '' \in \{ 0, 1/2\}$. The genus two $\tet$-function with characteristic $\kappa$
has the following expansion in powers of $\tau$, for fixed $\zeta=(\zeta_1, \zeta_2)^t$ (not to be confused with the 
odd moduli),
\bea
\label{4c1}
\tet [\kappa] (\zeta, \Omega) = 
\sum _{p=1} ^\infty { 1 \over p!} \left ( { \tau \over 2 \pi i} \right )^p \p_{\zeta _1} ^p \tet [\kappa _1] (\zeta _1, \tau_1)
\p_{\zeta _2} ^p \tet [\kappa _2] (\zeta _2, \tau_2)
\eea
For generic $\zeta $, the leading behavior is for $p=0$, and we find,
\bea
\label{4c2}
\tet [\kappa] (\zeta , \Omega)  ~ \to ~ 
 \tet [\kappa _1] (\zeta _1, \tau_1) \tet [\kappa _2] (\zeta _2, \tau_2) + \cO (\tau)
\eea
We shall also need the degeneration at special values of $\zeta $, such as even and odd spin structures.
The genus one odd spin structure is denoted $\nu_0$, and any of the three even spin structures
is denoted $\mu$. For the 10 even spin structures, we then have, 
\bea
\label{4c3}
\tet \left [ \matrix{ \mu_1 \cr \mu_2 } \right ] (0,\Omega) 
& = &  \tet [\mu _1] (0, \tau_1) \tet [\mu _2] (0, \tau_2) + \cO (\tau^2 )
\no \\
\tet \left [ \matrix{ \nu_0 \cr \nu_0 } \right ] (0,\Omega) 
& = &  {\tau \over 2 \pi i} \tet _1 ' (0, \tau_1) \tet _1 ' (0, \tau_2) + \cO (\tau^3 )
\eea
while for the six odd spin structures we have,
\bea
\label{4c4}
\p_1 \tet \left [ \matrix{ \mu \cr \nu_0 } \right ] (0,\Omega) 
& = & 2 \tau \p_{\tau_1}  \tet [\mu ] (0, \tau_1) \tet '_1 (0, \tau_2) + \cO (\tau^3 )
\no \\
\p_2 \tet \left [ \matrix{ \mu \cr \nu_0 } \right ] (0,\Omega) 
& = & \tet [\mu] (0, \tau_1) \tet '_1 (0, \tau_2) + \cO (\tau^2 )
\no \\
\p_1 \tet \left [ \matrix{ \nu_0 \cr \mu } \right ] (0,\Omega) 
& = & \tet _1' (0, \tau_1) \tet [\mu]  (0, \tau_2) + \cO (\tau^2 )
\no \\
\p_2 \tet \left [ \matrix{ \nu_0 \cr \mu } \right ] (0,\Omega) 
& = & 2 \tau  \tet '_1 (0, \tau_1) \p_{\tau_2}  \tet [\mu] (0, \tau_2) + \cO (\tau^3 )
\eea

\subsubsection{The $\tet$-divisor and the Riemann vector}

If $\zeta$ is an arbitrary point  in the $ \tet$-divisor, so that $\tet (\zeta, \Omega)=0$,
then we have the following asymptotics of the $\tet$-function, 
\bea
\label{4c5}
\tet (\zeta + x-y, \Omega) = \tet (\zeta _1+x-s_1, \tau_1) \tet (\zeta _2 + s_2 -y, \tau_2) + \cO (\tau)
\eea
with $\lim _{t\to 0} \zeta =(\zeta _1,\zeta _2)^t$,  and $x\in \Sigma _1$ and $y \in \Sigma _2$. 
Formula (\ref{4c5}) was derived as Proposition 3.6 in 
\cite{went}. The associated formula for $x,y $ in the same component
will not be needed here. Degeneration limits of other quantities which do not fit this formula
will also be needed. To this end, we give next a careful derivation of the 
separating degeneration limits of the $\tet$-divisor and the Riemann vector $\Delta _I$.

\sm

Consider a general element of the genus two $\tet$-divisor,
\bea
\label{4e8}
(p-\Delta) _I = \int _{z_0} ^p \omega_I - \Delta _I (z_0)
\eea
where the Riemann vector is given by,
\bea
\label{4e9}
\Delta _I (z_0) = \half - \half \Omega _{II} 
+ \sum _{K\not = I} \oint _{A_K} \omega_K (z) \int ^z _{z_0} \omega_I
\eea
The combination $p-\Delta$ is independent of the base point $z_0$. 
It will be convenient to separate its components $\Sigma _I$ of the degeneration, 
\bea
\label{4e11}
(p-\Delta )_1 = - \half + \half \Omega _{11} 
+  \oint _{A_2} \omega_2 (z) \int ^p _z \omega_1
\no \\
(p-\Delta )_2 = - \half + \half \Omega _{22} 
+  \oint _{A_1} \omega_1 (z) \int ^p _z \omega_2
\eea
In the separating degeneration limit, to leading order, we have $\Omega _{II} \to \tau _I$,
and  the genus one Riemann vectors of $\Sigma _I$ become, 
\bea
\label{4e14a}
\Delta ^{(I)} = \half - { \tau _I \over 2} 
\eea
For $ p \in \Sigma _1$, the path of integration from $z$ to $p$ in the  integral term for $(p-\Delta )_2$
in (\ref{4e11}) lies entirely inside $\Sigma _1$ where $\omega_2$ is of order $\cO(t)$. As a result, the 
integral term vanishes to leading order in $t$. Similarly, for $p \in \Sigma _2$, the integral
term in $(p-\Delta)_1$ vanishes to this order. In $(p-\Delta)_1$, the integral may be split up as follows,
\bea 
\label{4e12}
\int ^p _z \omega_1 = \int ^p _{s_1} \omega_1 + \int _z ^{s_2} \omega_1
\eea
In the second integral on the rhs above, the path lies entirely inside $\Sigma _2$, where $\omega_1$
is of order $\cO(t)$, and thus vanishes to leading order in $t$. Only the first integral on the rhs survives, 
and we find for $p \in \Sigma _1$, 
\bea
\label{4e13}
(p-\Delta )_1 & = & - \Delta ^{(1)} + p-s_1 + \cO(t)  
\no \\
(p-\Delta )_2 & = & - \Delta ^{(2)}  + \cO(t)
\eea
Similarly, for $p \in \Sigma _2$, we have,
\bea
\label{4e14}
(p-\Delta )_1 & = & - \Delta ^{(1)}  + \cO(t)
\no \\
(p-\Delta )_2 & = & - \Delta ^{(2)} + p-s_2 + \cO(t)
\eea
It is clear that for $p \in \Sigma _1$, the quantity $(p-\Delta)_2$ tends to the genus one $\tet$-divisor,
while for $p\in \Sigma _2$, it is the component   $(p-\Delta)_1$ that tends to the genus one $\tet$-divisor.

\subsubsection{The prime form and related quantities}

We shall also need the degeneration of the prime form $E(z,w)$ of the genus 2 surface. 
Special care needs to be taken when one of the points is in the degeneration funnel,
but we shall not need such behavior here. The remaining degeneration limits are as follows \cite{Fay}, 
\bea
\label{4c6}
z \in \Sigma _I, ~ w \in \Sigma _I & \hskip 0.8in & 
E(z,w) ~ \to ~ E^{(I)} (z,w)
\no \\
z \in \Sigma _1, ~ w \in \Sigma _2 &  & 
E(z,w) ~ \to ~ { 1 \over \sqrt{t}} \, E^{(1)} (z,s_1) \, E^{(2)} (s_2, w)
\eea
where $E^{(I)}(z,w)$ is the prime form on $\Sigma _I$, given by,
\bea
\label{4c7}
E^{(I)} (z,w) = { \tet _1 (z-w, \tau_I) \over \tet _1 ' (0, \tau_I)}
\eea
Finally, we shall need the degeneration of the form $\sigma (z)$ of weight $(h/2,0)=(1,0)$ for genus~2.
This may be obtained from the defining formula in the second line of (\ref{4a4}).
When $z, w \in \Sigma _1$, we choose $w_1\in \Sigma _1$ and $w_2 \in \Sigma _2$.
The leading behavior of the $\tet$-functions is obtained by using (\ref{4c2}) and (\ref{4e14}),
or equivalently (\ref{4c5}) with $\zeta= w_1 - \Delta$, and we find, 
\bea
\label{4c8}
\tet ( w_1+w_2-z -\Delta, \Omega ) = \tet (w_1-z - \Delta ^{(1)}, \tau_1 ) 
\tet (w_2 - s_2 - \Delta ^{(2)}, \tau_2) + \cO (\tau)
\eea
where the genus one Riemann vectors $\Delta ^{(I)} $ were introduced in (\ref{4e14a}).
As a result, we find,  
\bea
\label{4c9}
{ \sigma (z) \over \sigma (w) } = 
{ \sigma ^{(1)} (z) E^{(1)} (w,s_1)  \over  \sigma ^{(1)}  (w)  E^{(1)} (z,s_1) } + \cO (\tau)
 \eea
where the genus one $\sigma$ function is given by, 
\bea
\label{4c10}
{ \sigma ^{(1)} (z) \over \sigma ^{(1)}  (w) } = 
{ \tet (w_1 - z - \Delta ^{(1)}, \tau_1) E^{(1)} (w,w_1)  
\over 
 \tet (w_1 - w - \Delta ^{(1)}, \tau_1) E^{(1)} (z,w_1)  }  = e^{ i \pi (z-w)}
 \eea
Here, we have used  (\ref{A12}) and (\ref{4c7})  to reach the final simplified form on the right.
Putting all ingredients together produces the final degeneration formula for $\sigma$, 
\bea
\label{4c12}
{ \sigma (z) \over \sigma (w) }
= e^{i \pi (z-w)} \, { \tet _1 (w-s_1, \tau_1) \over  \tet _1 (z-s_1, \tau_1)} + \cO (\tau)
\eea
These degeneration formulas also agree with \cite{Verlinde:1986kw}, though care is needed
for the special circumstance of the degeneration components being tori.

\subsection{The degeneration limit of $\cZ_0$}
\label{sec73}

As explained in section 3.3.2 of \cite{Witten:2013cia}, the only natural way of choosing 
the points $q_I$ along the separating node  is to have $q_1$ and $q_2$ 
lie on opposite genus 1 components $\Sigma _I$. This is because each $\Sigma _I$ 
is a torus with a single puncture, and affords precisely one odd modulus. 
Indeed,  the supersymmetry variation equation $ \p_{\bar z} \xi ^+ = \chi _{\bar z} ^+$ can always
be solved on the torus, but the solution does not allow one to set $\xi ^+( s)=0$.
Thus on the torus with one puncture, there is one mode of $\chi _{\bar z} ^+$ 
which cannot be gauged away, and results in a single odd modulus.

\sm

Without loss of generality, we shall choose $q_I \in \Sigma _I$ for $I=1,2$.
Since the branch points $p_I$ may be chosen arbitrarily, we set $p_I \in \Sigma _I$.
The corresponding odd spin structures $\nu_I$  then take the form, 
\bea
\label{4g1}
\nu _1 = \left [ \matrix{ \mu_1 \cr \nu_0 \cr } \right ]
\hskip 1in 
\nu _2 = \left [ \matrix{ \nu_0 \cr \mu_2 \cr } \right ]
\eea
with $\mu_1$ and $\mu_2$ even.
We have the following asymptotic behavior,
\bea
\label{4g2}
\cC_1 & = & \exp { \pi i \over 2} \Big \{  \tau_1 \left (1 - 6 \mu_1' \right ) +  \tau_2 \left ( 1 - 6 \mu_2' \right ) \Big \}
\no \\ && \no \\
(\cM _{\nu _1 \nu_2})^2 & = &  
\tet_1' (0, \tau_1)^2 \tet '_1 (0, \tau_2)^2 \tet [\mu_1] (0, \tau_1)^2 \tet [\mu_2] (0, \tau_2)^2
+ \cO(\tau^2)
\eea
as well as
\bea
\label{4g3}
{ \sigma (p_1)^2 \sigma (p_2)^2 \over \sigma (q_1)^2 \sigma (q_2)^2} 
=  \cC_2 \, { \tet _1 (q_1-s_1, \tau_1)^2 \, \tet _1 (q_2-s_2, \tau_2)^2
\over \tet _1 (p_1-s_1, \tau_1)^2 \, \tet _1 (p_2-s_2, \tau_2)^2} + \cO(\tau)
\eea
with 
\bea
\label{4g4}
\cC_2 = e^{2\pi i (p_1 +p_2 -q_1-q_2 )}
\eea
To leading order the prime forms in (\ref{4g7a}) degenerate as follows, 
\bea
\label{4g5}
E(p_1,p_2)^4 & = & { 1 \over t^2} \, { \tet _1 (p_1-s_1, \tau_1)^4  \tet _1 (p_2-s_2, \tau_2)^4
\over \tet _1 '(0, \tau_1)^4 \, \tet _1 '(0, \tau_2)^4}
\no \\
E(q_1,q_2) & = & { 1 \over \sqrt{t}} \, { \tet _1 (q_1-s_1, \tau_1)  \tet _1 (q_2-s_2, \tau_2)
\over \tet _1 '(0, \tau_1) \, \tet _1 '(0, \tau_2)}
\eea
Combining these factors gives the following leading order behavior of $\cZ_0$,
\bea
\label{4g7}
\cZ_0 = { 1 \over t^{3/2}} \, \cC_1 \cC_2 { 
\tet _1 (p_1-s_1, \tau_1)^2 \, \tet _1 (p_2-s_2, \tau_2)^2
\tet _1 (q_1-s_1, \tau_1) \, \tet _1 (q_2-s_2, \tau_2)
\over 
\tet _1 '(0, \tau_1)^5 \, \tet _1 '(0, \tau_2)^5 \, 
 \tet [\mu_1] (0, \tau_1)^2 \, \tet [\mu_2] (0, \tau_2)^2}
\eea
To simplify, we apply the limits of $p-\Delta$ obtained in (\ref{4e13}) and (\ref{4e14}) to $p- \Delta$, and we find,
\bea
\label{4g8}
p_1-s_1 - \Delta ^{(1)} & = & \mu_1
\no \\
p_2 - s_2 - \Delta ^{(2)} & = & \mu_2
\eea
The remaining equations resulting from (\ref{4e13}) and (\ref{4e14a}) state that $\nu_0 ^{(1)} = - \Delta ^{(1)}$, and 
$\nu_0 ^{(2)} = - \Delta ^{(2)}$, which are manifestly obeyed, up to an immaterial shift by 
the period $1$ in each torus. Considering the $p_I$-dependent factors in $\cZ_0$, we have, 
\bea
\label{4g9}
 { \tet _1 (p_I-s_I, \tau_I)^2 \over  \tet [\mu_I] (0, \tau_I)^2} \, 
 e^{ i \pi \tau _I  (1/2 -3\mu_I') + 2 \pi i (p_I-s_I)}
= - e^{- i \pi \tau_I}
\eea
where the corresponding simplifications have been carried out with the help
of the formulas (\ref{A4}) of Appendix \ref{secA}.  Putting together the remaining pieces in $\cZ_0$, we find, 
\bea
\label{4g10}
\cZ_0 = { \cC_3 \over t^{3/2}} \, 
{  \tet _1 (q_1-s_1, \tau_1) \, \tet _1 (q_2-s_2, \tau_2)
\over 
\tet _1 '(0, \tau_1)^5 \, \tet _1 '(0, \tau_2)^5}
\eea
where 
\bea
\label{4g11}
\cC_3 = e^{2 \pi i ( s_1-q_1+s_2-q_2 -\tau_1/2-\tau_2/2)} 
\eea
As expected, this complete expression for $\cZ_0$ is independent of the branch points $p_I$.

\subsection{The degeneration limit of $\cZ[\delta]$}
\label{sec74}

We can now determine the degeneration limit of $\cZ[\delta]$ in (\ref{4g6}), 
which combines the factor $\cZ_0$ with a factor involving the spin structure $\delta$.
We shall work to leading order in $t$. Two cases need to be distinguished, according 
to how $\delta$ reduces onto the genus 1 components $\Sigma _I$. 

\subsubsection{$\delta \to $ even - even}

We first consider the case of the 9 even spin structures
which reduce to even spin structures on both components $\Sigma _I$.
The restriction of the spin structure may then be written as,
\bea
\label{4g12}
\delta = \left [ \matrix{ \delta _1 \cr \delta _2 \cr } \right ]
\eea
where both $\delta _1$ and $\delta _2$ are even. Using (\ref{4c1})  yields the following degeneration, 
\bea
\label{4g13}
\tet [\delta ] (0, \Omega ) ^5 = \tet [\delta _1] (0, \tau_1)^5 \tet [\delta _2] (0, \tau_2)^5 + \cO(\tau^2)
\eea
To compute the $\delta$-dependent contribution in the denominator, 
we   use the calculations of the Riemann vector in (\ref{4e12}) and (\ref{4e14a}), 
and we find, 
\bea
\label{4g14}
(q_1+q_2-2 \Delta)_I = q_I - s_I - 2 \Delta ^{(I)} 
\eea
so that
\bea
\label{4g15}
\tet [\delta ] (q_1+q_2-2 \Delta, \Omega ) 
= 
 \cC_3 \, \cC_4 \prod _{I=1}^2 \tet [\delta _I] (q_I-s_I , \tau_I) + \cO(\tau)
\eea
where $\cC_3$ was defined in (\ref{4g11}), and $\cC_4$ is given by,
\bea
\cC_4 = (-)^{2 \delta _1 ' + 2 \delta _1 '' + 2 \delta _2 ' + 2  \delta _2 ''} 
 = \< \nu_0 | \delta _1 \>  \< \nu_0 | \delta _2 \> 
\eea
Using now the result of (\ref{4g6}), we find to leading order in $t$, 
\bea
\label{4g17}
\cZ [\delta] =  { 1 \over t^{3/2}} \,   \prod _{I=1,2} \< \nu_0 | \delta _I \>
{\tet _1 (q_I-s_I, \tau_I) \, \tet [\delta _I] (0, \tau_I)^5
\over 
\tet [\delta _I] (q_I - s_I, \tau_I) \,  \tet _1 '(0, \tau_I)^5 } 
\eea
Next, we use the fact that the Szeg\"o kernel on each genus 1 component $\Sigma _I$ 
for even spin structure $\delta _I$ is given by,
\bea
\label{4g18}
S_{\delta _I} (z,w, \tau_I) 
= { \tet [\delta _I] (z-w, \tau_I) \tet _1 '(0, \tau_I) 
\over \tet [\delta _I] (0, \tau_I) \tet _1(z-w, \tau_I)} 
\eea
In summary, we assemble all the parts, and restore the original notation $\hat \Omega, \hat \tau$, and $\hat t$
to clearly exhibit  the dependence  on the super-period matrix, and we find, 
\bea
\label{4g19}
\cZ [\delta] (\hat \Omega) =  { 1 \over \hat t^{3/2}} \,  \prod _{I=1,2}
{ \< \nu_0 | \delta _I \>  \,   \tet [\delta _I] (0, \hat \tau_I)^4
\over 
S_{\delta _I} (q_I - s_I, \hat \tau _I) \,  \tet _1 '(0, \hat \tau_I)^4 } 
\eea

\subsubsection{$\delta \to $ odd - odd}
\label{sec:ABC}

Next, we consider the case of $\delta = \delta _0$ reducing to odd spin structures on both components.
The restriction of the spin structure $\delta _0$, and the limit of the $\tet$-constant are as follows,
\bea
\delta_0 = \left [ \matrix{ \nu_0 \cr \nu_0 \cr } \right ]
\hskip 1in 
\tet [\delta_0 ](0, \Omega)^5 = \left ( { \tau \over 2 \pi i} \right )^5 
\tet _1 '(0,\tau_1)^5 \tet _1 '(0,\tau_2)^5 
\eea
while we also need, 
\bea
\tet [\delta_0 ] (q_1+q_2-2 \Delta, \Omega ) 
= \prod _{I=1}^2 \tet _1 (q_I-s_I  + \tau_I, \tau_I)  + \cO(\tau)
\eea
Putting all together, all dependence on $q_1, q_2 $ drops out. Restoring the original notation 
$\hat \Omega, \hat \tau$, and $\hat t$ to exhibit  the dependence  on the super-period matrix,  we find, 
\bea
\cZ[\delta_0] = { 1 \over \hat t^{3/2} } \left ( { \hat \tau \over 2 \pi i} \right )^5 
\eea
This contribution vanishes as $\hat t \to 0$, will not contribute to the vacuum energy.

\subsection{Regularization near $\hat \tau=0$}
\label{sec:TZ}

Near $\hat \tau=0$, we shall follow the prescription developed in \cite{Witten:2013cia}, and interpolate between
matching $\hat \Omega =\Omega_R$ in the bulk of supermoduli space and matching $\Omega = \Omega _R$ 
at the boundary. Thus we need  to evaluate the difference $\hat \Omega - \Omega$ 
in the degeneration limit. This may be achieved  using the formula 
for the super-period matrix written to 2-loop order \cite{D'Hoker:1989ai},
\bea
\label{4h1}
\hat \Omega _{IJ} = \Omega _{IJ} - { i \over 8 \pi} \int _\Sigma d^2 z \int _\Sigma d^2 w
\omega_I (z) \chi _{\bar z} ^+ S_\delta (z,w) \chi _{\bar w} ^+ \omega_J (w) 
\eea
with worldsheet gravitini supported at $q_I$, 
\bea
\chi _{\bar z} ^+ = \sum _{I=1,2} \zeta ^I \delta (z, q_I)
\eea
For spin structure $\delta$ given by (\ref{4g12}),  the limit of the genus two Szego kernel $S_\delta$
as $z \in \Sigma _1$ and $w \in \Sigma _2$  is given in terms of the Szego kernels $S_{\delta _I}$ on the genus one 
components by,
\bea
\label{4h3}
S_\delta (z,w) = t^{1/2} \, S_{\delta _1} (z-s_1, \tau_1) S_{\delta _2} (s_2-w, \tau_2)
+ \cO(t^{3/2})
\eea
Thus, the asymptotic behavior of $\hat \Omega _{IJ} - \Omega _{IJ}$ is given by, 
\bea
\label{4h4}
\hat \Omega _{IJ} = \Omega _{IJ} - i \, t^{1/2} \, { \zeta ^1 \zeta ^2 \over 4 \pi} 
\varpi_I (q_1)  S_{\delta_1} (q_1-s_1, \tau_1)  S_{\delta_2} (q_2-s_2, \tau_2) \varpi_J (q_2) 
\eea
Using the fact that the genus one holomorphic Abelian differentials $\varpi_I$ are constant 
on their respective components, and may be set equal to 1, we see that
the degeneration parameters $t$ and $\hat t$ are related as follows 
(the differences $\hat \tau_I - \tau_I$ play no role here and may be omitted),
\bea
\label{4h8}
\hat t = t - t^{1/2} \, { \zeta ^1 \zeta ^2 \over 2 \pi^2} 
 S_{\delta_1} (q_1-s_1, \tau_1)  S_{\delta_2} (q_2-s_2, \tau_2) 
\eea
To regularize  the integrals, we follow \cite{Witten:2013cia}
and parametrize the  integration cycle $\Gamma$ near the separating node by (a more complete
prescription will be given in Section \ref{sec:81}),
\bea
\label{4z2}
\hat t^{1/2} =  t^{1/2} - h(t, \bar t) { \zeta ^1 \zeta ^2 \over 4 \pi^2} 
 S_{\delta_1} (q_1-s_1, \tau_1)  S_{\delta_2} (q_2-s_2, \tau_2) 
\eea
Here, $h(t, \bar t)$ is the regularization function which is subject to the following conditions, 
\bea
h (0,0)=1 \hskip 1in h (t, \bar t)=0 ~~ \hbox{for} ~~ |t| >1
\eea
Keeping $t$ fixed while taking $\hat t \to 0$ is carried out by eliminating $\hat t$ in 
(\ref{4h6}), in favor of $t$ and $ \zeta ^1 \zeta ^2$, and we find, 
\bea
\label{4h6}
\cZ [\delta] =  { 1 \over t^{3/2}} \,   \prod _{I=1,2}
{ \< \nu_0 | \delta _I \>  \,  \tet [\delta _I] (0, \tau_I)^4 \over 
S_{\delta _I} (q_I - s_I) \,  \tet _1 '(0, \tau_I)^4 } 
+ 3\,  { \zeta ^1 \zeta ^2 \over 4 \pi^2} \, { h(t, \bar t)  \over t^2} \,  \prod _{I=1,2}
{  \< \nu_0 | \delta _I \>  \,  \tet [\delta _I] (0, \tau_I)^4 \over  \tet _1 '(0, \tau_I)^4 } 
\eea
The second term on the rhs is the one of interest, as it will produce a non-zero contribution
upon integrating out the odd moduli. Notice that all dependence on $q_I$ has cancelled out,
so that this boundary contribution is properly slice-independent.

\subsection{Irreducible orbits under $SL(2,\bZ)_1 \times SL(2, \bZ)_2$}
\label{sec:QB}

To study the contributions  from different twists  in the separating degeneration limit,
it will be convenient to separate them into irreducible orbits under the modular subgroup
$SL(2,\bZ)_1 \times SL(2,\bZ)_2 \times \bZ_2 \subset Sp(4, \bZ)$ which leaves the separating degeneration invariant.
The subgroup $SL(2,\bZ)_I$ transforms the component $\Sigma _I$, while $\bZ_2$ exchanges the two components. 

\sm

$\bullet $ Under $SL(2,\bZ)_1 \times SL(2,\bZ)_2$, the twists in  $\cO_\g$ for $\g=1,2,3$ transform into 
three irreducible orbits, depending on whether the twist $\ep$ reduces to zero on component $\Sigma _2$,
on component $\Sigma _1$, or on neither component. We shall denote the corresponding sets of 
twists respectively by $\cO _\g ^1$, $\cO_\g^2$ and $\cO_\g^0$. By inspection of Table \ref{table:1}, we have,
\bea
\label{7a1}
\cO _\g =  \cO_\g^0 \cup \cO_\g^1 \cup \cO_\g^2   \hskip 0.7in \# \cO_\g^0 = 9 \hskip 0.7in  \# \cO_\g ^1= \# \cO_\g^2 = 3
\eea
Each one of the orbits $\cO^0_\g,\cO^1_\g,\cO^2_\g$  transforms irreducibly under $SL(2,\bZ)_1 \times SL(2,\bZ)_2$.

\sm

$\bullet$ Under $SL(2,\bZ)_1 \times SL(2,\bZ)_2$, the twists in $\cO_+$ transform into two irreducible orbits.
The first orbit $\cO_+^e$ contains all twists $\ev$ for which all spin structures in 
$\cD [\ev]$ descend to even-even in the separating degeneration. The second orbit $\cO_+^o$ 
contains all twists $\ev$ for which one spin structure in  $\cD [\ev]$ descend to the odd-odd
spin structure in the separating degeneration. By inspection of Table \ref{table:1}, we have,
\bea
\label{7a2}
\cO _+ = \cO_+^e \cup \cO_+^o  \hskip 1in  \# \cO_+^e = 9 \hskip 1in  \# \cO_+^e = 6
\eea
Each one of the orbits $\cO_+^e$ and $\cO_+^o$ transforms irreducibly under $SL(2,\bZ)_1 \times SL(2,\bZ)_2$.
The contents of the orbits has been listed in Table \ref{table:1}.

\subsection{Contributions from the twists in orbits $\cO_0, \cO_1,\cO_2,\cO_3, \cO_-$}

In this subsection, we shall show that the contributions from the separating node for the 
twist orbits $\cO_0, \cO_1,\cO_2,\cO_3, \cO_-$ all vanish. This is clear for the untwisted
orbit $\cO_0$, as well as for the orbit $\cO_-$ whose contributions cancel in view of the fact
that the  spin structures associated with any twist in $\cO_-$ can never be all even. For the orbits  $\cO_1,\cO_2,\cO_3$
we shall show that the contributions from the separating node cancel in the left sector by itself.

\sm

For a twist in one of the orbits $\cO_1, \cO_2, \cO_3$, the contribution of the two
pairs of fields in the left sector twisted by $\ep$ is given by (\ref{2d3}),  and we have, 
\bea
\label{2d3bis}
{Z_C[\delta;\ep ](\Omega) 
\over
Z_M[\delta] (\Omega) }
=
{\tet [\delta_i^+](0, \Omega)^2 \, \tet [\delta_i^-](0, \Omega)^2 \, \tet [\delta+\ep ] (0, \Omega)^2
\over
\tet_\g  (0, \tau_\ep)^4 \, \tet [\delta](0, \Omega)^2}
\eea
Recall that, for a fixed non-trivial twist $\ep$,  there are 6 even spin structures $\delta $ 
for which $\delta+\ep$ is even, forming the set $\cD[\ep]$, defined in (\ref{2d20}). 
They can be listed as, $\delta_i^+, \delta_i^-$ for $ i=2,3,4$, and 
with $\delta_i ^+ + \delta^-_i =\ep$.

\subsubsection{Twists in $\cO_\g ^0$}

With the help of an $SL(2,\bZ)_1 \times SL(2,\bZ)_2$  transformation, any twist in $\cO_\g^0$ may be rotated 
to a reference twist $\ep=\ep_4$. The six associated even spin structures are, 
\bea
\ep= \bigg[{0|{1 \over 2} \atop 0|{1\over 2}}\bigg]
& \hskip 1in &
\delta_2^+=\bigg[{\mu_2 \atop \mu_2 }\bigg]
\hskip 0.6in 
\delta_3^+=\bigg[{\mu_3 \atop \mu_3 }\bigg]
\hskip 0.6in 
\delta_4^+=\bigg[{\mu_3 \atop \mu_4 }\bigg]
\no \\ &&
\delta_2^-=\bigg[{\nu_0  \atop \nu_0 }\bigg]
 \hskip 0.6in 
\delta_3^-=\bigg[{\mu_4  \atop \mu _4 }\bigg]
\hskip 0.6in 
\delta_4^-=\bigg[{\mu_4  \atop \mu _3 }\bigg]
\eea
where $\mu_i $ are the even genus $1$ spin structures. The degeneration limits of the products of  $\tet$-constants
that enter into $\cZ[\delta] Z_C[\ep, \delta] /Z_M[\delta]$ is as follows, 
\bea
\tet [\delta _2^+]^2 \tet [\delta _2^+]^2  & = & \cO(\tau^2)
\no \\
\tet [\delta _3^+]^2 \tet [\delta _3^+]^2  & = & 
\tet _3 (0,\tau_1)^2 \tet _4 (0,\tau_1)^2  \tet _3 (0,\tau_2)^2 \tet _4 (0,\tau_2)^2 + \cO(\tau^2)
\no \\
\tet [\delta _4^+]^2 \tet [\delta _4^+]^2  & = & 
\tet _3 (0,\tau_1)^2 \tet _4 (0,\tau_1)^2  \tet _3 (0,\tau_2)^2 \tet _4 (0,\tau_2)^2 + \cO(\tau^2)
\eea
Clearly, the pair $\delta _2 ^\pm$ does not contribute as $\tet [\delta _2^-]=0$
in the separating degeneration limit. The remaining spin structures $\delta _3 ^\pm$ and $\delta _4^\pm$
contribute with pairwise opposite signs and the same $\tet$-function factors in the separating degeneration
 limit, so the sum over $\delta$ cancels.

\subsubsection{Twists in $\cO_\g ^1$ and $\cO^2_\g$}

Consider first the case $\ev \in \cO_\g ^2$, so that $\ep $ reduces to the zero twist on component $\Sigma _1$.
(The case $\ev \in \cO_\g ^1$ is analogous by the action of $\bZ_2$.) Under $SL(2,\bZ)_1 \times SL(2,\bZ)_2$, 
the twist may be rotated to a standard twist, which we choose to be $\ep=\ep_2$ in the notations of
Table (\ref{1f1}). Below we also list the six associated even spin structures of $\cD[\ep]$, 
\bea
\ep= \bigg[{0|0\atop 0|{1\over 2}}\bigg]
\hskip 1in 
\delta_i^+=\bigg[{\mu_i \atop 0|0}\bigg]
\hskip 1in 
\delta_i^-=\bigg[{\mu_i \atop 0|{1\over 2}}\bigg]
\eea
where $\mu_i $ are the even genus $1$ spin structures.
In the separating degeneration limit, we have
\bea
\tet[\delta_i ^+]&=&\tet_i (0,\tau_1)\tet_3(0,\tau_2) +\cO(\tau^2)
\nonumber\\
\tet[\delta_i ^-]&=&\tet_i (0,\tau_1)\tet_4(0,\tau_2)+\cO(\tau^2)
\eea
We can quote now the result for the limit of the other $\tet$-constants obtained 
in \cite{ADP}, eq. (7.4), where it is also shown that $\tau_\ep = \tau_1 + \cO(\tau)$, 
\bea
{\tet[\delta_i ^+](0,\Omega)^2 \tet[\delta_i ^-](0,\Omega)^2
\over
\tet_i (0,\tau_\e)^4}
=
\tet_3(0,\tau_2)^2\tet_4(0,\tau_2)^2+\cO(\tau^2)
\eea
to arrive at
\bea
{Z_C[\delta_i^+;\ep ](\Omega)  \over Z_M[\delta^+_i ] (\Omega) } 
= \tet_4(0,\tau_2)^4+\cO(\tau^2)
\hskip 0.7in 
{Z_C[\delta_i^-;\ep ](\Omega)  \over Z_M[\delta^- _i ] (\Omega) }
= \tet_3(0,\tau_2)^4+\cO(\tau^2)
\eea
Putting all pieces together, integrating over the odd moduli $\zeta ^1, \zeta ^2$, 
and summing over the spin structures $\delta$ gives, to leading order in $t$,
\bea
\sum _\delta \int \! d^2 \zeta \, \cZ [\delta ] \, { Z_C [\delta; \ep ] \over Z_M[\delta] } 
=  { 3 h(t, \bar t) \over 4 \pi^2 t^2} \sum _{i =2,3,4}   \< \nu_0 | \mu_i \>    \sum _{a=3,4}  \< \nu_0 | \mu_a \> \,  
{\tet [\mu_i ] (0, \tau_1)^4   \over  \tet _2(0, \tau_1)^4  \tet _1 '(0, \tau_2)^4} 
\eea
up to corrections which are of order $\cO(t^0)$.
The contribution vanishes for two different reasons. First, since the argument under the 
sums is independent of $a$, the sum over $a$ vanishes since $\< \nu_0 | \mu_3 \> = - \< \nu_0 | \mu_4 \> $. 
Second, the summation
over $i$ also vanishes independently for each $a$ in view of the genus 1 Riemann idenity.

\newpage

\section{Boundary of Supermoduli Space, Part II}
\setcounter{equation}{0}
\label{sec6}

In this final section, we shall calculate the contribution  to the vacuum energy from the boundary part of 
supermoduli space for the orbit $\cO_+$. As in the case of the contributions from the interior of 
supermoduli space, it is the orbit $\cO_+$ that 
contains all the characteristically $\cN=1$ supersymmetry effects. Non-zero contributions from the 
boundary arise only from the  separating degeneration node \cite{Witten:2013cia}. We shall 
parametrize the supermoduli integration cycle $\G$ near this node, identity the possible non-zero 
contributions, and calculate the total boundary contribution for both  $E_8 \times E_8$ and 
$Spin (32) / \bZ_2$ Heterotic strings. In a last subsection, we shall also confirm that the boundary
contributions vanish for the  case of Type superstrings.

\subsection{Parametrizing the supermoduli integration cycle $\G$}
\label{sec:81}

In a canonical homology basis $A_I, B_I$, the separating node decomposes $\Sigma$ 
into the genus one components $\Sigma _I$ for $I=1,2$. To parametrize the supermoduli 
integration cycle $\Gamma$ for the Heterotic strings, we introduce the following notation 
for the components of the super-period matrix $\hat \Omega$ for left chirality, and the 
bosonic period matrix $\Omega _R$ for right chirality,
\bea
\label{9a1}
\hat \Omega = \left ( \matrix{ \hat \tau_1 & \hat \tau \cr \hat \tau &  \hat \tau_1 \cr} \right )
\hskip 1in 
\bar \Omega_R  = \left ( \matrix{ \tilde \tau_1 & \tilde \tau \cr \tilde  \tau &  \tilde \tau_2 \cr} \right )
\eea 
It will be convenient to use a parametrization in terms of the natural degeneration 
parameters $ \hat t$ and $\tilde t$ defined by $\hat \tau = i \pi \hat t /2$ and $\tilde \tau = -i \pi \tilde t /2$.
To leading order  as $\hat \tau, \tilde \tau \to 0$, this parametrization is equivalent to the parametrization 
using $\hat \tau, \tilde \tau$ in view of the asymptotics of the period matrix given in (\ref{4b4}). 
Following \cite{Witten:2013cia}, a suitable  integration cycle $\Gamma$ is parametrized in terms of 
local complex coordinates $t, \bar t$,  as was already given in part in equation (\ref{4z2}),  
\bea
\label{9a2}
\tilde  \tau _I = \bar \tau_I & \hskip 0.8in & \tilde t = \bar t 
\no \\
\hat \tau _I = \tau_I 
& & \hat t ^\half 
= t ^\half - h(t,\bar t)  { \zeta ^1 \zeta ^2 \over 4 \pi^2} \prod _{I=1,2} S_{\delta _I} (q_I-s_I, \tau_I)  
\eea
Here,  $h(t , \bar t) $ is a regularization function, and the remaining ingredients were explained in 
Section \ref{sec:TZ}.  As long as $h(t , \bar t)$ satisfies the boundary condition $h(0,0)=1$,
and vanishes outside an open set containing $t=0$ (such as for example $h(t,\bar t)=0 $ for $|t|>1$),
the precise shape of $h(t , \bar t)$ will be immaterial in view of the fact that the supermoduli integral
is independent of changes in $\Gamma$ thanks to a superspace version of Stokes' theorem
\cite{Witten:2012ga,Witten:2012bh}. It is also  this freedom that we have used to set the 
diagonal parts of $\hat \Omega$ and $\Omega _R$ equal to one another,  as no contribution arises 
from the non-separating degeneration node. Note that in the interior of supermoduli space,
where $h(t , \bar t)=0$, the cycle $\G$ agrees with the prescription $\hat \Omega = \Omega _R$
used in \cite{D'Hoker:2001nj,D'Hoker:2002gw}, for which cancellation was established in earlier sections.

\subsection{Form of the boundary contributions}

Using the parametrization of the integration cycle $\G$ given in (\ref{9a2}), the integrals
of the boundary contribution $\cV ^{\rm bdy}_G$ of (\ref{2t1}) may be written out more explicitly,
\bea
\label{9a3}
\cV^{\rm bdy} _G = g_s^2 \, \mN \! \prod _{I=1,2} \int _{\cM_1 } \! \!  \! \! \! d^2 \tau_I \! \!
\int \! \! d^4 p_I \sum _{\ev , \, p_{\ev},\, \delta} \! \! C_\delta [\ev] 
\hat \cQ [\ev] \bar \cQ [\ev] \! \int _D \!  \! d\hat t \, d \tilde t \! \int_\zeta  \! d \zeta ^1 d \zeta ^2 
\cL [\delta ; \ev] (\tau_I, \hat t) \overline{ \cR_\mn [\ev] (\tau_I, t)} \quad
\eea
Here, $\cM_1$ is the moduli space of genus 1 curves; $D$ is the unit disk 
$D =\{ |t|<1 \}$; the twists $\ev$ are summed over the orbit $\cO_+$;
the $p_I$-integrals are over the 4 uncompactified dimensions; the sum over $p_\ev$ runs over the 
internal loop momenta of the 6 compactified dimensions (which are all twisted when $\ev \in \cO_+$);
the internal loop momentum factors $\hat \cQ [\ev]$ and $\bar \cQ [\ev]$ are respectively evaluated
for $\hat \Omega$ and $\Omega _R$; and it is understood that $\hat t$ is defined in terms of 
$t, \bar t, \zeta^1, \zeta ^2$ by the  shape of the cycle given in (\ref{9a2}). Finally, 
$\cL [\delta ; \ev] (\tau_I, \hat t)$ and $\overline{ \cR_\mn [\ev] (\tau_I, t)}$ collect the  integrands
corresponding to left and right chiralities, and the index $\mn$ distinguishes the two Heterotic string gauge groups.
The general expressions for these functions are as follows,
\bea
\label{9a4}
\cL  [\delta ; \ev] (\tau_I, \hat t) & = & 
Z_B [\ev] (\hat \Omega)  { \cZ [\delta ] (\hat \Omega) \over \tet [\delta ] (0, \hat \Omega)^4}
 \prod _{\kappa \in \cD [\ev]}   \tet [\kappa ] (0, \hat \Omega) 
 \no \\
\cR_\mn [\ev] (\tau_I, t) & = &  
Z_B [\ev] (\Omega)  { \Psi _4 (\Omega) ^{(3-\mn)/2} \over \Psi _{10} (\Omega)}  
\sum _{\delta _R  \in \cD [\ev] } C_{\delta _R} [\ev] \, \tet [\delta _R] (0, \Omega)^{4\mn}
 \prod _{\kappa \in \cD [\ev]}   \tet [\kappa ] (0, \Omega) \qquad
\eea
where the index $\mn$ takes to following values,
\bea
\label{9a5}
\mn =1 & \hskip 1in & E_8 \times E_8
\no \\
\mn = 3 && Spin (32)/\bZ_2
\eea
and $Z_B$ is the contribution from twisted bosons, defined already in (\ref{5a3}).

\subsection{Identifying non-zero boundary contributions}

The starting point is the near-boundary expression for the flat-Minkowski 
space left chiral factor $\cZ [\delta]$ which, to leading order in $\hat t$, is given by (\ref{4g19}),
\bea
\label{6a1}
\cZ [\delta] =  { 1 \over \hat t^{3/2}} \,   \prod _{I=1,2}
{ \< \nu_0 | \delta _I \>  \,  \tet [\delta _I] (0, \tau_I)^4
\over 
S_{\delta _I} (q_I - s_I) \,  \tet _1 '(0, \tau_I)^4 } 
\eea
The next order corrections are in $\cO (1/\hat t^{1/2})$, $\cO(\hat t^{1/2})$, and so on, but these
will not be needed for reasons we shall now explain. To obtain the separating degeneration limit
of the full left chiral measure in (\ref{4z1}), we shall need also the contributions from the 
ratio $Z_C [\delta, \ev]/Z_M [\delta]$. Before working out these limits in detail, we shall
first carry out a general analysis of the orders in $\hat t$ and $\tilde t$ that can produce 
non-zero contributions at the boundary.

\sm 

We use the scaling arguments of \cite{Witten:2013cia} to determine which behavior in 
$\hat t$ will produce non-vanishing contributions. In terms of $\hat t$ and $\tilde t$,
the contributions to the measure are,
\bea
\label{6a4}
{ d \tilde t \over \tilde t^2} \cdot {d \hat t \over \hat t^{3/2}} \, d \zeta ^1 d \zeta ^2
\eea
The factor $\tilde t^{-2}$ is from the $\bar \Psi_{10}$ denominator for the right chirality, 
while the factor $\hat t ^{-3/2}$ is from the factor $\cZ[\delta]$. 
The variables suitable to this degeneration are $\tilde t$ and $\hat \rho = \hat t ^{1/2}$,
($\hat \rho$ corresponds to the variable $\ep$ used in \cite{Witten:2013cia}) in terms of which the measure becomes,
\bea
\label{6a5}
{ d \tilde t \over \tilde t^2} \cdot {d \hat \rho \over \hat \rho^2} \, d \zeta ^1 d \zeta ^2
\eea
Next, we need to include the small $\tilde t$ and $\hat \rho$ contributions from the 
twisted boson and fermion fields. We shall derive their asymptotic behaviors in the subsequent two sections.

\subsubsection{Contributions from twisted boson fields}

The contribution from the left  twisted bosons for a twist $\ev \in \cO_+$ is given by 
\bea
\label{6a6}
Z_B [\ev] =  \prod _{\g =1}^3 
 { \tet [ \delta ^{\g +} _j ]  \, \tet [ \delta ^{\g -} _j ] \,  \over  \tet _j (0,\tau_\g)^2 } 
\eea
evaluated on the super-period matrix $\hat \Omega$ of (\ref{9a1}). The contribution from the 
right twisted bosons is given by the complex conjugate of (\ref{6a6}), evaluated 
on the period matrix $\Omega _R$ of (\ref{9a1}). 
In each case, the corresponding $\tau_\g$ may be determined from the Schottky relations in 
(\ref{2d5}). A key observation is that the leading order is $\hat t^0 \, \tilde t^0$, 
with corrections of order $\hat t ^2$ and $\tilde t^2$, but not of order $\hat t$ and $\tilde t$.

\subsubsection{Contributions from fermion fields}

The contribution of twisted fermion fields to left chiral measure is through a factor,
\bea
\label{6a7}
\prod _{\g=1}^3 { \tet [\delta + e^\g] \over \tet [\delta ]}= { 1 \over \tet [\delta ]^4} \prod _{\kappa \in \cD [\ev]}
\tet [\kappa ]
\eea
multiplying the Minkowski-space factor $\cZ[\delta]$. We have  argued in Section \ref{sec:ABC}
that the spin structure $\delta = \delta _0$ (which decomposes to odd -- odd under separating degeneration)
does not contribute. For the remaining 9 even spin structures, we shall make use of the decomposition 
of $\cO_+$ into irreducible orbits $\cO_+^e$ and $\cO_+^o$,  under the $SL(2,\bZ) _1 \times SL(2,\bZ)_2$ 
modular subgroup introduced in Section \ref{sec:QB}. In terms of these orbits, we have the following behavior. 

\begin{itemize}
\itemsep=-0.05in
\item For $\ev \in \cO_+^e$ the leading order is $\hat t^0$, while the next order is $\hat t ^2$;
\item For $\ev \in \cO_+^o$ however, the leading order is $\hat t $, while the next order is $\hat t ^3$.
The extra power of $\hat t $ comes from the presence of the factor $\tet [\delta _0]$
in the product over $\kappa$; the separating degeneration limit for this spin structure 
produces an extra $\hat t $, see (\ref{4c3}).
\end{itemize}
The contribution of twisted fermions and non-twisted fermions from the right chiral measure, 
respectively for the $E_8 \times E_8$ for $\mn=1$ and $Spin(32)/Z_2$ for $\mn=3$, 
\bea
\label{6a8}
\left ( \bar \Psi _4 \right ) ^{(3-\mn)/2}  \, \left ( \overline{\tet [\delta _R]} \right )^{4 \mn} 
\prod _{\kappa \in \cD [\ev]} \overline{\tet [\kappa] }
\eea
Again, the spin structure $\delta_R = \delta _0$ does not contribute in the separating degeneration. 
The behavior for the remaining 9  even spin  structures is again arranged by orbits $\cO_+^e$ and $\cO_+^o$,  
\begin{itemize}
\itemsep=-0.05in
\item For $\ev \in \cO_+^e$ the leading order is $\tilde t ^0$, while the next order is $\tilde t ^2$;
\item For $\ev \in \cO_+^o$ however, the leading order is $\tilde t $, while the next order is $\tilde t ^3$.
\end{itemize}

\subsection{Summary of behavior by orbits $\cO_+^e$ and $\cO_-^o$}

In summary for the Heterotic string, combining left and right contributions we have the following behavior
arising from the twisted fields, arranged by orbits $\cO_+^e$ and $\cO_-^o$,
\begin{itemize}
\itemsep=-0.05in
\item For $\ev \in \cO_+^e$ the leading order is $\hat t^0 \, \tilde t ^0$, while the next orders are $\hat t ^2$
and $\tilde t ^2$;
\item For $\ev \in \cO_+^o$ the leading order is $\hat t \, \tilde t $, while the next orders are $\hat t ^3 \, \tilde t $
and $\hat t \,  \tilde t ^3$.
\end{itemize}
assembling these contributions gives the following,
\bea
\label{6a10}
\ev \in \cO_+^e & \hskip 1in & 
{ d \tilde t \over \tilde t^2} \cdot {d \hat \rho \over \hat \rho^2} \, d \zeta ^1 d \zeta ^2 \left ( 
1 + c \, \hat t ^2 + \tilde c \, \tilde t^2 + \cdots \right )
\no \\
\ev \in \cO_+^o & \hskip 1in & 
{ d \tilde t \over \tilde t} \cdot d \hat \rho  \, d \zeta ^1 d \zeta ^2 \left ( 
1 + c \, \hat t ^2 + \tilde c \, \tilde t^2 + \cdots \right )
\eea
We see that twists in $\cO_+^e$ single out the identity operator as leading contribution.
Spin structure summation in the left chiral blocks cancels the leading contribution as 
pointed out in section 3.2.5 of \cite{Witten:2013cia}. The higher order corrections lead to convergent integrals
and produce no boundary contributions.

\sm

Twists in $\cO_+^o$ produce precisely the scaling structure explained in \cite{Witten:2013cia}. Thus,
it is the contributions from orbit $\cO_+^o$ that need to be collected. The factor $\cZ [\delta]$ will 
contribute only to the leading order in $\hat t$ to which it has been computed; no higher orders are needed.

\subsection{Simplifications in the  orbit $\cO_+^o$}

For $\ev $ belonging to the orbit $\cO^o_+$, the following combination, 
\bea
Z_B [\ev] \prod _{\kappa \in \cD [\ev]} \tet [\kappa] 
=  \prod _{\g =1}^3 
 { \tet [ \delta ^{\g +} _j ]  \, \tet [ \delta ^{\g -} _j ] \,  \over  \tet _j (0,\tau_\g)^2 } \prod _{\kappa \in \cD [\ev]} \tet [\kappa] 
\eea
which is common to $\cL$ and $\cR_\mn$ in (\ref{9a4}), permits an important simplification.
To see this, we shall choose the index $j$ for each value of $\g$ in a particularly useful way.
We use the fact that $\cD [e^\g]$ has 6 elements, 4 of which also belong to $\cD [\ev]$, and 2 
of which do not. We shall label these two spin structures in a manner that exposes the dependence
of $j$ on $\gamma$, 
\bea
\label{6b3}
\cD [e^\g] \setminus \cD [\ev] = \Big \{ \delta ^{\g +} _{j (\g)} , \delta ^{\g -} _{j (\g)} \Big \}
\eea
The above relation uniquely defines  $j (\g)$, and this choice is canonical. 
By construction, as $\g $ ranges over the values $1,2,3$, the resulting 3 pairs of spin structures will
produce 6 distinct spin structures, none of which belongs to $\cD [\ev]$. This means 
that the six spin structures in question are precisely {\sl all} even spin structures that are not in $\cD [\ev ]$.
Thus we have,
\bea
\label{6b4}
Z_B [\ev] \prod _{\kappa \in \cD [\ev]} \tet [\kappa] 
=  { 1 \over \cP} \prod _{\kappa} \tet [\kappa ]   =  {1 \over \cP} \, (\Psi _{10} )^\half
\hskip 1in 
\cP \equiv  \prod _{\g =1}^3 \tet _{j (\g)} (0,\tau_\g)^2 
\eea

\subsubsection{Calculation of the Prym $\tet$-functions for twists in $\cO^o_+$}
\label{sec:PT}

Next, we shall calculate the combination $\cP$ of Prym $\tet$-functions defined in (\ref{6b4}),
to leading order in the separating degeneration limit. This will also require computing the 
Prym period $\tau_\g$ for each one of the twists $e^\g$ belonging to $\ev \in \cO^o_+$.

\sm

The Prym period $\tau_\g$ is associated with a single twist $e^\g$. The orbit under $Sp(4, \bZ)$ 
of 15 non-zero twists decomposes into three irreducible orbits under $SL(2,\bZ)_1 \times SL(2, \bZ)_2$.
Decomposing the twist under this product group, $e^\g = (\ep _\ell, \ep _r)$,
we see that $\ep_\ell$ and $\ep_r$ cannot both vanish, since then the genus 2
twist itself would vanish. The three reduced orbits are,
\bea
\label{6e1}
\cT _\ell & = & \{ ( \ep _\ell \not = 0, \ep _r=0) \}
\no \\
\cT _r & = & \{ ( \ep _\ell  = 0, \ep _r \not =0) \}
\no \\
\cT  & = & \{ ( \ep _\ell  \not = 0, \ep _r \not =0) \}
\eea
with $\# \cT_\ell= \# \cT_r =3$ and $\# \cT =9$, totaling 15 non-zero genus 2 twists.
Each one of these orbits is irreducible under $SL(2,\bZ)_1 \times SL(2, \bZ)_2$.

\sm

It is a characteristic of the orbit $\cO_+^o$ that all of its twists belong to $\cT$. The list 
of the 9 twists $e^\g=\ep _a$ of $\cT$ is given by $a=4, 6, 8, 9, 10, 13, 14, 15, 16$
in the notation of Table (\ref{1f1}).
We shall now evaluate the Prym period $\tau_\g$ by using the Schottky relations,
\bea
\label{6e2}
{\tet [\delta ^+ _i ] (0, \Omega)^2 \, \tet [\delta ^- _i ] (0, \Omega) ^2
\over \tet [\delta ^+ _j ] (0, \Omega)^2 \, \tet [\delta ^- _j ](0, \Omega)^2}
={ \tet _i (0, \tau_\g)^4 \over \tet _j (0, \tau_\g)^4}
\eea
such that $\delta ^+ _i + \delta ^- _i = \delta ^+ _j + \delta ^- _j = e^\g$, 
and $i,j$ take on three possible values. Only two of the ratios above are independent.
Since all twists in $\cT$ are mapped into one another under $SL(2,\bZ)_1 \times SL(2, \bZ)_2$,
it suffices to work out these relations for a single representative twist, say $e^\g = \ep_4$, for 
which we have,
\bea
\label{6e4}
\ep _4 =  \delta _1 + \delta _4 = \delta _2 + \delta _3 = \delta _9 + \delta _0
\eea
The two independent Schottky combinations of genus two $\tet$-functions may be expressed 
as follows (we suppress the $\Omega$-dependence)
\bea
\label{6e5}
{\tet [\delta _2 ] ^4 \, \tet [\delta _3 ]  ^4
\over \tet [\delta _1 ] ^4 \, \tet [\delta _4 ]^4}
= 
{ \tet _i (0, \tau_\g)^8 \over \tet _j (0, \tau_\g)^8}
\hskip 1in 
{\tet [\delta _2 ] ^4 \, \tet [\delta _3 ]  ^4
\over \tet [\delta _9 ] ^4 \, \tet [\delta _0 ]^4}
= { \tet _i (0, \tau_\g)^8 \over \tet _k (0, \tau_\g)^8}
\eea
where the assignments of $i,j,k$ are to be determined.

\sm

In the separating degeneration,  the left equation in (\ref{6e5})  tends to 1, while the right tends to $\infty$. 
On the standard genus 1 fundamental domain 
for $\M_1$, namely $|\tau_\g | \geq 1$ and $-1\leq 2\Re (\tau_\g ) \leq 1$, the $\tet$-constants 
$\tet _i ( 0 , \tau_\g )$ for $i=3,4$ are bounded away from 0 by,
\bea
\left | \tet _i (0, \tau_\g ) -1 \right | \leq 2 \sum _{n=1}^\infty e^{- \pi \sqrt{3}\, n^2 /2} = 0.13169 \cdots
\eea
Therefore, $k$ in (\ref{6e5}) can equal neither  3 nor 4, so we must have $k =2$, 
for which we have the asymptotics $\tet _2 (0, \tau_\g )^8 \sim 2^8 \, e^{2 \pi i \tau_\g}$ as $\tau_\g \to i \infty$.
As a result, we then have $\tet _i(0 , \tau_\g) \to 1$ and $\tet _j(0 , \tau_\g) \to 1$ as $\tau_\g \to i \infty$.
More generally, the product of genus two $\tet$-functions that will produce a zero limit correspond to the 
combination that contains $\tet [\delta _0]$, and this product will map to $\tet _2$, while the other combinations 
map to $\tet _3$ and $\tet_4$ which both tend to 1.

\begin{table}[htb]
\begin{center}
\begin{tabular}{|c||c|c|c|c|c|}   \hline
twist $\ev$ & $\cD [\ev] $ & $e^1$ & $e^2$  & $e^3$
\\ \hline  \hline
$(\ep_4, \ep_9, \ep_{10})$ & $\{ \delta _1, \delta _4, \delta _9, \delta _0\} $ & $\ep_4=\delta _2 + \delta _3 $
& $\ep_9 = \delta _5 + \delta _7 $	  & $\ep_{10}=\delta _6 + \delta _8 $
\\ \hline 
$(\ep_4, \ep_{15}, \ep_{16})$ & $\{ \delta _2, \delta _3, \delta _9, \delta _0\} $ & $\ep_4=\delta _1 + \delta _4 $
& $\ep_{15} = \delta _5 + \delta _8 $	  & $\ep_{16}=\delta _6 + \delta _7 $
\\ \hline 
$(\ep_6, \ep_8, \ep_{10})$ & $\{ \delta _1, \delta _6, \delta _8, \delta _0\} $ & $\ep_6=\delta _3 + \delta _5 $
& $\ep_8 = \delta _2 + \delta _7 $	  & $\ep_{10}=\delta _4 + \delta _9 $
\\ \hline 
$(\ep_6, \ep_{14}, \ep_{15})$ & $\{ \delta _3, \delta _5, \delta _8, \delta _0\} $ & $\ep_6=\delta _1 + \delta _6 $
& $\ep_{14} = \delta _4 + \delta _7 $	  & $\ep_{15}=\delta _2 + \delta _9 $
\\ \hline 
$(\ep_8, \ep_{13}, \ep_{16})$ & $\{ \delta _2, \delta _6, \delta _7, \delta _0\} $ & $\ep_8=\delta _1 + \delta _8 $
& $\ep_{13} = \delta _4 + \delta _5 $	  & $\ep_{16}=\delta _3 + \delta _9 $
\\ \hline 
$(\ep_9, \ep_{13}, \ep_{14})$ & $\{ \delta _4, \delta _5, \delta _7, \delta _0\} $ & $\ep_9=\delta _1 + \delta _9 $
& $\ep_{13} = \delta _2 + \delta _6 $	  & $\ep_{14}=\delta _3 + \delta _8 $
\\ \hline 
\end{tabular}
\end{center}
\caption{Table of twists $\ev$, $\cD[\ev]$ and pairs of even spin structures in $\cD [e^\g] \setminus \cD[\ev]$}
\label{table:11}
\end{table}

In Table \ref{table:11} we list all the pairs of spin structures in $\cD [e^\g] \setminus \cD[\ev]$
corresponding to the twists $ e^\g \in \cO_+^o$. By inspection, we see that  the set $\cD[\ev]$ 
always contains $\delta_0$, and thus the corresponding $\cD [e^\g] \setminus \cD[\ev]$ never 
contains $\delta _0$. As a result, in the separating degeneration limit $t \to 0$, 
the functions $\tet _{j (\g)}( 0, \tau _\g)$ tend to 1 for all $\g \in \{ 1,2,3\}$,  and we have,
\bea
\label{9a7}
\cP =  \prod _{\g =1}^3 \tet _{j (\g)} (0,\tau_{e^\g})^2   \to 1
\eea
As a result, the chiral factors $\cL [ \delta , \ev] $ and $\cR_\mn [\ev] $ in (\ref{9a4}) simplify considerably, 
and we find the following leading behavior as $t \to 0$, 
\bea
\label{9a6}
\cL  [\delta ; \ev] (\tau_I, \hat t) & = & 
\Psi _{10}  (\hat \Omega)  ^\half  \,  { \cZ [\delta ] (\hat \Omega) \over \tet [\delta ] (0, \hat \Omega)^4}
 \no \\
\cR_\mn [\ev] (\tau_I, t) & = &  
 { \Psi _4 (\Omega) ^{(3-\mn)/2} \over \Psi _{10}   (\Omega)^ \half }  
\sum _{\delta _R  \in \cD [\ev] } C_{\delta _R} [\ev] \, \tet [\delta _R] (0, \Omega)^{4\mn}
\eea
where $\hat \Omega$ and $\Omega = \Omega _R$ are understood to be  given in terms of 
$\tau_I, \hat t$, and $t$ by (\ref{9a1}) and (\ref{9a2}).

\subsection{Integration near the separating mode}

In this section, we proceed to integrating over odd moduli, as well as over the even moduli $t, \bar t$ 
which control the degeneration. To do so, we compute the leading $t \to 0$ 
behavior of $\cL [ \delta , \ev] $ and $\cR_\mn [\ev] $ in (\ref{9a6}), using (\ref{6a1})
as well as the limit of  $\Psi _{10}$ provided by (\ref{psi10}), 
\bea
\label{9b1}
{ \Psi_{10} (\hat \Omega)^\half  \over \overline{\Psi_{10} (\Omega)} ^\half }  
= {\hat t \over  \tilde t} \times  { \eta (\tau_1)^{24} \, \eta (\tau_2)^{24} 
\over | \eta (\tau_1) \, \eta (\tau_2)|^{24}  } + \cO ( \hat t^3, \tilde t)
\eea
while all other factors have finite, or vanishing, limits. Neglecting all terms in the integrand which 
do not contribute to the boundary integral, the integrations over odd moduli and 
over the even moduli $t, \bar t$ governing the  separating node,  reduce as follows,
\bea
\label{9b2}
\int _D  \int _\zeta d \zeta ^1 d \zeta ^2 { d \tilde t \, d\hat t \over \tilde t \, \hat t ^{1/2}} 
= 2 \int _D  \int _\zeta d \zeta ^1 d \zeta ^2 { d \tilde t \, d( \hat t^{1/2}) \over \tilde t }
\eea
to be carried out with the integration cycle of (\ref{9a2}).  The integration over $\zeta$ gives rise to
\bea
\label{9b3}
\int _D  \int _\zeta d \zeta ^1 d \zeta ^2 { d \tilde t \, d\hat t \over \tilde t \, \hat t ^{1/2}} 
=  - { 1 \over 4 \pi^2} \prod _{I=1,2} S_{\delta_I} (q_I-s_I, \tau_I)   \int _D { d\bar t \, d h \over \bar t}
\eea
The last integral is evaluated by picking up the pole at $t=0$, and gives  \cite{Witten:2013cia},
\bea
\label{9b4}
\int _D  \int _\zeta d \zeta ^1 d \zeta ^2 { d \tilde t \, d\hat t \over \tilde t \, \hat t ^{1/2}} 
=   { 1 \over 2 \pi} \prod _{I=1,2} S_{\delta_I} (q_I-s_I, \tau_I)
\eea
The product of Szego-kernels cancels that same product in the denominator of the 
limit of $\cZ[\delta]$, given in (\ref{6a1}). The result further simplifies to become,
\bea
\label{9b5}
\int _D \!  d\hat t \, d \tilde t  \int_\zeta  \! d \zeta ^1 d \zeta ^2 
\cL [\delta ; \ev] (\tau_I, \hat t) \overline{ \cR_\mn [\ev] (\tau_I, t)}
=
{ \< \delta _0 | \delta \> \over (2 \pi)^9 } \prod _{I=1,2} { \overline{\psi_4 (\tau_I)} ^{(3-\mn)/2} 
\over \overline{ \eta (\tau _I)} ^{12}}  \! \! \!
\sum _{\delta _R  \in \cD [\ev] } \! \! \! C_{\delta _R} [\ev] \, \overline{ \tet [\delta _R] }^{4\mn} 
\quad
\eea
Here, $\tet [\delta _R]= \tet [\delta _R^{(1)} ] (0, \tau_1 ) \tet [\delta _R^{(2)} ] (0, \tau_2 )$  represents the 
genus two $\tet$-function in the degeneration limit, and $\delta _R ^{(I)}$ stand for the genus 1 spin 
structure restrictions of $\delta _R$ on the components $\Sigma _I$ of the separating degeneration.
Also, we have used the relation $\< \nu_0 | \delta _1\> \< \nu_0 | \delta _2 \> = \< \delta _0 | \delta \>$.

\subsection{Summation over spin structures and  twists}

Very little in (\ref{9b5}) still depends upon the spin structures $\delta$ and $\delta _R$,
and upon the twists $\ev \in \cO^o_+$. The factors $\hat \cQ$ and $\tilde \cQ$  in (\ref{9a3}) 
appear to still depend on $\ev$ through their Prym period $\tau_\g$, but as we have shown in Section \ref{sec:PT},
these Prym periods all diverge $\tau_\g \to i \infty$, and localize the sum over the lattice of internal momenta 
in the 6 compactified and twisted directions to just the zero momentum term. Thus, 
$\hat \cQ$ and $\tilde \cQ$ reduce to their 4-dimensional uncompactified form, which
is independent of the twist $\ev$. 

\sm

As a result, the only remaining contributions in the sums over $\delta, \delta _R$ and $\ev \in \cO_+^o$
may be collected in the following factor $\cS$, 
\bea
\label{9c1}
\cS = \sum _{\ev \, \in \cO_+^o} ~ \sum _{\delta _R \in \cD [\ev]} ~ \sum _{\delta \in \cD [\ev]} C_\delta [\ev] C_{\delta_R} [\ev] 
 \< \delta _0 | \delta \> \,  \overline{ \tet [\delta _R]}^{4\mn} 
\eea
where we recall that $\mn=1$ for $E_8 \times E_8$ and $\mn=3$ for $Spin (32)/\bZ_2$,
as was stated earlier in (\ref{9a5}).  We shall now carry out these sums.

\sm

Since for given $\ev$, the summations over $\delta$ and $\delta _R$ are both over the 
set $\cD[\ev]$, we choose the  same reference spin structure $\delta_*$ in (\ref{5e5})
and (\ref{5e4}), and we have,  
\bea
\label{6c3}
C_\delta [\ev] & = & C_{\delta _* } [\ev] \, \< \delta_* | \delta  \>
\no \\
C_{\delta_R} [\ev] & = & C_{ \delta _*} [\ev]\,  \< \delta_* | \delta _R \>
\eea
where $C_{\delta _*} [\ev]= \pm 1$. As a result of (\ref{9c1}), the sum $\cS$ is given by,
\bea
\label{6c4}
\cS =
\sum _{\ev \, \in \cO_+^o} ~  \sum _{\delta _R \in \cD [\ev]} ~ \sum _{\delta \in \cD [\ev]}
 \<  \delta_*  | \delta \> \<  \delta_* | \delta _R \> \< \delta _0 | \delta \> \,
\overline{ \tet [\delta _R]}^{4 \mn }
\eea
Now any triplet in $\cD[\ev]$ is syzygous (as was shown in Section 4.1),
so that we have, 
\bea
\label{6c5}
 \<  \delta_* | \delta \> \< \delta _0 | \delta \> & = & \< \delta _0 | \delta_* \>
\no \\
\< \delta _0 |  \delta_* \> \<  \delta_* | \delta _R \> & = & \< \delta _0 | \delta _R \>
\eea
Combining the two relations in (\ref{6c5}), and substituting the result into (\ref{6c4}), 
transforms the expression into a sum over $\delta$ whose argument no longer depends on $\delta$, 
simply giving a factor of 4. The result is as follows,
\bea
\label{6c6}
\cS =4 \sum _{\ev \, \in \cO_+^o}  \sum _{\delta _R \in \cD [\ev]}  
 \< \delta_0 | \delta _R \> \overline{ \tet [\delta _R]}^{4 \mn } 
 \eea
We need this quantity in the separating degeneration limit only. Thus, the contribution from
$\delta _R = \delta _0$ drops out in the limit. To write the other contributions, we shall use the 
shorthands, 
\bea
\label{6c7}
s_i = \overline{ \tet [\mu_i](0, \tau_1)}^4
\hskip 1in
t_i = \overline{ \tet [\mu_i](0, \tau_2)}^4
\eea
with $i =2,3,4$. Inspection of Table \ref{table:1} allows us to label  each twist in $\cO_+^o$ 
by its associated four spin structures, and carry out the corresponding sum in $\cS$ over $\delta _R \in \cD [\ev]$,
and we have, 
\bea
\label{6c8}
(\delta _1, \delta _4, \delta _9, \delta _0) & \hskip 1in & +s_2^\mn t_2^\mn + s_3^\mn t_3^\mn + s_4^\mn t_4^\mn
\no \\
(\delta _1, \delta _6, \delta _8, \delta _0) & \hskip 1in & +s_2^\mn t_4^\mn + s_3^\mn t_3^\mn + s_4^\mn t_2^\mn
\no \\
(\delta _2, \delta _3, \delta _9, \delta _0) & \hskip 1in & +s_2^\mn t_2^\mn - s_3^\mn t_4^\mn - s_4^\mn t_3^\mn
\no \\
(\delta _2, \delta _6, \delta _7, \delta _0) & \hskip 1in & -s_2^\mn t_3^\mn - s_3^\mn t_4^\mn + s_4^\mn t_2^\mn
\no \\
(\delta _3, \delta _5, \delta _8, \delta _0) & \hskip 1in & +s_2^\mn t_4^\mn - s_3^\mn t_2^\mn - s_4^\mn t_3^\mn
\no \\
(\delta _4, \delta _5, \delta _7, \delta _0) & \hskip 1in & -s_2^\mn t_3^\mn - s_3^\mn t_2^\mn + s_4^\mn t_4^\mn
\eea
As a result, $\cS$ is given by,
\bea
\label{6c9}
\cS  & = &  8 \Big ( s_2^\mn t_2^\mn + s_3^\mn t_3^\mn + s_4^\mn t_4^\mn
- s_2^\mn t_3^\mn + s_2^\mn t_4^\mn - s_3^\mn t_2^\mn - s_3^\mn t_4^\mn + s_4^\mn t_2^\mn - s_4^\mn t_3^\mn \Big )
\no \\
& = & 8 \Big ( s_3^\mn - s_2^\mn - s_4^\mn \Big ) \Big ( t_3^\mn - t_2^\mn - t_4^\mn \Big )
\eea
For the gauge group $E_8 \times E_8$ we have $\mn=1$, in which case we have $\cS=0$
by the genus 1 Jacobi identity. For the gauge group $Spin (32)/\bZ_2$, we have $\mn=3$, 
and we use the Jabobi identity again to show that, 
\bea
\label{6c10}
s_3^3 - s_2^3 - s_4^3 = 3 s_2 s_3 s_4
\eea
and analogously for $t_i$. As a result, we have,
\bea
\label{6c12}
\cS = 72 s_2 s_3 s_4 t_2 t_3 t_4 = 2^{11} \cdot 3^2 \cdot \overline{\eta (\tau_1)}^{12} \, \overline{\eta (\tau_2)}^{12} 
\eea
which does not vanish for $Spin (32)/\bZ_2$. 

\subsection{Integrations over $\tau_I$  for $Spin (32)/\bZ_2$} 

Combining the results of (\ref{6c12}), (\ref{9a3}), and (\ref{9b5}), we are left with the remaining 
integrations over $\tau_I$ and four-dimensional internal loop momenta $p_I$,
\bea
\label{9k1}
\cV^{\rm bdy} _G =  g_s^2 \, \mN \,  { 36 \over \pi ^9} \,  \prod _{I=1,2} \int _{\cM_1 } \! \!  \! \! d^2 \tau_I 
\int \! \! d^4 p_I \, e^{ - 2 \pi p_I^2 \Im (\tau_I)} 
\eea
Each 4-dimensional  loop momentum integral giving a factor $(2 \, \Im (\tau_I))^{-2}$ so that,
\bea
\label{9k2}
\cV^{\rm bdy} _G =  g_s^2 \, \mN \, { 9 \over 4 \pi ^9} \,  \prod _{I=1,2} \int _{\cM_1 } { d^2 \tau_I   \over (\Im \tau_I)^2 }
\eea
The volume of $\cM_1= \{ \tau_I \in \bC, \, |\tau_I|>1, \, | \Re (\tau_I) | < 1/2 \}$, in the Poincar\'e metric is 
given by,\footnote{Our conventions here follow those
of  \cite{D'Hoker:2005ht}, so that $d^2 \tau = 2 d \Re (\tau) \, d \Im (\tau)$.}
\bea
\int _{\cM_1} { d^2 \tau_I \over (\Im \tau _I )^2} = { 2 \pi \over 3}
\eea
Restoring also  the $\alpha'$-dependence from our earlier
choice of units $\alpha '=2$, we obtain the following expression for the two-loop vacuum energy,
\bea
 \cV_G ^{\rm bdy}= { 4 g_s^2 \, \mN \over \pi ^7 (\alpha ')^2}  
\eea

\subsection{Comments on the overall normalization for $Spin (32)/\bZ_2$}

In this paper, the contribution to the two-loop vacuum amplitude from the bulk of supermoduli 
space was shown to vanish for both Heterotic strings. It follows from the results of \cite{Witten:2013cia} 
that the entire two-loop vacuum energy arises from the boundary contributions. For $Spin (32)/\bZ_2$, 
this result is non-zero and given, on the one hand by \cite{Witten:2013cia} in terms of the $D$-term 
one-loop tadpole $\< V_D \>$, on the other hand by our present calculations, so that,
\bea
\label{9u1}
\cV_G = 2 \pi g_s^2 \< V_D \> ^2 = { 4 g_s^2 \, \mN \over \pi ^7 (\alpha ')^2} 
\eea
The one-loop tadpole $\< V_D \>$  was computed in \cite{Atick:1987gy} in terms of a 
quantity $c$ (the model considered here has only a single $U(1)$ factor).
Adapting normalizations of \cite{Atick:1987gy}, $\< V_D \>$ is given by, 
\bea
\label{9u2}
\< V_D \> = 2 \pi c /\hat g = { 1 \over 96 \pi} \sum _i n_i q_i h_i
\eea
in units chosen so that $\alpha '=2$. Only a single multiplet contributes in the present model, 
so that $i=1$, for which the charge $q_i$ and helicity $h_i$ obey $q_ih_i=1$. The number of 
generations  $n_i$ is given in terms of the Euler number $\chi (Y)$ of the orbifold $Y$ by $n_i = \chi (Y)/2$, 
and was determined in \cite{Donagi:2004ht} to be given by $n_i=48$.  Putting all this together, 
and exhibiting the dependence on $\alpha '$ explicitly, we find $\< V_D \> = 1/( \pi \alpha ' )$. In view of 
(\ref{9u1}), we conclude that the overall normalization factor of the 2-loop vacuum energy 
$\mN$ should obey $\mN = \pi^6 /2$.

\sm

The value of $\mN$ may be computed from first principles following the techniques used in \cite{D'Hoker:2005ht}
for the Type IIB superstring. Such calculations require a painstaking effort to achieve consistent overall normalizations
throughout,  and will not be pursued further here.

\subsection{Vanishing of boundary contributions for Type II superstrings}

For the Type II superstrings, the non-trivial structure of the cycle $\Gamma$ of (\ref{9a2}) must be 
implemented on both  left and right chiralities. We shall denote by $\hat t_L$ and $\hat t_R$ the corresponding 
parameters in the left and right chirality super-period matrices, and by $\zeta _L^{1,2}$ and $\zeta _R^{1,2}$
the corresponding odd moduli. The asymptotic expansion of the right chirality sector now
mirrors the one of the left chirality sector, both of which may be parametrized by 
the bosonic moduli $t , \bar t$ of the cycle $\Gamma$, and their odd counterparts $\zeta _L^{1,2}, \zeta _R^{1,2}$,
\bea
\label{9m1}
\hat \tau _{LI} = \tau_I & \hskip 0.6in &
\hat \rho _L = \hat t_L ^\half 
= t ^\half - h_L(t,\bar t)  { \zeta ^1_L \zeta ^2_L \over 4 \pi^2} \prod _{I=1,2} S_{\delta _I} (q_I-s_I, \tau_I)
\no \\
\hat \tau _{RI} = \bar \tau_I 
& & \hat \rho _R = \hat t_R ^\half 
= \bar t ^\half - h_R(t,\bar t)  { \zeta ^1_R \zeta ^2_R \over 4 \pi^2} \prod _{I=1,2} \overline{S_{\delta _I} (q_I-s_I, \tau_I)  }
\eea
The regularization functions $h_L$ and $h_R$ are subject to the same boundary conditions as 
were given in Section \ref{sec:81}
for their  counterpart $h$ in the Heterotic string, but are otherwise arbitrary.
Combining left and right chirality contributions, we have the following behavior, 
arranged according to whether the twist $\ev$ belongs to orbit $\cO_+^e$ or to orbit $\cO_-^o$,
\bea
\label{9m2}
\ev \in \cO_+^e & \hskip 1in & 
{ d \hat \rho_L \over \hat \rho _L^2} \cdot {d \hat \rho_R \over \hat \rho^2_R} \, d \zeta_L ^1 
d \zeta_L ^2 d \zeta_R ^1 d \zeta_R ^2 \left ( 
1 + c_L \hat t_L ^2 +  c_R \hat t_R^2 + \cdots \right )
\no \\
\ev \in \cO_+^o & \hskip 1in & 
 d \hat \rho_L  \cdot d \hat \rho_R\, d \zeta_L ^1 
d \zeta_L ^2 d \zeta_R ^1 d \zeta_R ^2 \left ( 
1 + c_L \hat t_L ^2 +  c_R \hat t_R^2 + \cdots \right )
\eea
The twists in $\cO_+^e$ again single out the identity operator as leading contribution.
Spin structure summation in the left and right chiral blocks cancels this leading contribution as 
pointed out in section 3.2.5 of \cite{Witten:2013cia}. The higher order corrections,
as well as the behavior in the orbit $\cO_+^o$, lead to convergent integrals
and produce no boundary contributions. Hence the boundary contributions for Type II vanish.

\newpage

\appendix

\setcounter{equation}{0}

\section{Spin structures and theta functions at genus 1}
\setcounter{equation}{0}
\label{secA}

Let $\tau\in \bC$, $\Im\,\tau>0$.
The general $\tet$-function at genus 1 with characteristics is defined by
\bea
\tet [\kappa ] (z, \tau) = \sum _{n \in \bZ} \exp \Big \{ i \pi \tau (n+ \kappa ')^2 + 2 \pi i (n + \kappa ') (z+ \kappa '') \Big \}
\eea
where $\kappa = [\kappa ' | \kappa '']$ is the genus 1 half characteristic with $\kappa ', \kappa '' \in \{ 0, 1/2 \}$.
The above $\tet$-functions with characteristics may be recast in terms of the $\tet $-function
without characteristics $\tet (z, \tau) \equiv \tet [0] (z, \tau)$, as follows,
\bea
\tet [\kappa ] (z, \tau) = e^{i \pi  \tau (\kappa ' )^2 + 2 \pi i {\kappa '} (z + \kappa '')}
\tet ( z+ \tau \kappa ' + \kappa '', \tau )
\eea
To make contact with the standard old-fashioned notation for $\tet$-functions, 
we introduce the following conventions for genus 1 spin structures, 
\bea
\label{A3}
\nu_0  = \left [ \half  \Big | \half \right ] 
\hskip 0.7in 
\mu_2 = \left [ \half  \Big | 0 \right ] 
\hskip 0.7in 
\mu_3 =  [ 0  \big | 0 ] 
\hskip 0.7in 
\mu_4 = \left [ 0  \Big | \half \right ]  
\eea
The characteristic $\nu_0$ corresponds to the odd spin structure on the torus
$T=\bC/\bZ+\tau\bZ$, while the characteristics $\mu_2$, $\mu_3$, and $\mu_4$ 
correspond to the three even spin structures on $T$.
The standard  theta functions $\tet_j(z,\tau)$ are then related to those used in our 
notation as follows, 
\bea
\label{A4}
\tet  [\nu _0] (z, \tau ) & = & \tet _1 (z, \tau) 
\no \\
\tet  [\mu _i] (z, \tau ) & = & \tet _i (z,\tau) 
\hskip 1in i=2,3,4
\eea
Periodicity relations are as follows,
\bea
\tet  [\kappa]  (z+1,\tau) & = & (-)^{2\kappa'} \, \tet [\kappa] (z,\tau) 
\no \\
\tet  [\kappa]  (z+\tau,\tau) & = & (-)^{2 \kappa ''} \, e^{- \pi i \tau - 2\pi i z} \, \tet [\kappa] (z,\tau) 
\eea
Half-period shift relations are given by,
\bea
\tet [\kappa ] \left ( z + \half, \tau \right ) & = & 
(-)^{4 \kappa ' \kappa ''} \tet  [ \kappa ' | \tilde \kappa '' ] (z, \tau)
\no \\
\tet [\kappa ] \left ( z + {\tau \over 2} , \tau \right ) & = & 
(-i)^{2 \kappa ''} \, e^{- \pi i \tau/4 - i \pi z} \,  \tet  [ \tilde \kappa '  | \kappa ''  ] (z, \tau)
\eea
where 
\bea
\tilde \kappa ' = \half +\kappa ' \qquad ( \hbox{mod} \, 1)
 \hskip 1in 
\tilde \kappa '' = \half +\kappa '' \qquad ( \hbox{mod} \, 1)
\eea
In particular, the genus 1 Riemann vector is given by $\Delta = \half - {\tau \over 2}$,
so that we have,
\bea
\label{A12}
\tet ( u - \Delta, \tau ) = -i \, e^{- i \pi \tau /4 - i \pi u} \, \tet_1 (u, \tau )
\eea
Another useful identity is,
\bea
\label{A12a}
\tet _1 ' (0, \tau) = - \pi \tet _2 (0, \tau) \tet _3 (0, \tau) \tet _4 (0, \tau) 
= - 2 \pi \eta (\tau)^3
\eea
The well-known genus 1 formula for differentiation with respect to the modulus is given by,
\bea
\label{A13}
\p_{\tau_1} \ln {\tet [\kappa_i] (0,\tau_1) \over \tet [\kappa_j] (0,\tau_1)}
= i { \pi \over 4} \sigma (\kappa_i, \kappa_j) \tet [\kappa _k] (0,\tau_1)^4
\eea
where $\kappa_i + \kappa_j + \kappa_k = \nu_0$, and 
$\sigma (\kappa _j, \kappa _i) = - \sigma (\kappa _i, \kappa _j)$ with the following
values on the canonical six spin structures of (\ref{A3}), 
\bea
\label{A14}
\sigma (\mu_2, \mu_3) = \sigma ( \mu_3, \mu_4) = \sigma (\mu_2, \mu_4) =1
\eea

\section{Spin structures and theta functions at genus 2}
\setcounter{equation}{0}
\label{secB}

In this appendix, we review some fundamental facts about
genus 2 Riemann surfaces, their spin structures, $\tet$-functions, and 
modular properties (see also \cite{ADP}).

\subsection{Spin Structures}

Each spin structure $\kappa$ can be
identified with a $\tet$-characteristic $\kappa = (\kappa '|\kappa'')$,
where $\kappa ', \kappa '' \in \{0,\half\}^2$, represented here by {\sl
column matrices}. The parity of the spin structure $\kappa$ is that of the
integer $4\kappa' \cdot \kappa''$. There are 6 odd spin structures  and 10 are even. 
The symplectic pairing mod 2 between any two  
spin structures $\kappa$ and $\lambda$ is defined by
\be
\label{B1}
\< \kappa | \lambda \> \equiv \exp \{4 \pi i (\kappa ' \lambda '' -
\kappa '' \lambda ') \}
\ee
It will be convenient to use a definite basis for the spin structures,
in a given  homology basis, as in \cite{D'Hoker:2001qp}, and used again in \cite{ADP}.
The odd spin structures may be labeled by,
\bea
\label{B2}
2\nu_1 =\left (\matrix{0 \cr  1\cr} \bigg | \matrix{0 \cr 1\cr} \right )
\qquad
2\nu_3 =\left (\matrix{0 \cr  1\cr} \bigg | \matrix{1 \cr 1\cr} \right )
\qquad
2\nu_5 =\left (\matrix{1 \cr  1\cr} \bigg | \matrix{0 \cr 1\cr} \right )
\nonumber \\
2\nu_2 =\left (\matrix{1 \cr  0\cr} \bigg | \matrix{1 \cr 0\cr} \right )
\qquad
2\nu_4 =\left (\matrix{1 \cr  0\cr} \bigg | \matrix{1 \cr 1\cr} \right )
\qquad
2\nu_6 =\left (\matrix{1 \cr  1\cr} \bigg | \matrix{1 \cr 0\cr} \right )
\eea
The even spin structures may be labeled by,
\bea
\label{B3}
2\delta_1 =\left (\matrix{0 \cr 0\cr} \bigg | \matrix{0 \cr 0\cr} \right )
\quad
2\delta_2 =\left (\matrix{0 \cr 0\cr} \bigg | \matrix{0 \cr 1\cr} \right )
\quad
2\delta_3 =\left (\matrix{0 \cr 0\cr} \bigg | \matrix{1 \cr 0\cr} \right )
\quad
2\delta_4 =\left (\matrix{0 \cr 0\cr} \bigg | \matrix{1 \cr 1\cr} \right )
\quad
2\delta_5 =\left (\matrix{0 \cr 1\cr} \bigg | \matrix{0 \cr 0\cr} \right )
\nonumber \\
2\delta_6 =\left (\matrix{0 \cr 1\cr} \bigg | \matrix{1 \cr 0\cr} \right )
\quad
2\delta_7 =\left (\matrix{1 \cr 0\cr} \bigg | \matrix{0 \cr 0\cr} \right )
\quad
2\delta_8 =\left (\matrix{1 \cr 0\cr} \bigg | \matrix{0 \cr 1\cr} \right )
\quad
2\delta_9 =\left (\matrix{1 \cr 1\cr} \bigg | \matrix{0 \cr 0\cr} \right )
\quad
2\delta_0 =\left (\matrix{1\cr 1\cr} \bigg | \matrix{1\cr 1\cr} \right)
\eea

\subsection{Twists}

We  label the twists by half-characteristics following \cite{ADP}, 
\bea
\label{1f1}
2\ep_1 =\left (\matrix{0 \cr 0\cr} \bigg | \matrix{0 \cr 0\cr} \right )
\qquad \
2\ep_2 =\left (\matrix{0 \cr 0\cr} \bigg | \matrix{0 \cr 1\cr} \right )
\qquad \
2\ep_3 =\left (\matrix{0 \cr 0\cr} \bigg | \matrix{1 \cr 0\cr} \right )
\qquad \
2\ep_4 =\left (\matrix{0 \cr 0\cr} \bigg | \matrix{1 \cr 1\cr} \right )
\nonumber \\
2\ep_5 =\left (\matrix{0 \cr 1\cr} \bigg | \matrix{0 \cr 0\cr} \right )
\qquad \
2\ep_6 =\left (\matrix{0 \cr 1\cr} \bigg | \matrix{1 \cr 0\cr} \right )
\qquad \
2\ep_7 =\left (\matrix{1 \cr 0\cr} \bigg | \matrix{0 \cr 0\cr} \right )
\qquad \
2\ep_8 =\left (\matrix{1 \cr 0\cr} \bigg | \matrix{0 \cr 1\cr} \right )
\nonumber \\
2\ep_9 =\left (\matrix{1 \cr 1\cr} \bigg | \matrix{0 \cr 0\cr} \right )
\qquad
2\ep_{10} =\left (\matrix{1\cr 1\cr} \bigg | \matrix{1\cr 1\cr} \right)
\qquad
2\ep _{11} =\left (\matrix{0 \cr  1\cr} \bigg | \matrix{0 \cr 1\cr} \right )
\qquad
2\ep_{12} =\left (\matrix{1 \cr  0\cr} \bigg | \matrix{1 \cr 0\cr} \right )
\no \\
2\ep_{13} =\left (\matrix{0 \cr  1\cr} \bigg | \matrix{1 \cr 1\cr} \right )
\qquad
2\ep_{14} =\left (\matrix{1 \cr  0\cr} \bigg | \matrix{1 \cr 1\cr} \right )
\qquad
2\ep_{15} =\left (\matrix{1 \cr  1\cr} \bigg | \matrix{0 \cr 1\cr} \right )
\qquad
2\ep_{16} =\left (\matrix{1 \cr  1\cr} \bigg | \matrix{1 \cr 0\cr} \right )
\eea

\subsection{$\tet$-functions}

The $\tet$-function is an entire function in the period matrix $\Omega$
and $\zeta \in {\bf C}^2$, defined by
\be
\label{B4}
\tet [\kappa ] (\zeta, \Omega)
\equiv
\sum _{n \in {\bf Z}^2} \exp \{\pi i (n+\kappa ')\Omega (n+\kappa')
+ 2\pi i (n+\kappa ') (\zeta + \kappa '') \}
\ee
Here, $\tet$ is even or odd in $\zeta$ depending on the parity of the spin
structure. The following useful periodicity relations hold, in which
$N,M \in {\bf Z}^2$ and  $\lambda ', \lambda '' \in {\bf C}^2$,
\bea
\label{B5}
\tet [\kappa ] (\zeta + M + \Omega N, \Omega )
&=&
\tet [\kappa ](\zeta , \Omega) \ \exp \{ -i \pi N \Omega N
- 2 \pi i N (\zeta +\kappa '') + 2 \pi i \kappa ' M \}
\nonumber \\
\tet [\kappa ' +N, \kappa '' +M ] (\zeta , \Omega)
&=&
\tet [\kappa ',\kappa ''] (\zeta, \Omega) \ \exp \{ 2\pi i \kappa ' M \}
 \\
\tet [\kappa + \lambda ] (\zeta , \Omega )
&=&
\tet [\kappa ](\zeta + \lambda '' + \Omega \lambda' , \Omega) \ \exp \{ i
\pi \lambda ' \Omega \lambda ' + 2 \pi i \lambda ' (\zeta +\lambda '' +
\kappa '')  \}
\no
\eea

\subsection{The Action of Modular Transformations}

Modular transformations $M$ form the infinite discrete group $Sp(4,{\bf Z})$, defined by
\be
\label{B7}
M=\left ( \matrix{A & B \cr C & D \cr} \right )
\qquad \qquad
M \left ( \matrix{0 & I \cr -I & 0 \cr} \right ) M^t= \left ( \matrix{0 & I \cr -I & 0 \cr} \right )
\ee
where $A,B,C,D$ are integer valued $2 \times 2$ matrices.
To exhibit the action of the modular group on 1/2 characteristics,
it is convenient to assemble the 1/2 characteristics into a single column of 4 entries.
In this notation, the action  on spin structures $\kappa$ is given by \cite{Igusa}
\bea
\label{B8}
\left (\matrix{ \tilde \kappa' \cr \tilde \kappa ''\cr}  \right )
=
\left ( \matrix{D & -C \cr -B & A \cr} \right )
\left ( \matrix{ \kappa ' \cr \kappa '' \cr} \right )
+ \half \ {\rm diag}
\left ( \matrix{CD^T  \cr AB^T \cr} \right )
\eea
Here and below, ${\rm diag} (M)$ of a $n \times n$ matrix $M$ is an
$1\times n$ column vector whose entries are the diagonal entries on $M$.
The action of modular transformations on twists is as follows, 
\bea
\label{B8t}
\tilde \ep = M[\ep] 
\hskip 1in 
\left (\matrix{ \tilde \ep'  \cr \tilde \ep ''\cr}  \right )
=
\left ( \matrix{D & -C \cr -B & A \cr} \right )
\left ( \matrix{ \ep ' \cr \ep '' \cr} \right )
\eea
On the period matrix, modular transformations act by
\bea
\label{B9}
\tilde \Omega = (A\Omega + B ) (C\Omega + D)^{-1}
\eea
 while on the Jacobi $\tet$-functions,  we have
\bea
\label{B10}
\tet [\tilde \kappa ] \biggl ( \{(C\Omega +D)^{-1} \}^T  \zeta , \tilde
\Omega \biggr ) =
\epsilon (\kappa, M) \det (C\Omega + D) ^\half
e^{ i \pi \zeta  (C\Omega +D)^{-1} C \zeta }
\tet [ \kappa ] (\zeta, \Omega)
\quad
\eea
where $\kappa = (\kappa ' |\kappa '')$ and $\tilde \kappa = (\tilde
\kappa ' | \tilde \kappa '')$. The phase factor $\epsilon (\kappa, M)$
depends upon both $\kappa $ and the modular transformation $M$ and obeys
$\epsilon (\kappa , M )^8=1$. Its expression was calculated in \cite{Igusa},
and is given by,
\bea
\label{B11}
\epsilon (\delta, M) & = & \epsilon_0 (M) \exp \{ 2 \pi i \phi  (\kappa, M)\}
\\
\epsilon _0 (M)^2 & = & \exp \{ 2 \pi i \, {1 \over 8} \, \tr (M-I) \}
\no \\
\phi  (\kappa, M) & = & - \half \kappa ' D^T B \kappa '
+ \kappa ' B^T C \kappa '' - \half \kappa '' C^T A \kappa ''
+ \half (\kappa ' D^T - \kappa '' C^T) {\rm diag} (AB^T)
\no
\eea
The modular group is generated by the
following elements\footnote{The sign of the lower right entry of the matrix $\Sigma$
has been corrected and reversed compared to \cite{ADP}.}
\bea
\label{B12}
M_i = \left ( \matrix{I & B_i \cr 0 & I \cr} \right )
\qquad \  
S = \left ( \matrix{0 & I \cr -I & 0 \cr} \right )
\qquad \  
\Sigma = \left ( \matrix{\sigma & 0 \cr 0 & \sigma \cr} \right )
\qquad \  
T = \left ( \matrix{\tau _+ & 0 \cr 0 & \tau _- \cr} \right )
\eea
where we use the following notations,
\bea
B_1 = \left ( \matrix{1 & 0 \cr 0 & 0 \cr} \right )
\qquad \ \ \
B_2 = \left ( \matrix{0 & 0 \cr 0 & 1 \cr} \right )
\qquad \ \ \
B_3 = \left ( \matrix{0 & 1 \cr 1 & 0 \cr} \right )
\no \\
\sigma = \left ( \matrix{0 & 1 \cr -1 & 0 \cr} \right )
\qquad \
\tau _+ = \left ( \matrix{1 & 1 \cr 0 & 1 \cr} \right )
\qquad \ \
\tau _- = \left ( \matrix{1 & 0 \cr -1 & 1 \cr} \right )
\eea
The transformation laws for even spin structures under these generators
are given in Table~3; those for odd spin structures will not be
needed here and may be found in \cite{D'Hoker:2001qp}; those for twists
are listed in Table~4, where we also list the actions of the composite
generators $T_2, S_2$ which leave the twist $\ep_2$ invariant.
These generators are defined as follows,
\bea
S_2 = SM_1 SM_1
\hskip 1in 
T_2 = \Sigma T \Sigma
\hskip 1in 
S_3 = S_2^{-1} T_2 S_2 T_2
\eea
and take the following matrix form, 
\bea
S_2= - \left ( \matrix{ I & B_1 \cr -B_1 & B_2 \cr} \right )
\hskip 0.5in 
T_2= - \left ( \matrix{ \tau_- & 0 \cr 0 & \tau_+ \cr} \right )
\hskip 0.5in 
S_3=  \left ( \matrix{ \tau_-^2  & -B_2-B_3 \cr 0 & \tau_+^2 \cr} \right )
\eea

\begin{table}[htb]
\begin{center}
\begin{tabular}{|c||c|c|c|c|c|c||c|c||c|c|c|c|c|} \hline
 $\delta$ & $M_1$  & $M_2$ & $M_3$ & $S$ & $T$ & $\Sigma$ & $T_2$ & $S_2$ & $\epsilon ^2
(M_1)$ & $\epsilon ^2(M_2)$ & $\epsilon ^2 (M_3)$ & $\epsilon ^2 (T_2)$
& $\epsilon^2 (S_2)$
                \\ \hline \hline
 $\delta _1$
            & $\delta _3$
            & $\delta _2$
            & $\delta _1$
            & $\delta _1$
            & $\delta _1$
            & $\delta _1$
            & $\delta _1$
            & $\delta _3$
            & 1 &  1 & 1  & 1 & $i$
 \\ \hline
 $\delta _2$
            & $\delta _4$
            & $\delta _1$
            & $\delta _2$
            & $\delta _5$
            & $\delta _4$
            & $\delta _3$
            & $\delta _2$
            & $\delta _4$
             & 1 & 1 & 1 &  1 & $i$
 \\ \hline
 $\delta _3$
            & $\delta _1$
            & $\delta _4$
            & $\delta _3$
            & $\delta _7$
            & $\delta _3$
	     & $\delta _2$
            & $\delta _4$
            & $\delta _7$
             & 1 & 1 &  1 &  1 & 1
 \\ \hline
 $\delta _4$
            & $\delta _2$
            & $\delta _3$
            & $\delta _4$
            & $\delta _9$
            & $\delta _2$
            & $\delta _4$
            & $\delta _3$
            & $\delta _8$
             & 1 & 1 &  1 &  1 & 1
 \\ \hline
 $\delta _5$
            & $\delta _6$
            & $\delta _5$
            & $\delta _6$
            & $\delta _2$
            & $\delta _5$
            & $\delta _7$
            & $\delta _9$
            & $\delta _6$
            & 1 & $i$  &  1 & 1 & $i$
 \\ \hline
 $\delta _6$
            & $\delta _5$
            & $\delta _6$
            & $\delta _5$
            & $\delta _8$
            & $\delta _6$
            & $\delta _8$
            & $\delta _0$
            & $\delta _9$
            & 1 & $i$ & 1 & 1 & 1
 \\ \hline
 $\delta _7$
            & $\delta _7$
            & $\delta _8$
            & $\delta _8$
            & $\delta _3$
            & $\delta _9$
            & $\delta _5$
            & $\delta _7$
            & $\delta _1$
            & $i$ & 1 & 1 & 1 & $i$
 \\ \hline
 $\delta _8$
            & $\delta _8$
            & $\delta _7$
            & $\delta _7$
            & $\delta _6$
            & $\delta _0$
            & $\delta _6$
            & $\delta _8$
            & $\delta _2$
            & $i$ & 1 & 1 & 1 & $i$
 \\ \hline
 $\delta _9$
            & $\delta _9$
            & $\delta _9$
            & $\delta _0$
            & $\delta _4$
            & $\delta _7$
            & $\delta _9$
            & $\delta _5$
            & $\delta _5$
            & $i$ & $i$ & $-1$ & 1 & $i$
 \\ \hline
 $\delta _0$
            & $\delta _0$
            & $\delta _0$
            & $\delta _9$
            & $\delta _0$
            & $\delta _8$
            & $\delta _0$
            & $\delta _6$
            & $\delta _0$
            & $i$ & $i$  & $-1$ & 1 & $-1$
 \\ \hline
\end{tabular}
\end{center}
\label{table:4}
\caption{Modular transformations acting on even spin structures,  under the generators 
$M_1, M_2, M_3, S, T, \Sigma$ of the modular group $Sp(4,\bZ)$, 
and under the composite generators $S_2, T_2$ of the subgroup $H_{\ep_2}$, 
together with the non-trivial  phase factors of $\tet$-functions. }
\end{table}

We shall be most interested in the modular transformations of
$\tet$-constants $\tet ^2 [\delta]$ and thus in even spin structures
$\delta$ and the squares of $\epsilon$, which are given by
\bea
\label{eps}
\left \{ \matrix{
\epsilon (\delta , M_1) ^2 =
\exp \{ 2 \pi i \delta _1' (1 - \delta _1') \}
\cr
\epsilon (\delta , M_2) ^2 =
\exp \{ 2 \pi i \delta _2' (1 - \delta _2') \}
\cr
\epsilon (\delta , M_3) ^2 =
\exp \{ -4 \pi i \delta _1 ' \delta _2' \}
\cr} \right .
\hskip 1in
\left \{ \matrix{
\epsilon (\delta , S) ^2 = - 1
\cr
\epsilon (\delta , \Sigma) ^2 = -1
\cr
\epsilon (\delta , T) ^2 = + 1\cr} \right .
\eea
The non-trivial entries for $\epsilon ^2$ are listed in Table 3.

\begin{table}[htb]
\begin{center}
\begin{tabular}{|c||c|c|c|c|c|c|c|c|c|c|c|c|c|c|c|c|}   
\hline
$M$  &
$\ep_1$  & $\ep_2$  & $\ep_3$  & $\ep_4$  &
$\ep_5$  & $\ep_6$  & $\ep_7$  & $\ep_8$  & $\ep_9$  &
$\ep_{10}$  & $\ep_{11}$  & $\ep_{12}$  & $\ep_{13}$  & $\ep_{14}$  &
$\ep_{15}$   & $\ep_{16}$
\\ \hline \hline
$M_1$  &
$\ep_1$  & $\ep_2$  & $\ep_3$  & $\ep_4$  &
$\ep_5$  & $\ep_6$  & $\ep_{12}$  & $\ep_{14}$  & $\ep_{16}$  &
$\ep_{15}$  & $\ep_{11}$  & $\ep_7$  & $\ep_{13}$  & $\ep_8$  &
$\ep_{10}$   & $\ep_9$
\\ \hline
$M_2$  &
$\ep_1$  & $\ep_2$  & $\ep_3$  & $\ep_4$  &
$\ep_{11}$  & $\ep_{13}$  & $\ep_7$  & $\ep_8$  & $\ep_{15}$  &
$\ep_{16}$  & $\ep_5$  & $\ep_{12}$  & $\ep_6$  & $\ep_{14}$  &
$\ep_9$   & $\ep_{10}$
\\ \hline
$M_3$  &
$\ep_1$  & $\ep_2$  & $\ep_3$  & $\ep_4$  &
$\ep_6$  & $\ep_5$  & $\ep_8$  & $\ep_7$  & $\ep_{10}$  &
$\ep_9$  & $\ep_{13}$  & $\ep_{14}$  & $\ep_{11}$  & $\ep_{12}$  &
$\ep_{16}$   & $\ep_{15}$
\\ \hline
$S$  &
$\ep_1$  & $\ep_5$  & $\ep_7$  & $\ep_9$  &
$\ep_2$  & $\ep_8$  & $\ep_3$  & $\ep_6$  & $\ep_4$  &
$\ep_{10}$  & $\ep_{11}$  & $\ep_{12}$  & $\ep_{15}$  & $\ep_{16}$  &
$\ep_{13}$   & $\ep_{14}$
\\ \hline
$\Sigma$  &
$\ep_1$  & $\ep_3$  & $\ep_2$  & $\ep_4$  &
$\ep_7$  & $\ep_8$  & $\ep_5$  & $\ep_6$  & $\ep_9$  &
$\ep_{10}$  & $\ep_{12}$  & $\ep_{11}$  & $\ep_{14}$  & $\ep_{13}$  &
$\ep_{11}$   & $\ep_{13}$
\\ \hline
$T$  &
$\ep_1$  & $\ep_4$  & $\ep_3$  & $\ep_2$  &
$\ep_5$  & $\ep_6$  & $\ep_9$  & $\ep_{10}$  & 
$\ep_7$  & $\ep_8$  & $\ep_{13}$  & $\ep_{16}$  & 
$\ep_{11}$  & $\ep_{15}$  & $\ep_{14}$   & $\ep_{12}$
\\ \hline \hline \hline
$T_2$  &
$\ep_1$  & $\ep_2$  & $\ep_4$  & $\ep_3$  &
$\ep_9$  & $\ep_{10}$  & $\ep_7$  & $\ep_8$  & $\ep_5$  &
$\ep_6$  & $\ep_{15}$  & $\ep_{14}$  & $\ep_{16}$  & $\ep_{12}$  &
$\ep_{11}$   & $\ep_{13}$
\\ \hline
$S_2$  &
$\ep_1$  & $\ep_2$  & $\ep_{12}$  & $\ep_{14}$  &
$\ep_5$  & $\ep_{16}$  & $\ep_3$  & $\ep_4$  & $\ep_6$  &
$\ep_{15}$  & $\ep_{11}$  & $\ep_7$  & $\ep_{10}$  & $\ep_8$  &
$\ep_{13}$   & $\ep_9$
\\ \hline \hline
\end{tabular}
\end{center}
\caption{Modular transformations acting on twists, 
under the generators $M_1, M_2, M_3, S, T, \Sigma$ of the modular group $Sp(4,\bZ)$, 
and under the composite generators $S_2, T_2$ of the subgroup $H_{\ep_2}$ defined to
leave the twist $\ep_2$ invariant. }
\label{table:5}
\end{table}

\subsection{The Riemann relations}

At various times, we shall make use of the Riemann relations. They may be
expressed as the following quadrilinear sum over all spin structures
\be
\sum _\kappa \<\kappa | \lambda \>
\tet [\kappa ](\zeta _1 ) \tet [\kappa ](\zeta _2)
\tet [\kappa ](\zeta _3 ) \tet [\kappa ](\zeta _4)
= 4 
\tet [\lambda ] (\zeta _1 ') \tet [\lambda ] (\zeta _2 ')
\tet [\lambda ] (\zeta _3 ') \tet [\lambda ] (\zeta _4 ')
\ee
where the signature symbol $\< \kappa | \lambda \>$ was introduced
earlier. There is one Riemann relation for any spin
structure $\lambda$ and we have the following relations between the
vectors $\zeta$ and $\zeta '$, expressed in terms of a matrix $\Lambda$,
which satisfies $\Lambda ^2 = I$ and $2 \Lambda$ has only integer entries,
\be
\left ( \matrix{
\zeta _1 ' \cr  \zeta _2 ' \cr \zeta _3 ' \cr \zeta _4 ' \cr} \right )
= \Lambda 
\left ( \matrix{
\zeta _1  \cr  \zeta _2  \cr \zeta _3  \cr \zeta _4  \cr} \right )
\qquad \qquad
\Lambda =
\half \left (\matrix{
 1 &  1 &  1 &  1 \cr 
 1 &  1 & -1 & -1 \cr 
 1 & -1 &  1 & -1 \cr 
 1 & -1 & -1 &  1 \cr} \right )
\ee 
When $\zeta_1=0$ and $\zeta _2 + \zeta _3 + \zeta _4=0$ then $\zeta '_1=0$
and the right hand side of the relation vanishes.

\section{The sign factor  for $\Gamma[\delta;\ep]$ for $\ep=\ep_2$}
\setcounter{equation}{0}
\label{secQ}

In \cite{ADP}, the expression $\Gamma[\delta_i^{\gamma \alpha};\ep]$ was obtained with
a sign factor $\xi$,
\bea
\Gamma[\delta_i^{\gamma\alpha};\ep]
=
i \xi
{\<\nu_0|\mu_i\>\tet_i(0,\tau_\gamma)^4
\over (2\pi)^7\eta(\tau_\gamma)^{12}}
{\tet[\delta_i^{\gamma\alpha}](0,\Omega)^2
\over \tet[\delta_i^{\gamma(-\alpha)}](0,\Omega)^2}
{\tet_j(0,\tau_\gamma)^4
\over
\tet[\delta_j^{\gamma+}](0,\Omega)^2
\tet[\delta_j^{\gamma-}](0,\Omega)^2}
\eea
where $\alpha$ can be $+$ or $-$.
The sign factor $\xi $ was calculated in \cite{ADP}, but the final result is not 
correct\footnote{The sign $\xi$ turned out to be immaterial for the conclusions of \cite{ADP}.}.
Here, we shall present a simplified and corrected version of the calculation of $\xi$, first for
the reference twist $\ep=\ep_2$ and then, by modular transformation, for general $\ep$. 

\subsection{Parametrization of spin structures in $\cD[\ep]$}

The full left block formula is intrinsic and invariant under shifts of the even spin 
structures by full periods. But the individual factors involve single powers of $\tet$
and are not intrinsic. Thus, we need to fix a convention for the even spin structures 
which includes any added full periods.  We follow the conventions of (\ref{B2}) and 
denote the six distinct odd spin structures by $\nu_a, \nu_b, \nu_c, \nu_d, \nu_e, \nu_f$
where $(a,b,c,d,e,f)$ is a permutation of $(1,2,3,4,5,6)$.
Any non-zero twist $\ep$ may be parametrized as the difference between two odd spin structures,
\bea
\label{Q1}
\ep = \nu _b - \nu_a
\eea
For  $\ep=\ep_2$ these are are $(a,b)=(2,4)$ or $(a,b)=(4,2)$.
Next, we parametrize the six even spin structures $\delta \in \cD [\ep_2]$ in terms of 
sums of three odd spin structures as follows,
\bea
\label{Q2}
\delta ^+ _i = \nu_a + \nu _c + \nu _d & \hskip 1in & \delta ^- _i = \nu_b + \nu _c + \nu _d
\no \\
\delta ^+ _j = \nu_a + \nu _c + \nu _e & \hskip 1in & \delta ^- _j = \nu_b + \nu _c + \nu _e
\no \\
\delta ^+ _k = \nu_a + \nu _c + \nu _f & \hskip 1in & \delta ^- _k = \nu_b + \nu _c + \nu _f
\eea
As was the case in \cite{D'Hoker:2001qp}, it is also here convenient to use a set of genus 1
even spin structures $\kappa_a$ for $a=1, \cdots, 6$, which is well-adapted to the 
parametrization that we use for the odd spin structures. In terms of the standard 
$\mu_2, \mu_3, \mu_4$ the spin structures $\kappa_a$  are defined by,
\bea
\kappa _1 = \kappa _2 = \mu_3
\hskip 0.5in 
\kappa _3=\kappa _4=\mu_4
\hskip 0.5in 
\kappa _5=\kappa _6=\mu_2 
\eea
We shall use both notations interchangeably, preferring whichever is more convenient
for the computation at hand. The odd spin structures may then be labeled as follows,
\bea
\nu_q = \left [ \matrix{ \kappa _q \cr \nu_0 \cr } \right ] 
\hskip 1in 
\nu_m = \left [ \matrix{ \nu_0 \cr \kappa_m \cr } \right ] 
\eea
where $q=1,3,5$ and $m=2,4,6$.
For the case at hand with $\ep=\ep_2$, we have $c=6$ in the conventions of (\ref{B2}), and we set,
\bea
\label{Q3}
\nu_d = \left [ \matrix{ \mu_i \cr \nu_0} \right ]
\hskip 1in 
\nu_e = \left [ \matrix{ \mu_j \cr \nu_0} \right ]
\hskip 1in 
\nu_f = \left [ \matrix{ \mu_k \cr \nu_0} \right ]
\eea 
where $(i,j,k)$ is a permutation of $(2,3,4)$. The $(i,j,k)$ notation allows us to consider different 
spin structures all at once.

\subsection{Formula for $\Gamma [\delta^+_i; \ep]$ up to overall sign}

The starting point is an expression for $\Gamma [\delta^+_i; \ep]$ obtained by combining 
formulas (5.13), (5.14), (5.15), and (5.16) of \cite{ADP}. The expression is formulated in split 
gauge where the points $q_1, q_2$ are related by $S_{\delta ^+ _i} (q_1, q_2)=0$, and is  given by,  
\bea
\label{Q4}
\Gamma [\delta^+_i , \ep] & \equiv &
- {i \over 4 \pi^3}  {\sigma (\mu_i,\mu_j) \over \tet _k ^4} 
\Z  S_{\delta _i ^-} (q_1,q_2)  S _{\delta_j ^+} (q_1,q_2)  S_{\delta _j ^-} (q_1,q_2)
\nonumber \\
\Z & = & - {C \over C_r ^2 C_s ^2} \cdot
{\tet [\delta ]^5 E(p_r ,p_s )^4 \sigma (p_r )^2 \sigma (p_s )^2
\over
\tet [\delta] (q_1 + q_2 - 2 \Delta) E(q_1,q_2) \sigma (q_1)^2 \sigma
(q_2)^2} \cdot {1 \over \M_{rs}^2}
\eea
Here,  $p_r = \nu _r + \Delta$ and $p_s = \nu _s + \Delta$ are two
arbitrary branch points, and $\nu_r, \nu_s$ their associated odd spin structures,
and $\cM_{rs}= \cM_{\nu _r \nu_s}$ was introduced in \cite{D'Hoker:2001qp}, and given here in (\ref{4d3}).
The exponential factors were also introduced in
\cite{ADP} and are given by,
\bea
\label{Q5}
C & = &   - \exp \{ -8 \pi i \nu _s '  \Omega \nu _r '  \}
\no \\
C _r^2 & =  & - \exp \{ - 2 \pi i \nu _r' \Omega \nu _r'  \}
\no \\
C _s^2 & =  & - \exp \{ - 2 \pi i \nu _s' \Omega \nu _s'  \}
\eea
Expressions in terms of $\tet$-constants are most easily obtained by placing
insertion points at branch points. If $q_1,q_2$ are in split gauge, then their
limits to branch points are such that  $\tet [\delta^+_i ](q_1+q_2-2 \Delta)$ vanishes. 
The Szego kernel $S_{\delta ^-_i}(q_1,q_2)$ then also vanishes, 
but their ratio remains  finite. The calculation of this limit was carried out
in \cite{ADP} as well, by choosing $q_1 \to p_r=p_c$ and $q_2 \to p_s=p_d$.
The result for $\Gamma [\delta^+_i; \ep]$ is as follows,
\bea
\label{Q6}
\Gamma [\delta ,\ep] =  i \, \kappa (i,j)  \, \kappa ' (i,j) \,  {\sigma (\mu_i,\mu_j) \over 8 \pi ^7 \tet _k ^4} 
{\tet [\delta^+_i ]^4
\over
\tet [\delta _i^+]^2 \tet [\delta _i^-]^2 \tet [\delta _j^+]^2 \tet [\delta _j^-]^2 }
\eea
where $\kappa$ and $\kappa '$ have been defined as follows,
\bea
\label{Q7}
\kappa (i,j) & \equiv &   {C K_1 K_3 K_4 \over C_c ^2 C_d ^2 K_2}
\no \\
\kappa ' (i,j)& \equiv &
 {\pi ^4 \M_{ab}  \over \M_{ac}  \M_{cd} \M_{da} } \cdot
\tet [\delta^+_i ]^3  \tet [\delta _i ^-]  \tet [\delta _j ^+]  \tet [\delta _j ^-] \tet [\delta _k ^+]  \tet [\delta _k ^-]
\eea
The   factors $K_1, K_2, K_3, K_4$ are defined by,
\bea
\label{Q8}
 \p_I \tet [\delta^-_i ] (\nu _c - \nu _d) & = & K_1 \p_I \tet [\nu_b] (0)
\no \\
 \p_I \tet [\delta^+_i  ] (\nu _c + \nu _d) & = & K_2 \p_I \tet [\nu_a] (0)
\no \\
\tet [\delta _j ^+] (\nu _c - \nu _d) & = & K_3 \ \tet [\delta _k ^-](0)
\no \\
\tet [\delta _j ^-] (\nu _c - \nu _d) & = & K_4 \ \tet [\delta _k ^+](0)
\eea
Next, we shall  calculate $\kappa$ and $\kappa '$.

\subsection{Calculation of $\kappa$}

The $K$-factors may be computed starting from the following basic formula,
\bea
\label{Q9}
\tet [\delta ](\Omega \rho ' + \rho '' ) =  \tet [\delta + \rho](0)
\exp \Big \{
- i \pi \rho ' \Omega \rho ' - 2 \pi i \rho ' ( \delta '' + \rho '') \Big  \}
\eea
One finds,
\bea
\label{Q10}
K_1 & = &  \exp \{
- i \pi (\nu _c - \nu _d)' \Omega (\nu _c - \nu _d)'
+ 2 \pi i (\nu _c - \nu _d)' \nu _b '' + 4 \pi i \nu _c ' \nu _d '' + 4 \pi i \nu _b ' \nu _d ''\}
\nonumber \\
K_2 & = & - \exp \{ - i \pi (\nu _c + \nu _d)' \Omega (\nu _c + \nu _d)'
+ 2 \pi i (\nu _c + \nu _d )' \nu _a ''\}
\nonumber \\
K_3 & = & \exp \{
- i \pi (\nu _c - \nu _d)' \Omega (\nu _c - \nu _d)'
+ 2 \pi i (\nu _c - \nu _d)' (\nu _a + \nu _d + \nu _e)'' \}
\nonumber \\
K_4 & = & \exp \{
- i \pi (\nu _c - \nu _d)' \Omega (\nu _c - \nu _d)'
+ 2 \pi i (\nu _c - \nu _d)' (\nu _b + \nu _d + \nu _e)'' \}
\eea
All dependence on $\Omega$ cancels, and we find,
\bea
\label{Q11}
\kappa (i,j) =  
- \exp  4 \pi i \Big \{ (\nu _c - \nu_d)' ( \nu _b + \nu_e)'' + \nu _d ' \nu _a ''
+ \nu _b ' \nu _d ''  \Big \} 
\eea
which takes the values $\pm 1$. To evaluate $\kappa$ further, we use the
parametrization of the odd spin structures given in (\ref{Q3}), and distinguish the 
case $(a,b)=(2,4)$ and $(a,b)=(4,2)$,
\bea
\label{Q12}
(a,b)=(2,4) & \hskip 0.7in &
\kappa (i,j) = + \exp \{ 2 \pi i (\mu_i'' + \mu_j'') + 4 \pi i \mu_i ' \mu_j '' \}
\no \\
(a,b)=(4,2) & \hskip 0.7in &
\kappa (i,j) = - \exp \{ 2 \pi i (\mu_i'' + \mu_j'') + 4 \pi i \mu_i ' \mu_j '' \}
\eea

\subsection{Calculation of   $\kappa'$}

The expressions given in \cite{ADP} are in terms of the even spin structures normalized 
as in (\ref{B2}). We shall work out here the cases when $(a,b)=(2,4)$ and $(a,b)=(4,2)$.
The first group of $\cM$-factors needed is as follows, 
\bea
\label{Q13}
\cM _{mn}  =  \pi^2  \sigma (\kappa_m, \kappa_n) 
\tet \left [ \matrix{\nu_0 \cr \nu_0 \cr } \right ] \prod _{i=2,3,4}  \tet \left [ \matrix{\mu _i \cr \kappa_p \cr } \right ]
\eea
where $(m,n,p)$ is a permutation of $(2,4,6)$, and by construction we have $\cM_{mn}=-\cM_{nm}$.
The second group of $\cM$-factors is given by,
\bea
\label{Q14}
\cM _{q m} = - \pi^2 
\tet \left [ \matrix{\kappa_q \cr \kappa_n \cr } \right ] \, \tet \left [ \matrix{\kappa_q \cr \kappa_p \cr } \right ]
\tet \left [ \matrix{\kappa_r \cr \kappa_m \cr } \right ] \, \tet \left [ \matrix{\kappa_s \cr \kappa_m \cr } \right ]
\eea
where $(q,r,s)$ is a permutation of $(1,3,5)$ and  $(m,n,p)$ is a permutation of $(2,4,6)$.
To obtain the relations between $\tet$-constants shifted by full periods, we use,
\bea
\label{Q15}
\tet [\delta ^+_i]  
& = & \tet \left [ \matrix{\mu_i + 2 \nu_0 \cr \kappa_b + 2 \kappa_a + 2 \kappa_6 \cr } \right ] 
= \tet \left [ \matrix{\mu_i \cr \kappa_b \cr } \right ] \, \exp \{ 2 \pi i \mu_i'  \} 
\no \\ 
\tet [\delta ^-_i]  
& = & \tet \left [ \matrix{\mu_i + 2 \nu_0 \cr \kappa_a + 2 \kappa_b + 2 \kappa_6 \cr } \right ] 
= \tet \left [ \matrix{\mu_i \cr \kappa_a \cr } \right ] \, \exp \{ 2 \pi i \mu_i' \} 
\eea
As a result, we find,
\bea
\label{Q16}
\tet [\delta^+_i ]^3  \tet [\delta _i ^-]  \tet [\delta _j ^+]  \tet [\delta _j ^-] \tet [\delta _k ^+]  \tet [\delta _k ^-]
= 
\tet \! \left [ \matrix{\mu_i \cr \kappa_b \cr } \right ]^3 \tet \! \left [ \matrix{\mu_i \cr \kappa_a \cr } \right ]
\tet \! \left [ \matrix{\mu_j \cr \kappa_b \cr } \right ] \tet \! \left [ \matrix{\mu_j \cr \kappa_a \cr } \right ]
\tet \! \left [ \matrix{\mu_k \cr \kappa_b \cr } \right ] \tet \! \left [ \matrix{\mu_k \cr \kappa_a \cr } \right ]
\eea
In computing $\kappa'$, all factors of $\tet [\delta]$-constants with $\delta$ normalized as in (\ref{B3})  
cancel, and only the overall sign of each factor needs to be retained. We find,
\bea
\label{Q17}
(a,b)=(2,4) & \hskip 0.7in & \kappa '(i,j) = +1 
\no \\
(a,b)=(4,2) &  & \kappa ' (i,j)= -1 
\eea
independently of $i,j$.

\begin{table}[htb]
\begin{center}
\begin{tabular}{|c|c||c|c|c|c|}   \hline
$i$ & $j$ & $\kappa (i,j) \kappa ' (i,j)$ & $\sigma (\mu_i, \mu_j) $ & $ \< \mu_i | \nu_0 \> $ & $\xi$
\\ \hline  \hline
3 & 4 & $-$ & + & + & $-$
\\ \hline 
4 & 3 & $-$ & $-$ & $-$ & $-$
\\ \hline 
3 & 2 & + & $-$ & + & $-$
\\ \hline 
2 & 3 & + & + & $-$ & $-$
\\ \hline 
4 & 2 & $-$ & + & + & $-$
\\ \hline 
2 & 4 & + & + & $-$ & $-$
\\ \hline 
\end{tabular}
\end{center}
\caption{Evaluation of $\xi$ }
\label{table:6}
\end{table}

\subsection{The final sign}

Combining the results of the preceding two subsections, we find that while $\kappa$ and $\kappa '$
separately change sign under the interchange of $a$ and $b$, their product is invariant, and we have,  
\bea
\label{Q18}
\kappa (i,j) \, \kappa ' (i,j) =  \exp \{ 2 \pi i (\mu_i'' + \mu_j'') + 4 \pi i \mu_i ' \mu_j '' \}
\eea
This sign is not intrinsic, but neither is the sign of $\sigma (\mu_i, \mu_j)$ which 
enters into the formula for $\Gamma [\delta ^+_i; \ep]$ of (\ref{Q6}). The following
combination, however, is intrinsic,
\bea
\xi = \kappa (i,j) \kappa ' (i,j) \sigma (\mu_i, \mu_j) \, \< \mu _i | \nu_0 \>
\eea
To evaluate $\xi$, we simply list its 6 possible values in Table \ref{table:6}. Thus, our final formula 
for $\Gamma [\delta^+_i ,\ep]$ is as follows,
\bea
\label{Q6a}
\Gamma [\delta^+_i ,\ep] =  - i \  {\< \mu _i | \nu_0\> \over 8 \pi ^7 \tet _k ^4} 
{\tet [\delta^+_i ]^4
\over
\tet [\delta _i^+]^2 \tet [\delta _i^-]^2 \tet [\delta _j^+]^2 \tet [\delta _j^-]^2 }
\eea


\end{document}